\documentclass[longauth]{aa}
\usepackage{natbib}
\bibpunct{(}{)}{;}{a}{}{,} % to follow the A&A style

\usepackage{gensymb}
\usepackage{lscape}
\defcitealias{Hollenbach.McKee1989}{HM89}
\usepackage{lastpage}

\newcommand*\kms{km s$^{-1}$ }

\newcommand{\msun}{\mbox{M}_\odot}
\newcommand{\macc}{\dot{M}_{\rm acc}}
\usepackage{graphicx}
\usepackage{rotating}

\usepackage[colorlinks=True,linkcolor=blue,citecolor=blue]{hyperref}
\usepackage{txfonts}

\begin{document} 
\title{JOYS: Launch and destruction of dust in protostellar jets. \\ The case of BHR71-IRS1 with JWST MIRI}
\titlerunning{JOYS: BHR71-IRS1 }
\authorrunning{Tychoniec et al.}
\author{Łukasz Tychoniec \inst{1}, 
Logan Francis \inst{1}, 
Maria Gabriela Navarro \inst{2},
Jakobus M. Vorster \inst{3}, 
Ewine F. van Dishoeck \inst{1, 4}, \\
Alessio Caratti o Garatti \inst{5}, 
Korash Assani \inst{6,7}, 
Valentin J.~M. Le Gouellec \inst{8,9}, 
Benoît Tabone \inst{10},
Pamela Klaassen \inst{11}, 
Adriaan G. M. Janssen \inst{1}, 
Kay Justtanont \inst{12},
Daniel Harsono \inst{13},
Pooneh Nazari \inst{14},
Simon D. Reyes-Reyes \inst{15}, \\
Katerina Slavicinska \inst{1,16},
Caroline Gieser \inst{15},
Tyler L. Bourke \inst{17},
Yao-Lun Yang \inst{18},
Brunella Nisini \inst{2}, 
Teresa Giannini \inst{2}, 
Henrik Beuther \inst{15},
R. Devaraj \inst{19},
Thomas~P.~Ray \inst{19},
Nashanty G. C. Brunken \inst{1},
Yuan Chen \inst{1}, 
Martijn L. van Gelder \inst{1}}
\institute{Leiden Observatory, Leiden University, PO Box 9513, 2300RA, Leiden, The Netherlands\\ \email{tychoniec@strw.leidenuniv.nl}
\and 
INAF - Osservatorio Astronomico di Roma, Via di Frascati 33, 00078 Monte Porzio Catone, Italy
\and 
Department of Physics, PO Box 64, FI-00014, University of Helsinki, Finland
\and
Max-Planck-Institut f{\"u}r Extraterrestrische Physik, Giessenbachstrasse 1, D-85748 Garching, Germany
\and 
INAF-Osservatorio Astronomico di Capodimonte, Salita Moiariello 16, I-80131 Napoli, Italy
\and
Department of Astronomy, University of Virginia, Charlottesville, VA 22903, USA
\and
Virginia Institute of Theoretical Astronomy, University of Virginia, Charlottesville, VA 22903, USA
\and
Institut de Cienci\`es de l’Espai (ICE-CSIC), Campus UAB, Carrer de Can Magrans S/N, E-08193 Cerdanyola del Vall\`es, Spain
\and
Institut d’Estudis Espacials de Catalunya (IEEC), c/Gran Capitá, 2-4, 08034 Barcelona, Spain
\and
Université Paris-Saclay, CNRS, Institut d’Astrophysique Spatiale, 91405, Orsay, France
\and
 UK Astronomy Technology Centre, Royal Observatory Edinburgh, Blackford Hill, Edinburgh EH9 3HJ, UK
\and
Department of Physics and Astronomy, Chalmers University of Technology, 412 96 Gothenburg, Sweden
\and
Institute of Astronomy, Department of Physics, National Tsing Hua University, Hsinchu, Taiwan
\and
European Southern Observatory, Karl-Schwarzschild-Strasse 2, 85748 Garching bei M\"unchen, Germany
\and 
Max Planck Institute for Astronomy, Königstuhl 17, 69117 Heidelberg, Germany
\and
Laboratory for Astrophysics, Leiden Observatory, Leiden University, PO Box 9513, NL 2300 RA Leiden, The Netherlands
\and
SKA Observatory, Jodrell Bank, Lower Withington, Macclesfield, Cheshire, SK11 9FT, UK
\and
Star and Planet Formation Laboratory, RIKEN Pioneering Research Institute, Wako-shi, Saitama, 351-0106, Japan
\and
School of Cosmic Physics, Dublin Institute for Advanced Studies, 31 Fitzwilliam Place, Dublin 2, Ireland
}
   \date{Received December 20, 2025 / Accepted April 13, 2026}

  \abstract
   {Protostellar winds can theoretically lift solids from the planet-forming disks, but direct evidence for launched dust has been scarce so far. Numerous atomic lines that are unique to mid-infrared (IR) wavelengths reveal refractories eroded from dust grains and provide information on wind properties in the earliest stages of the star formation process.}
   {We characterize the gas-phase composition, shock properties, and dust content of the jet from the Class 0 protostar BHR71-IRS1, one of the best cases of a resolved central jet inside a wide-angle wind.}
   {We present JWST MIRI-MRS spectral imaging of the inner 2000 au of the BHR71-IRS1 blueshifted side of the outflow. Atomic line intensities were compared to shock models to constrain the physical conditions and elemental abundances of the outflowing gas. Dust continuum maps were constructed from point spread function-subtracted cubes, and the spectral energy distribution of the dust was analyzed.}
   {The ionized central jet of BHR71-IRS1 is spatially resolved and imaged for the first time, revealing a unique inventory of refractory, volatile, and noble-gas fine-structure lines (Fe, Ni, Co, Cl, S, Ne, and Ar). The emission is concentrated along four bright knots that wiggle along the jet axis. Point spread function-subtracted continuum maps reveal extended mid-IR continuum emission co-spatial with the jet bullets and within the H$_2$-traced outflow cone. Spectral energy distributions along the jet were fit together with the extinction, revealing a warm (200-400 K) and a cold (70-90 K) dust component. The shock modeling constrained by the mid-IR lines indicates a decline in the shock velocity from 70 to 35 km s$^{-1}$ and in the pre-shock density from $>$10$^5$ to $ 4\times 10^4$ cm$^{-3}$ with distance from the protostar. Gas-phase Fe and Ni are measurably depleted relative to solar abundances. This is consistent with a substantial fraction of refractories remaining locked in grains in spite of the shocks.}
   {These JWST observations provide direct evidence that dust is launched in a Class 0 jet and at least partly survives shock processing. The richness of refractory tracers in the BHR71-IRS1 jet provides a window into the inner-disk composition at the onset of planet formation.}

\maketitle
   
\section{Introduction}

Protostellar winds serve as funnels through which angular momentum is removed from a system, aiding accretion onto the protostar-disk system \citep{Pudritz.Ouyed.ea2007, Frank.Ray.ea2014}. These winds are thought to arise from a magneto-centrifugal mechanism that is launched over a range of disk radii \citep[disk winds;][]{Pudritz.Norman1983a} or from the star-disk corotation radius \citep[X-winds;][]{Shu.Najita.ea1994a}. The innermost part of the wind constitutes a jet, that is, a fast $\gtrsim$100 \kms and highly collimated component of the wind, which is often seen as a series of emission spots that are referred to as bullets or knots \citep{Raga.Canto.ea1990}.

A jet can lift molecular, atomic, and partially ionized gas, and, where the conditions permit it, also dust from the disk surface. The survival of dust launched from the innermost part of the disk is limited by sublimation. As the wind propagates, it interacts with the interstellar medium (ISM), entraining ambient gas and dust. The protostellar outflow (a general term that describes launched and entrained material; these materials are often challenging to discern observationally) has two dust components: (i) dust lifted from the disk, and (ii) dust entrained from the surrounding envelope and the ISM. The launched dust in protostellar outflows, its properties, and its composition are topics of active research \citep[e.g.,][]{Nisini.CarattioGaratti.ea2002,Smith.Bally.ea2005, Dionatos.Nisini.ea2009, Cacciapuoti.Testi.ea2024}. 

The presence of dust in protostellar winds matters for the physics of the launched gas and for planet formation. For example, small dust can shield the gas from photodissociating radiation (UV) \citep{Tabone.Godard.ea2020} or serve as a formation surface for the H$_2$ molecules \citep{Hollenbach.McKee1979}. The study of dust grains in protostellar winds provides insights into the composition of the building blocks of planets. Some ejected grains can later rain down onto the disk and envelope, which has been suggested to explain thermally processed meteorites farther out in the disk \citep{Shu.Shang.ea2001,Davis.Alexander.ea2014}, and large grains in the outer envelope \citep{Galametz.Maury.ea2019, Tsukamoto.Machida.ea2021, Cacciapuoti.Testi.ea2024}. 

\begin{figure*}
    \begin{center}
\includegraphics[width=0.95\textwidth]{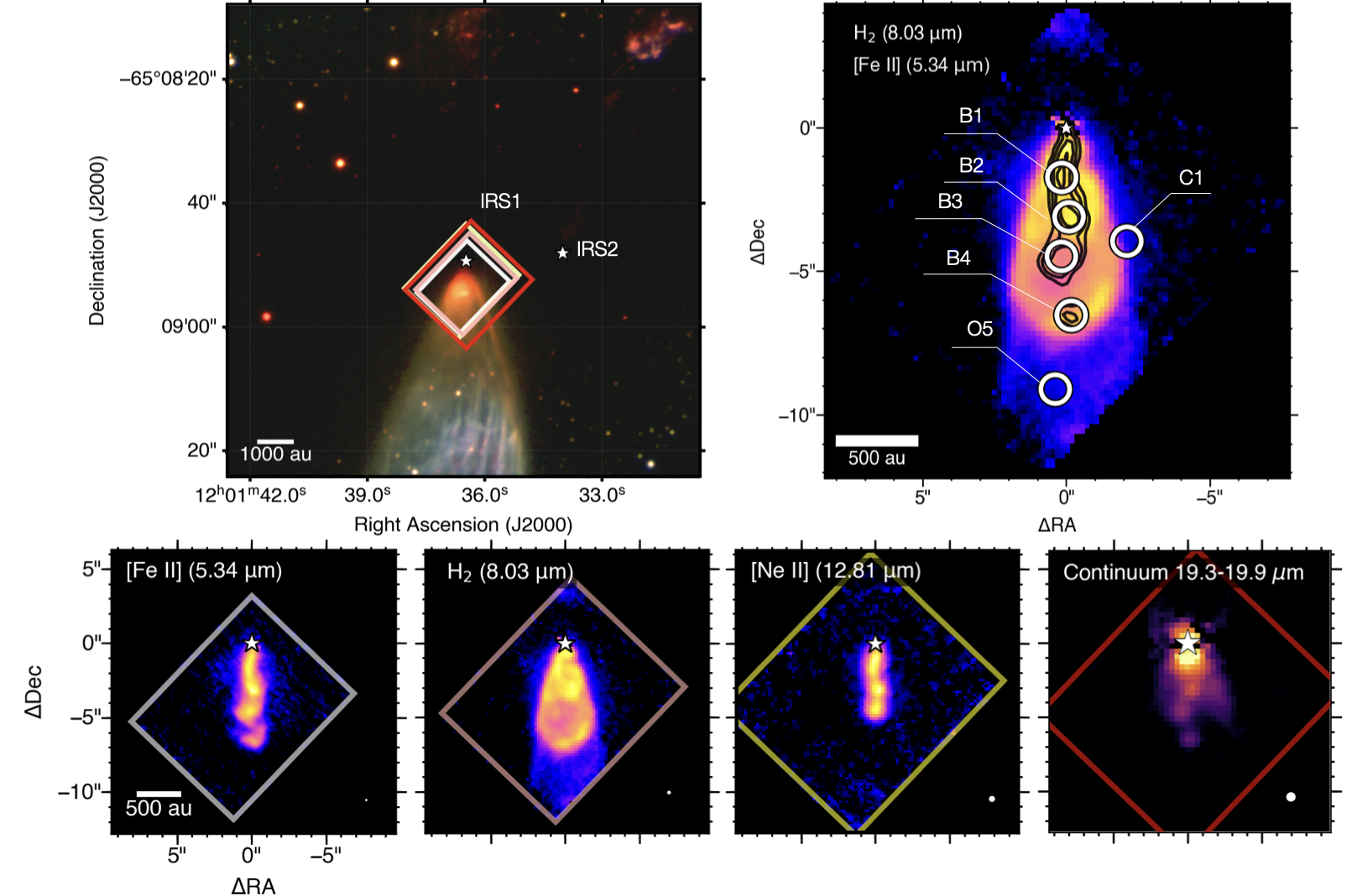}
        \caption{{\it Top left:} Large-scale view of the BHR71 globule in the $Ks$ (red) and $H$ (green) bands from the Persson Auxiliary Nasmyth Infrared Camera \citep[PANIC;][]{Martini.Persson.ea2004} taken on 2009 January 17 and 18 and in the $J$ (blue) band from the Infrared Side-Port Imager \citep[ISPI;][]{vanderBliek.Norman.ea2004} taken on 2009 June 11 (see also \citep{Tobin.Hartmann.ea2010,Tobin.Bourke.ea2019}). The stars mark the positions of the protostars from ALMA high-resolution images \citep{Ohashi.Tobin.ea2023}. 
   The colored rectangles highlight the field of view of the MIRI-MRS mosaics for channels 1 (white), 2 (pink), 3 (yellow), and 4 (red). {\it Top right:} MIRI-MRS 
        integrated Gaussian intensity map of H$_2$ S(4) (8.03 $\mu$m; color scale) and [Fe \textsc{ii}] a$^6$D$_{9/2}$-a$^4$F$_{9/2}$ (5.34 $\mu$m; black contours). 
        The circles show and label the regions we selected for spectral analysis. The coordinates are relative to the position of the IR protostar source, 12$^h$01$^m$36.454, $-65{^\circ}08\arcmin49\farcs267$ (J2000), which is indicated with the white star.
        {\it Bottom:}  Integrated Gaussian intensity maps of (from left to right:) [Fe \textsc{ii}] a$^6$D$_{9/2}$-a$^4$F$_{9/2}$ (5.34 $\mu$m; channel 1),
        H$_2$ S(4) (8.03 $\mu$m; channel 2), and [Ne \textsc{ii}] $^{2}$P$_{3/2}$-$^2$P$_{1/2}$ (12.81 $\mu$m; channel 3). The final panel on the right shows the thermal dust continuum emission from 19.3 to 19.9 $\mu$m (channel 4). The rectangles indicate the field of view of the MIRI-MRS mosaic, and the colors correspond to those in the top left plot. In the bottom right corners, the MIRI-MRS empirical FWHM of the PSF \citep{Law.Morrison.ea2023} is indicated as a white circle.}
        \label{fig:Figure1_overviewmaps}
        \vspace{-0.5cm}
\end{center}
\end{figure*}

The {\it James Webb} Space Telescope (JWST) Mid-Infrared Instrument (MIRI) enables us to directly detect faint thermal emission from warm dust. Crucially, the MIRI resolution and sensitivity enable us to decompose the entrained and launched grains. For example, {\it Spitzer} revealed warm dust (30--170~K) associated with shocked gas in the Class 0 HH-211 outflow \citep{Tappe.Lada.ea2008}, while MIRI has since shown continuum emission at the bow-shock position and directly in the jet beam \citep{CarattioGaratti.Ray.ea2024}.  Near- and mid-IR observations with JWST toward edge-on disks have also shown small grains launched with the disk wind and heated by stellar UV  \citep{Labdon.Kraus.ea2023,Duchene.Menard.ea2024,Dartois.Noble.ea2025}. 
Observations with the Atacama Large Millimeter/submillimeter Array (ALMA) revealed low spectral indices at millimeter wavelengths toward the outflow cavity walls, which were interpreted as grain growth. These grains can either grow at the overdensities in the cavity walls or be launched directly from the disk \citep{Sabatini.Podio.ea2024, Sabatini.Bianchi.ea2025}.

The refractory content of the gaseous jet revealed that the abundances of refractory species with respect to hydrogen are significantly lower than the solar abundances \citep{Nisini.CarattioGaratti.ea2002, Nisini.Bacciotti.ea2005, Podio.Bacciotti.ea2006, Giannini.Antoniucci.ea2015} and are consistent with model predictions in which grains survive typical shocks in protostellar jets \citep{Gusdorf.PineaudesForets.ea2008, Guillet.Jones.ea2009}. The refractory abundances also inform us about the composition of grain cores, such as Fe-Ni grains. This composition is otherwise difficult to determine. The depletion of refractory material in the gas phase depends on the shock velocity, which is consistent with the hypothesis that dust grains are destroyed more efficiently in faster shocks \citep{Giannini.Nisini.ea2019}. Alternatively, a low refractory abundance can be a sign of a low overall dust abundance because the dust was launched from a dust-free zone, or it might be the result of dust trapping in the disk by opened gaps \citep{Micolta.Calvet.ea2024, McClure.vantHoff.ea2025}.

Most previous studies of elemental abundances in jets were limited to Class II sources or studied the region along the jets of Class I protostars at large enough distances from the central source for the envelope to have been dissipated \citep[e.g.,][]{Nisini.CarattioGaratti.ea2002, Podio.Bacciotti.ea2006, Podio.Medves.ea2009, AgraAmboage.Dougados.ea2011}. JWST observations open the avenue for studying the gas content of the youngest jets in unprecedented detail because of the numerous atomic lines that uniquely occur at mid-IR wavelengths. Refractory species that are characterized by high sublimation temperatures ($>$~1000 K), such as Fe, Ni, and Co, are suitable tracers of dust-disrupting shocks, but can also be liberated from grains in the launch region, within the dust sublimation radius. Semi-refractory species such as Cl and S have lower sublimation temperatures, but can often exist in solids in various refractory forms (e.g., FeS), which are expected to be released much more easily into the gas phase. Of special interest are extremely volatile noble gases such as Ne and Ar, whose ionization potentials are higher than 13.6 eV; their presence in an ionized state indicates high temperatures and/or irradiation. JWST excels in detecting and mapping these tracers \citep{Tychoniec.vanGelder.ea2024, CarattioGaratti.Ray.ea2024,  Nisini.Navarro.ea2024, Assani.Harsono.ea2024, Narang.Manoj.ea2024, Federman.Megeath.ea2026}.

We present the first resolved images of the ionized jet of BHR71-IRS1 ([B2001b] IRS 1) \citep{Bourke.Hyland.ea1995, Bourke2001}. It is located in an isolated Bok globule associated with a larger region of Chamaeleon, whose distance is measured at 176$\pm7~$pc \citep{Voirin.Manara.ea2018}. The BHR71 globule is forming two stars: IRS1 and IRS2, which are separated by 16\arcsec or 2850 au. IRS1 is classified as a Class~0 protostar based on its bolometric temperature of 66 K, with a bolometric luminosity estimated at 10 L$_\odot$ \citep{Ohashi.Tobin.ea2023} and an envelope mass of $2.7\ {\rm M_{\odot}}$  \citep{Kristensen.vanDishoeck.ea2012}.
The inclination of the outflow with respect to the viewer's line of sight is estimated to be 50$\degree$ based on the modeled rotational axis of the envelope perpendicular to the outflow axis \citep{Yang.Evans.ea2020}.  The systemic velocity of cloud BHR71 is measured to be $-4.7 $~km~s$^{-1}$ \citep{Kristensen.vanDishoeck.ea2012}.

Previous works were based on modeling of {\it Herschel} observations and identified that bow shocks impacting on dense ambient material cannot explain the emission of [O \textsc{i}] and OH in this region. They also postulated that these species probably arise from a collimated central jet \citep{Benedettini.Gusdorf.ea2017}. However, the origin of these lines could not be discerned with the spatial resolution of {\it Herschel} ($\geq$~9\arcsec). Spectrally resolved {\it Herschel}-HIFI data showed extremely high-velocity emission from H$_2$O ($\pm$60 km s$^{-1}$), likely originating in a jet \citep{Kristensen.vanDishoeck.ea2012, Mottram.Kristensen.ea2014}. ALMA observations of BHR71 reveal a spectacular wide-angle bipolar outflow in low-$J$ CO and SiO lines \citep{Zapata.FernandezLopez.ea2018,Tobin.Bourke.ea2019, Gavino.Joergensen.ea2024}, but cannot reveal an ionized jet component and hot H$_2$ outflow. Detailed studies of the bow shock and outflow on larger scales have been conducted with {\it Spitzer} \citep{Neufeld.Nisini.ea2009, Giannini.Nisini.ea2011} and Stratospheric Observatory for Infrared Astronomy (SOFIA) \citep{Gusdorf.Riquelme.ea2015}.  

We present new JWST MIRI-MRS spectral imaging observations of the BHR71-IRS1 outflow at sub-arcsecond resolution (down to 50 au) that show remarkably bright refractory species and thermal dust emission along the jet axis. The H$_2$ emission shows a wide-angle component with a bow-like morphology around the collimated jet. The paper is organized as follows: Section \ref{sec:Methods} presents the JWST MIRI-MRS observations. Section \ref{sec:results} presents the results with emission line maps, dust continuum maps, spectra, and the extracted fluxes and line velocities. In Section \ref{sec:analysis} we provide an analysis of the dust emission in the jet, and we compare the atomic line properties to the shock models. In Section \ref{sec:discussion} we provide a discussion that includes the origin and properties of dust observed in the wind and a comparison of the jet properties with those of other protostars, and we place the refractory abundances in the context of the Solar System. In Section \ref{sec:summary} we present our conclusions. The appendix covers details of the PSF subtraction in Appendix \ref{sec:app_psf}, a basic model of the jet wiggling in Appendix \ref{sec:appendix_wiggle}, extinction and temperature measurements of the H$_2$ in Appendix \ref{sec:app_H2analysis}, details of the shock modeling in Appendix \ref{sec:appendix_hm89}, and additional figures and tables in \ref{sec:app_figs_tabs}.

\section{Methods}
\label{sec:Methods}

\subsection{Observations}
The data we present here were obtained with the Medium Resolution Spectroscopy mode \citep[MRS;][]{Wells.Pel.ea2015, Argyriou.Glasse.ea2023} of the Mid-InfraRed Instrument \citep[MIRI;][]{Wright.Wright.ea2015, Rieke.Ressler.ea2015, Wright.Rieke.ea2023} on board the {\it James Webb} Space Telescope \citep[JWST;][]{Rigby.Perrin.ea2023} as part of the program JWST Observations of Young ProtoStars (JOYS) \citep[PID 1290;][]{vanDishoeck.Tychoniec.ea2025}. 
The observations were taken on July 26, 2023, from 04:51 until 08:24 UTC. A $3\times3$ MIRI-MRS mosaic was obtained, centered on position 12$^h$01$^m$36$^s$.548 $-65{^\circ}08\arcmin53\farcs59$ (J2000), which was offset from the IRS1 source position on purpose and shifted on the less-extincted blueshifted lobe. The mosaic field of view was 11\farcs7 $\times$ 9\farcs5 in channel 1 and increased to 16\farcs0 $\times$ 13\farcs4 in channel 4 (see Fig. \ref{fig:Figure1_overviewmaps}, top left). The empirical point-spread function (PSF) full width at half maximum (FWHM) size increased from 0\farcs27 to 1\arcsec (48 to 176 au) based on in-flight measurements \citep{Law.Morrison.ea2023}. Three sub-bands were integrated with 36 groups in the FASTR1 integration mode, with two dithers in a negative direction optimized for the extended source. The integration time per sub-band was 200s, which means that 600s were needed to obtain a full spectrum from 4.9 $\mu$m to 28.6 $\mu$m for each mosaic pointing. This resulted in a total on-source time of 3 hours.
Prior to the science observation, a single-pointing sky background image was obtained on a nearby sky position of 12$^h$01$^m$30$^s$.72, $-65{^\circ}08\arcmin44\farcs60$ (J2000) on the same day from 03:38 to 04:51. Three sub-bands were integrated with 72 groups in the FASTR1 integration mode with no dithering.

\begin{figure*}
    \begin{center}
    \includegraphics[width=0.98\textwidth,trim={0.2cm 0.0cm 0cm 0cm}, clip]{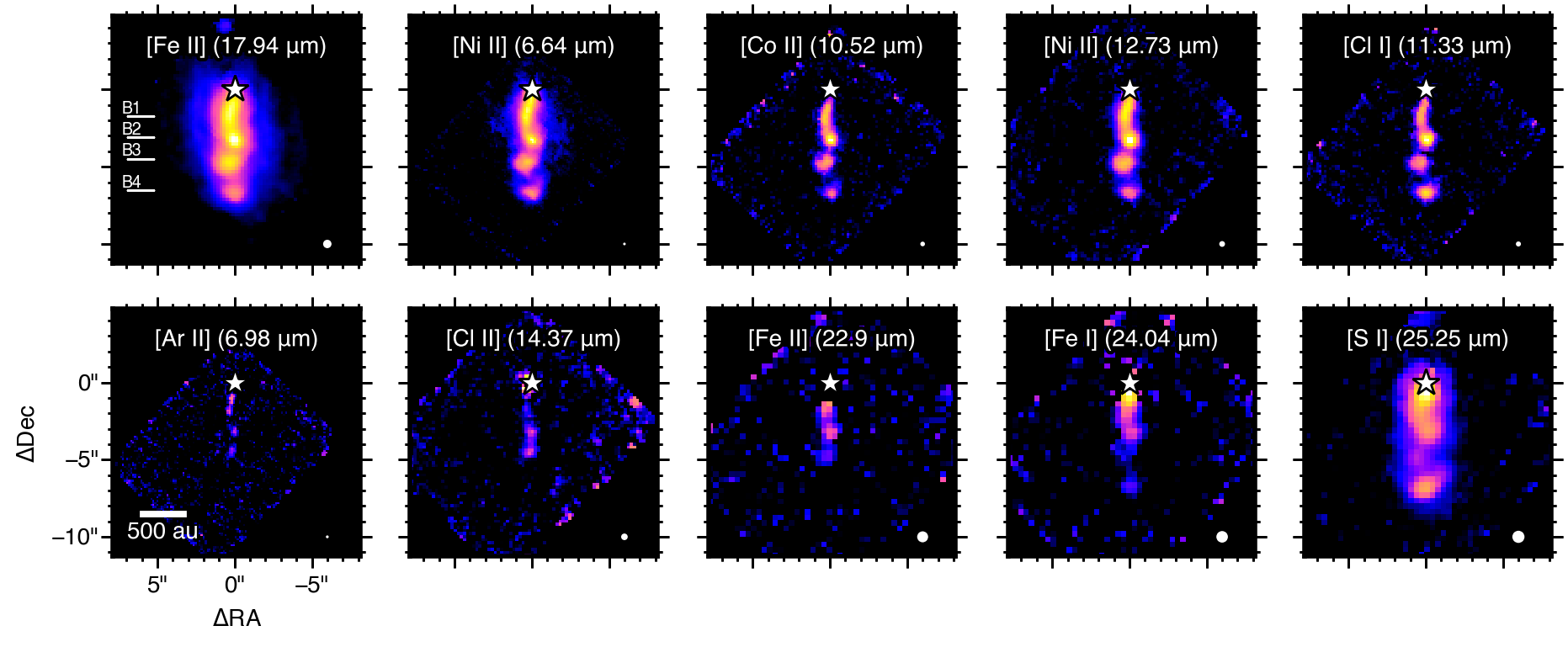}
        \caption{MIRI-MRS integrated Gaussian intensity maps of selected atomic and ionic emission lines. The coordinates are relative to the source position of the IR protostar, indicated with the white star. The line quantum identifiers are listed in Table \ref{tab:fluxes_atomic}. In the top left corner, regions identified as jet bullets are indicated. In the bottom right corners, the MIRI-MRS empirical FWHM of the PSF \citep{Law.Morrison.ea2023} is shown as a white circle.}
        \label{fig:Figure2_atomiclines}
\end{center}
\vspace{-0.2cm}
\end{figure*}

The data were reduced with the JWST pipeline version 1.17.1  \citep{Bushouse.Eisenhamer.ea2023} using the calibration reference data system \citep[CRDS;][]{Greenfield.Miller2016} context file \texttt{jwst$\_$1322.pmap}. We processed uncal files with the \texttt{Detector 1} pipeline with the default settings. 

The calibrated detector files were constructed in the \texttt{Spec2} step, including fringe flat for extended sources \citep{Crouzet.Mueller.ea2025} and residual fringe correction. At this stage, we assigned the world coordinate system (WCS) information. On the simultaneously obtained MIRI image, we measured offsets of the detected background stars using the Gaia DR3 catalog \citep{Gaia.DR3} and adjusted the MIRI-MRS WCS coordinates by the same offsets (0\farcs35 in R.A. and -0\farcs08 in Dec.). The background detector image was subtracted from the science detector images at this stage using dedicated background rate files. Faint H$_2$ emission lines detected in the background were masked and replaced using the package called vortex image processing (VIP) \citep{Christiaens.Gonzalez.ea2023}.
\texttt{Spec2.selfcal} was used to create a bad-pixel mask from science and background detector images.

The final data cubes for each channel and sub-band were created at the \texttt{Spec3} pipeline step with a drizzle algorithm \citep{Law.Morrison.ea2023}. Two types of cubes were created: one cube aligned with the R.A., Dec. coordinates for line images and analysis, and the other cube aligned with the detector (ifualign) for the optimal PSF removal and continuum analysis. The pipeline default \texttt{Spec3.outlier\_detection} step was used to mask any residual outliers in the spectra.

The absolute flux calibration uncertainty of MIRI-MRS was estimated from the flight performance at 5.6 $\pm$ 0.7 $\%$ \citep{Argyriou.Glasse.ea2023}. The cube is provided in barycentric reference frame, and therefore, a shift of $-$6.02 km s$^{-1}$ was applied to relate the observed velocities to the local standard of rest \citep{Schoenrich.Binney.ea2010}.

\subsection{Emission maps and spectral extraction}
\label{sec:spectralextraction}

To construct the line emission maps, we extracted spectra per pixel for a specified line and simultaneously fit a Gaussian function and a local continuum as a linear function. The resulting line flux per pixel (area under the fitted Gaussian) were plotted as integrated emission maps. For continuum emission maps, the point-like protostar and unresolved disk contributions were modeled with PSF supplied by the {\tt stpsf} package \citep{Perrin.Long.ea2025}. The details of the PSF subtraction are presented in Appendix \ref{sec:app_psf}. The continuum emission maps were produced by integrating a cube with a width of 0.6 $\mu$m centered on the specified wavelength. The full BHR71-IRS1 spectrum at the source position is presented in Fig. B.17 of \cite{vanGelder.Francis.ea2024}.

\section{Results}
\label{sec:results}

\subsection{Maps}

In Fig. \ref{fig:Figure1_overviewmaps} we present an overview map of BHR71-IRS1 as seen with MIRI-MRS, together with a larger near-IR view of the region. The BHR71 core contains two protostars, IRS1 and IRS2, which are marked in the composite near-IR image (Fig. \ref{fig:Figure1_overviewmaps}, top left). The narrow central emission prominent in the ionized species collimated component of the gas, which we refer to as jet, is resolved and imaged for the first time here. We refer in particular to [Fe \textsc{ii}] a$^6$D$_{9/2}-$a$^4$F$_{9/2}$ at 5.34 $\mu$m and to the [Ne~\textsc{ii}] $^2$P$_{1/2}$-$^2$P$_{3/2}$ line at 12.81 $\mu$m (bottom panel of Fig. \ref{fig:Figure1_overviewmaps}). The H$_2$ rotational transitions trace a wide-angle wind and/or entrained gas, which is shown in H$_2$~($0-0$)~S(4) in Fig. \ref{fig:Figure1_overviewmaps} (top right and bottom panel). The [Fe \textsc{ii}] jet does not extend as far south as the H$_2$ emission, which is clearly seen to extend beyond the field of view of the MIRI-MRS mosaic in Fig. \ref{fig:Figure1_overviewmaps} (top right). At the same time, [Ne \textsc{ii}] does not extend to the same distance as the [Fe \textsc{ii}] line. Extended continuum emission is detected toward the source, with a representative image at 19.6 $\mu$m included in Fig. \ref{fig:Figure1_overviewmaps} (bottom panel).

\begin{figure*}

\begin{center}
\includegraphics[width=0.99\textwidth]{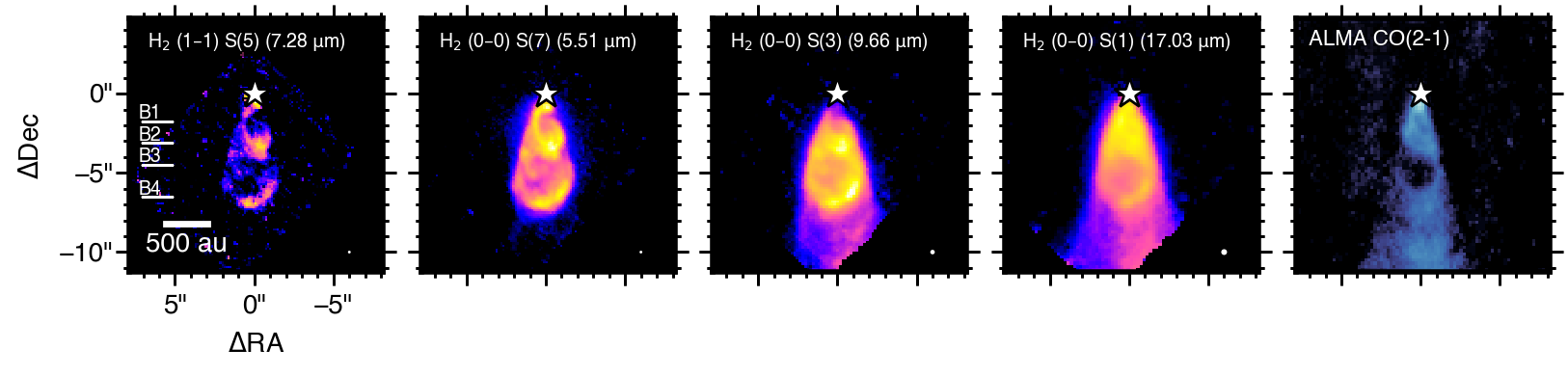}
 \caption{MIRI-MRS integrated Gaussian intensity maps of the representative H$_2$  emission lines of the BHR71-IRS1 outflow. From left to right, we show lines of increasing $E_{\rm up}$: H$_2$ $v=1$~--~$1$ S(5), 10340 K; H$_2$ $v=0$~--~$0$ S(7), 7197 K;  H$_2$ $v=0$~--~$0$ S(3), 2503 K; and H$_2$ $v=0$~--~$0$ S(1), 1015 K. In the bottom right corners, the MIRI-MRS empirical FWHM of the PSF \citep{Law.Morrison.ea2023} is shown as a white circle. In the rightmost image, ALMA CO (2-1) integrated emission over the entire blueshifted range (-85; 0 km s$^{-1}$) with respect to v$_{LSR}$ is shown in color scale.  The data presented in \cite{Gavino.Joergensen.ea2024} were taken in May 2021.}
        \label{fig:Figure3_h2maps}
\end{center}
\end{figure*}

\subsubsection{Atomic and ionic fine-structure lines}
\label{sec:finestructure_result}

We detected the nickel (Ni), iron (Fe), cobalt (Co), chlorine (Cl), sulphur (S), neon (Ne), and argon (Ar) atomic and ionic fine-structure lines in the BHR71-IRS1 jet. Selected lines of each detected species are presented in  Fig.~\ref{fig:Figure2_atomiclines}.

\begin{figure*}
    \begin{center}

    \includegraphics[width=0.85\textwidth]{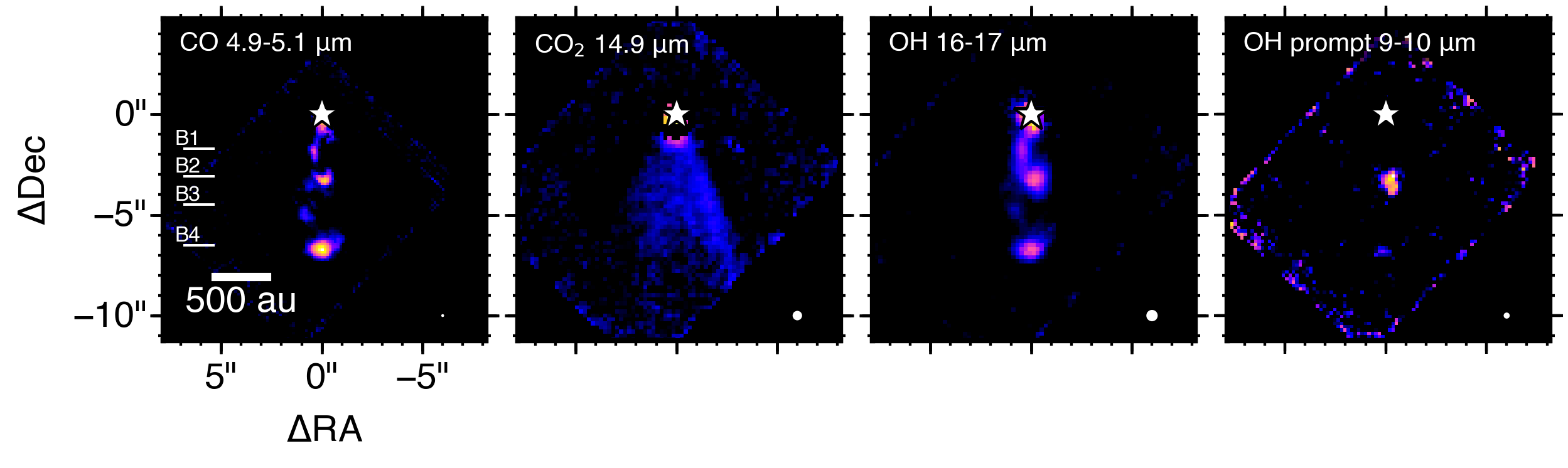}
        \caption{MIRI-MRS integrated Gaussian intensity maps of molecular lines. For molecules with multiple detected lines, the lines are stacked. {\it Left:} Rovibrational CO ($v=1$~--~$0$) map integrated from all unblended lines from channel 1 in 4.9 to the 5.1 $\mu$m range. In total, 14 lines are stacked, with $E_\textrm {up}=4725-7338$~K. {\it Left middle:} CO$_2$, only the Q branch at 14.98 $\mu$m ($E_{\textrm up}=964 $K)  is imaged. {\it Right middle:} OH rotational lines integrated over the 15.93-17.80 $\mu$m range. A total of 11 lines are stacked with $E_\textrm {up}=7097-14506$ K. {\it Right:}
        OH prompt emission lines from 9.13-9.92 $\mu$m. A total of en lines are stacked with $E_\textrm {up}=29624-48052$ K.}
        \label{fig:figure4_moleculesmaps}
\end{center}
\end{figure*}

The shape of all fine-structure lines is similar to that of the [Fe~\textsc{ii}] and [Ne \textsc{ii}] lines shown in Fig. \ref{fig:Figure1_overviewmaps}. They can be categorized into three groups: i) refractory or semi-refractory emission lines that are present along the full extent of the jet with a consistent trend of decreasing intensity away from the source ([Fe \textsc{ii}] 17.94 $\mu$m, [Ni \textsc{ii}] 6.64 $\mu$m, [Co~\textsc{ii}]~10.52 $\mu$m, [Ni \textsc{ii}] 12.73 $\mu$m, and [Cl \textsc{i}] 11.33 $\mu$m; Fig. \ref{fig:Figure2_atomiclines}, top row); ii) species that have a higher ionization potential than the other transitions covered here, with $\sim$13 eV, 16 eV, and 22 eV for Cl, Ar, and Ne, respectively \citep{NIST_ASD} that are only seen across the first three bullets and then terminate abruptly (e.g., [Cl~\textsc{ii}] 14.37 $\mu$m and [Ar \textsc{ii}] 6.98 $\mu$m, Fig. \ref{fig:Figure2_atomiclines}, bottom row; see also [Ne \textsc{ii}] in Fig. \ref{fig:Figure1_overviewmaps}, bottom panel).  [Fe \textsc{ii}] 22.90 $\mu$m follows a similar trend. While [Fe \textsc{ii}] has a much lower ionization potential of 7.9 eV, we note that this particular transition has $E_{\rm up}$ of 12073 K. This means that this group is characterized by a high ionization potential or high $E_{\rm up}$;
iii) lines detected at all bullets, but decreasing significantly in brightness at position B3 ([S \textsc{i}] 25.25 $\mu$m and [Fe \textsc{i}] 24.04 $\mu$m; Fig.~\ref{fig:Figure2_atomiclines}, bottom row). All tracers present in the jet oscillate along the symmetry axis of the outflow. They begin as continuous emission close to the source and then break up into a set of discrete emission features that we refer to as bullets B1-B4. The detailed analysis of the oscillation pattern is presented in Appendix \ref{sec:appendix_wiggle}.

\subsubsection{Molecular hydrogen (H$_2$)}

We detected 14 rotationally excited transitions of (H$_2$) of the $v=0$~--~$0$ and $v=1$~--~$1$ vibrational state. The four H$_2$ transitions presented in Fig. \ref{fig:Figure3_h2maps} were selected to show the change in line morphology with the excitation energy. The H$_2$ lines seen with MIRI-MRS have a wider opening angle than the collimated jet seen in the fine-structure lines. The H$_2$ images show two main features in their morphology: extended, diffuse emission that fills the outflow cavity, and spiraling, shell-like shapes that are highlighted in particular at the higher-excitation H$_2$ (1-1) S(5) (7.28 $\mu$m) line (Fig. \ref{fig:Figure3_h2maps}, leftmost panel). This is a unique feature of this flow that has not been seen before in H$_2$ in other sources \citep{Francis.Tychoniec.ea2026}, while high-velocity CO ALMA images of this outflow revealed similar shells \citep{Gavino.Joergensen.ea2024}. The peaks of these higher-energy lines coincide with bullets B2 and B4 of the jet (Fig. \ref{fig:Figure1_overviewmaps}, top right). The B3 position appears fainter across all H$_2$ transitions.

For the lines of high upper-energy levels $>$~7000 K, at around line S(7) at 5.51 $\mu$m (Fig. \ref{fig:Figure3_h2maps}, second from the left), the extent of the emission is confined to the southernmost shell-like structure, which is exactly at position B4, that is, the full extent of the collimated jet, while for lower-energy transitions, the H$_2$ emission extends beyond this and likely beyond the field of view, as shown by the near-IR image in Fig. \ref{fig:Figure1_overviewmaps} \citep{Tobin.Bourke.ea2019} and in {\it Spitzer} maps \citep{Neufeld.Nisini.ea2009, Giannini.Nisini.ea2011} (see also  Fig. \ref{fig:appE1_h2} for the remaining detected H$_2$ transitions).

\begin{figure*}
    \begin{center}
    \includegraphics[width=0.95\textwidth]{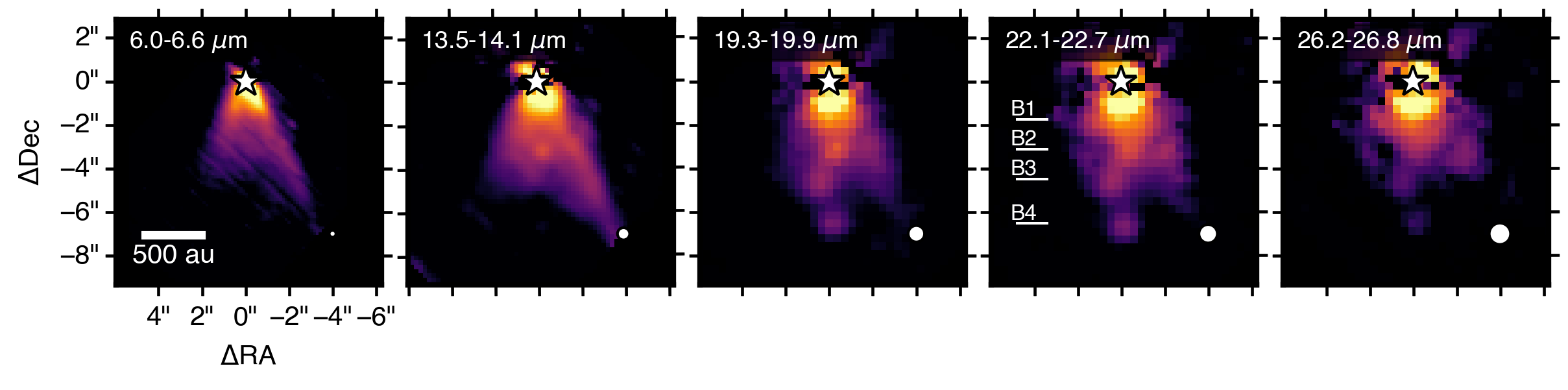}
        \caption{MIRI-MRS images of continuum emission from cubes centered on 6.3, 13.8, 19.6, 22.4, and 26.4 $\mu$m, from left to right. The images were generated by integrating cubes over 0.6 $\mu$m around the central wavelength. The cubes were PSF-subtracted before the imaging.   
        The white star marks the position of the protostar in the mid-IR.} 
        \label{fig:figure5_continuum_psfsub}
\end{center}
\end{figure*}

\subsubsection{Other molecular lines}

Several molecules were detected in the BHR71-IRS1 outflow with MIRI-MRS in their ro-vibrational and/or rotational transitions: CO, CO$_2$, and OH (Fig. \ref{fig:figure4_moleculesmaps}). CO ($v=1$~--~$0$) is detected at the brightest spots in the H$_2$ emission map and is notably strongest at the most distant shock position B4, where little ionized emission is present. CO$_2$ ($v=1$~--~$0$) traces a similar shape as the extended continuum emission, likely coming from sublimated ices from dust at the edges of the outflow cavity walls close to the protostar. The morphology of the OH rotational transitions is similar to that of CO and to fine-structure lines like [S \textsc{i}], with a decrease in brightness at B3, but a prominent presence in B4. The structure is similar overall to that of the spiraling jet and the H$_2$ shell emission. The OH prompt 9-10 $\mu$m emission, arising from the photodissociation of H$_2$O by UV photons and the subsequent instantaneous emission from the OH excited state \citep{Tabone2021}, is observed to be brightest at spot B2 and is far fainter in spot B4.  HCO$^+$ ($v=1$~--~$0$) 12.07 $\mu$m is also faintly detected at all positions except B3. Faint and cold (<$ 300$ K H$_2$O lines are also detected, but H$_2$O and HCO$^+$ are both too faint to image.

A detailed analysis of molecular lines at the source position, where SiO and H$_2$O are detected, was provided in \cite{vanGelder.Francis.ea2024}. SiO is not detected at the outflow positions.  HD lines have been detected in this source along the outflow and were analyzed in \cite{Francis.vanDishoeck.ea2025}. The ices along the line of sight will be analyzed elsewhere.

\subsubsection{Extended continuum emission maps}

One of the most striking features we found in the BHR71-IRS1 MIRI-MRS observations after we removed the protostellar PSF is the clear presence of the extended continuum emission (see Fig. \ref{fig:Figure1_overviewmaps}, bottom right). This emission could originate from heated dust or photons released by accretion, it might be scattered off dust, or a combination of the two. 

Fig. \ref{fig:figure5_continuum_psfsub}  shows the continuum maps at different wavelengths after the PSF subtraction. The maps were created by integrating 0.6 $\mu$m centered at 6.3, 13.8, 19.6, 22.4, and 26.5 $\mu$m. Two clear extended-emission components are detected: the first component, along the jet, is clumped into a series of bullets that is most clearly seen at longer wavelengths. The clumps are well aligned with positions B1-B4 seen in the gas lines of the jet. The second component is seen along the cavity wall and is asymmetric, brighter on the western side, similarly to the CO$_2$ emission, also in spatial extent (Fig. \ref{fig:figure4_moleculesmaps}, middle left). CO$_2$ and the corresponding continuum component both match the width of the low-excitation H$_2$ lines (Fig. \ref{fig:Figure3_h2maps}, right). Close to the source, the western side appears to be brighter in continuum than the eastern side, similarly to what is shown in the ro-vibrational CO map (Fig. \ref{fig:figure4_moleculesmaps}, left).
 
 At longer wavelengths, this component persists, suggesting that it is not only scattered light, but also thermal emission from warm dust. The component is also seen to fill the outflow cavity, and we therefore infer that this emission is not only warm dust shock- or UV-heated at the cavity walls, but also wind-entrained dust. The continuum emission along the jet is co-spatial with the bullets. Residual emission also appears to be close to the source and is not subtracted by the PSF removal. It might arise from a cold, extended envelope. 
\subsection{Spectra, line fluxes, and line ratios}
\label{sec:3.2.Spectra}

From the regions shown in Fig. \ref{fig:Figure1_overviewmaps} (top right), we extracted spectra using circular apertures with a diameter of 1\farcs0 at several positions: four bright ionized knots (bullets B1-B4), one bullet at the western side of the outflow cavity (CR1), and another farther along the outflow (O5), where no atomic jet is detected. The regions are presented in Fig. \ref{fig:Figure1_overviewmaps} (top right), and their coordinates are listed in Tab. \ref{tab:appE3_regions}. The resulting spectra are presented in Fig. \ref{fig:appE2_spectra} (for spectra extracted on-source, see \cite{vanGelder.Francis.ea2024}).

To obtain the total flux of the emission lines, we fit a linear function to the local continuum in the line-free channels to the 0.24 $\mu$m region, which is broad enough for a local continuum fit centered on the line rest wavelength, and a Gaussian profile to the emission line. From the area below the Gaussian divided by the aperture area, we obtained the line intensity. Since the region size was selected to fill the entire region, we assumed that no filling factor was required.  The Gaussian centroid provides information about the radial velocity offset of the line. In the case of a non-detection, we obtained the upper limit by using  three times the RMS derived from a line-free part of the spectrum, multiplied by the empirical FWHM of MIRI based on in-flight calibration data. This ranges from 75 to 294 km s$^{-1}$ for channels 1 to 4, respectively \citep{Argyriou.Glasse.ea2023}.
  In several cases, the emission lines are blended with the superimposed molecular emission (e.g., CO and H$_2$O) originating from the disk and/or outflow. There, we first fit a slab model to the continuum-subtracted spectra and subtracted it from the data before we fit the remaining emission lines (for details of the slab model fit, see \cite{vanGelder.Francis.ea2024}). 
The fluxes were corrected for extinction using H$_2$ derived values (see Appendix \ref{sec:app_H2analysis}). The measured line intensities and velocity offsets are reported in Table \ref{tab:fluxes_atomic}.

\begin{figure}
    \begin{center}
    \includegraphics[width=0.47\textwidth]{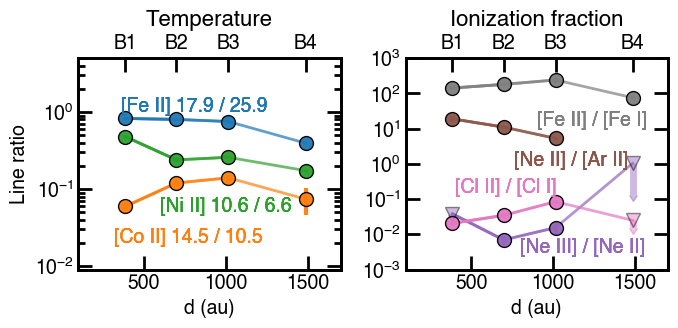}
    \caption{Line ratios of selected species at positions B1-B4 plotted as a function of the deprojected distance from the source.
    {\it Left:} Ratios of the same species sensitive to the gas temperature:
    [Co \textsc{ii}] 14.5 $\mu$m to [Co II] 10.5$\mu$m (orange),  [Fe \textsc{ii}] 17.9  $\mu$m  to [Fe \textsc{ii}] 25.9 $\mu$m (blue), and [Ni \textsc{ii}] 10.6  $\mu$m  to [Ni \textsc{ii}] 6.6 $\mu$m (green). {\it Right:} Ratios sensitive to the ionization fraction: 
    [Fe \textsc{ii}] 25.9$\mu$m to [Fe \textsc{i}] 24.0 $\mu$m (gray), [Ne \textsc{ii} 12.8 $\mu$m to [Ar II] $\mu$m 6.9 $\mu$m (brown), 
    [Cl \textsc{ii}] 11.3 $\mu$m to [Cl \textsc{i}] 14.4 $\mu$m. (pink), and [Ne \textsc{iii}] 15.55 $\mu$m to [Ne \textsc{ii}] 12.8 $\mu$m (purple).
    }
    \label{fig:fig6_ratios}
\end{center}
\end{figure}

 The variations in the physical conditions are reflected in the extinction-corrected line ratios for the selected species.  In Fig. \ref{fig:fig6_ratios} we compare the line ratios that are sensitive to the gas temperature for species with similar ionization potentials (left) and for the ionization fraction for species at comparable upper energy levels  (right). In addition, Fig. \ref{fig:appE0_peaks} presents the peak brightness as a function of distance from the protostar for selected lines and the continuum at 19.6 $\mu$m.

Fig. \ref{fig:fig6_ratios} (left) shows that the line ratio of two [Fe \textsc{ii}] lines at 17.9 $\mu$m ($E_{\rm up}=3496$ K) to 25.99 $\mu$m ($E_{\rm up}=554$ K)  is constant for bullets B1 to B3 and then decreases by a factor of 5 at the position of B4. For two transitions of [Ni \textsc{ii}] at 10.68 ($E_{\rm up}=13424$ K) and 6.63 $\mu$m ($E_{\rm up}=2168$ K), the ratio decreases from B1 to B2, then remains approximately constant from B2 to B3, and further decreases in B4. The [Co II] 14.56 $\mu$m ($E_{\rm up}$ = 5797 K) to [Co \textsc{ii}] 10.52 $\mu$m ($E_{\rm up}$ = 1368 K) ratio initially increases from B1 to B3 and then decreases in B4. Overall, this shows that the gas temperature remains roughly constant from B1 to B3, and it then drops significantly for B4. This is also supported by a non-LTE model, which for two ratios of [Fe II], 25.9 $mu$m to 17.9 $\mu$m and 17.9 $\mu$m to 5.3 $\mu$m, shows a gradual increase in the electron temperature from B1 to B3 and a sharp decrease for B4 (Fig. \ref{fig:D4})
  
 Fig. \ref{fig:fig6_ratios} (right) shows that three tracers of the ionization fraction, [Fe \textsc{ii}] (7.9eV) to [Fe~\textsc{i}], [Cl \textsc{ii}] (13 eV) to [Cl \textsc{i}], and [Ne \textsc{iii} (41 eV) to [Ne \textsc{ii}] (21.6 eV)
all increase from B1 to B3 and then decrease at B4. The exception is the line ratio of [Ne \textsc{ii}] (21.6 eV) to [Ar \textsc{ii}] (15.8 eV), which decreases from B1 to B3. We interpret this as still being consistent with the pattern of increasing ionization conditions from B1 to B3, since the decrease in [Ne \textsc{ii}] to [Ar \textsc{ii}] is likely an effect of [Ne \textsc{ii}] being further ionized to [Ne \textsc{iii}].
 
 We combined these pieces of information from the line ratios together and found that at the inner bullets (B1--B3), the gas experiences a high temperature: collisional excitation produces strong [Ne \textsc{ii}], [Ar \textsc{ii}], and high $E_{up}$ [Fe \textsc{ii}] emission. At B4, the conditions change abruptly and result in less excited gas and weaker emission of ions with respect to neutral species. The refractory variations along the jet are consistent with changes in the shock efficiency that disrupt grains. 

From the line brightness profiles shown in Fig. \ref{fig:appE0_peaks}, we also infer that closer to the protostar, that is, in the inner 300 au region, [Fe I] and [S I] are brightest, which suggests that the gas is launched in a neutral state, likely due to the high-density conditions, and is only ionized further at the internal working surface bullets. This increase in density is also supported by the electron density found from the [Fe II] line ratios in Fig. \ref{fig:D4}.

\begin{figure}
    \begin{center}
    \includegraphics[width=0.46\textwidth]{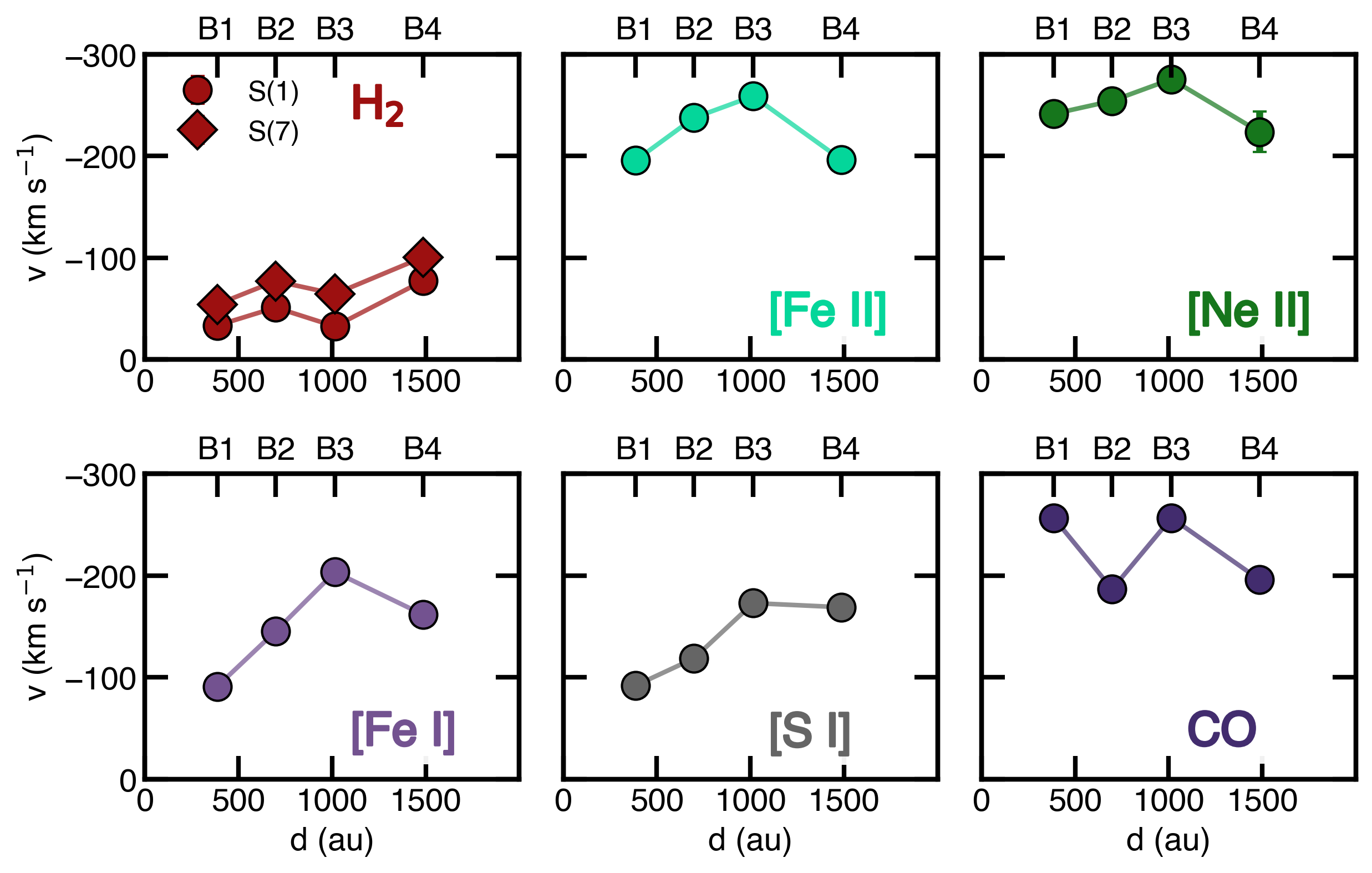}
    \caption{MIRI-MRS centroid velocity with respect to the $v_{LSR}$ of the protostar and corrected for source inclination for selected emission lines.}
    \label{fig:Fig8_atomic_vel}
\end{center}
\end{figure}

\begin{figure}
    \begin{center}
    \includegraphics[width=0.46\textwidth]{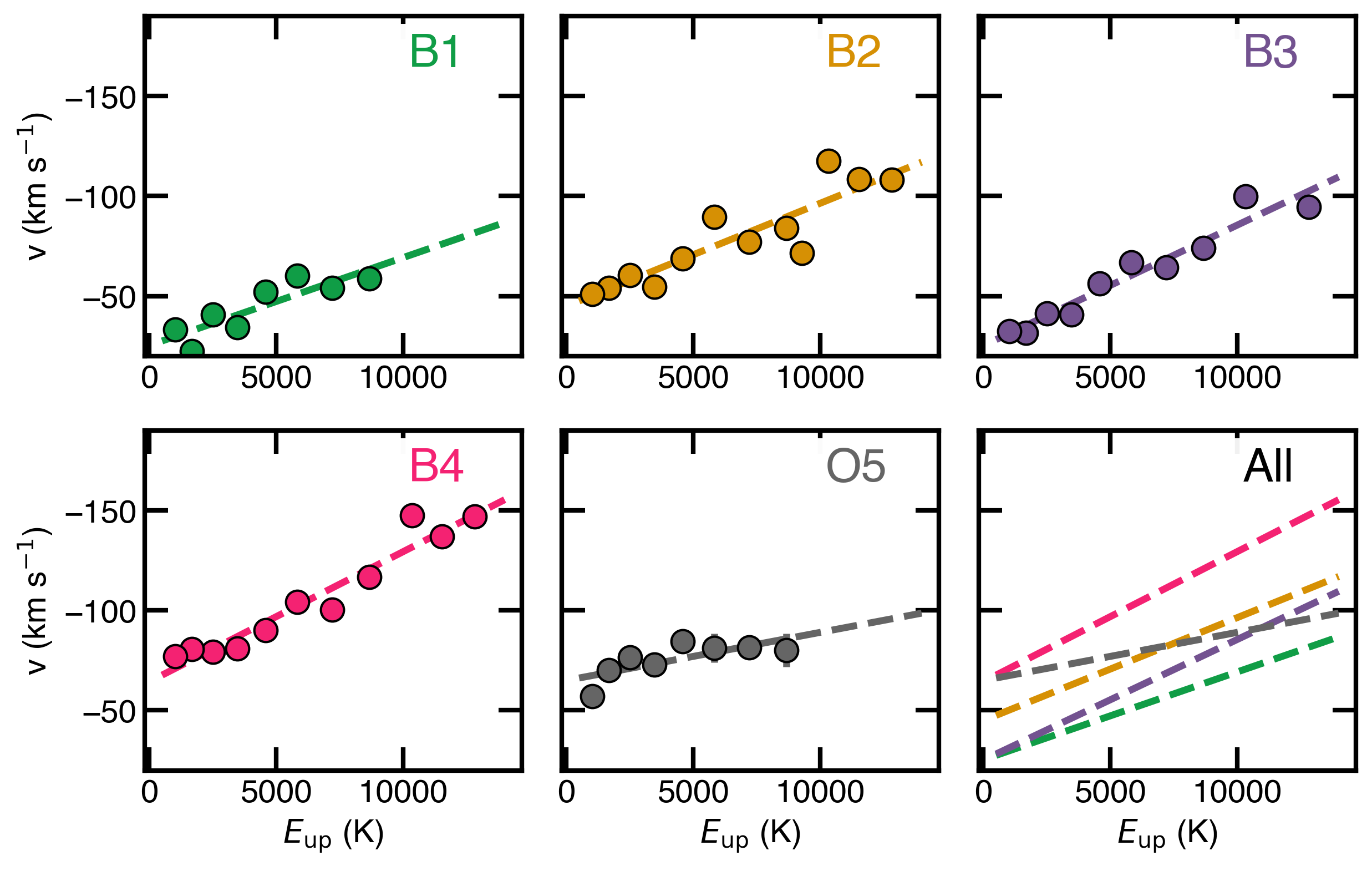}
    \caption{MIRI-MRS centroid velocity of H$_2$ rotational transitions with respect to the $v_{LSR}$ of the protostar and corrected for the source inclination as a function of $E_{ \rm up}$.  The color shows different positions along the jet (see Fig. \ref{fig:Figure1_overviewmaps}, top right).}
    \label{fig:Fig9_H2velocities}
\end{center}
\end{figure}

With the limited spectral resolution of MIRI, a detailed analysis of the kinematics is challenging. However, by fitting Gaussian profiles to the emission lines, we obtained information on the average velocity in the region from which the spectra were extracted. In Figure \ref{fig:Fig8_atomic_vel},we present the values of the centroid positions converted into the line velocity with respect to the rest wavelength of the transition measured at selected apertures B1-B4 along the jet for selected atomic, ionic, and molecular emission lines as a function of distance from the protostar. All values were corrected for inclination, set in reference to the LSR, and corrected for cloud velocity \citep[$v_{LSR}=$-4.7 km s$^{-1}$][]{Kristensen.vanDishoeck.ea2012}. For CO, we fit multiple emission lines from the ro-vibrational spectra, masking those contaminated by H$_2$ lines, for a total of 14 lines in 4.9 to 5.1 $\mu$m range, thereby improving the accuracy of this measurement. In Fig. \ref{fig:Fig9_H2velocities} we show for each bullet the H$_2$ rotational transition velocity as a function of the upper energy levels.

We identified two key features in the measured velocities. First, within a single aperture, the centroid velocity of the H$_2$ emission lines increases with the upper energy level of the transition (Fig. \ref{fig:Fig9_H2velocities}).  The second observed trend is an increase in velocity up to bullet B3, followed by a decrease in position B4 for [Ne \textsc{ii}] and [Fe \textsc{ii}], while [S \textsc{i}] remains constant (Fig. \ref{fig:Fig8_atomic_vel}). The average offset of CO lines shows a different pattern. The velocities oscillate between -240 km s$^{-1}$ and -180 km s$^{-1}$ in inclination-corrected radial velocities, and the trend of the oscillation is opposite to the trend seen in H$_2$. This can be also related to Fig. \ref{fig:fig6_ratios}, where B3 shows peak ionization, and \ref{fig:appE1_h2}, where [Ne \textsc{ii}]/[Fe \textsc{ii}] shows the highest ratio at this spot. All this indicates strong ionization.

The velocities of H$_2$ seem to oscillate from B1-B4, but show an overall increasing trend. The inclination is based on the overall source properties \citep{Yang.Evans.ea2020} and should therefore be treated as the average. The jet wiggling can affect the relative velocities. However, the absolute inclination uncertainty applies similarly to all lines and is therefore not relevant here.  

Taking the average flow velocity based on [Fe \textsc{ii}] 5.34 $\mu$m at all bullets, we can estimate the dynamical scale of the outflow. For the $v_{\rm mean}$ = $-222$ km s$^{-1}$ and distance to B4 of 1483 au, the averaged timescale of the jet is 32$\pm$1 years. On-source H$_2$O, CO$_2$, and SiO ro-vibrational lines are also detected with lower velocities of -42, -10, and -23 km s$^{-1}$, respectively, as reported in \cite{vanGelder.Francis.ea2024}.

\begin{figure}
\begin{center}
\includegraphics[width=0.46\textwidth]{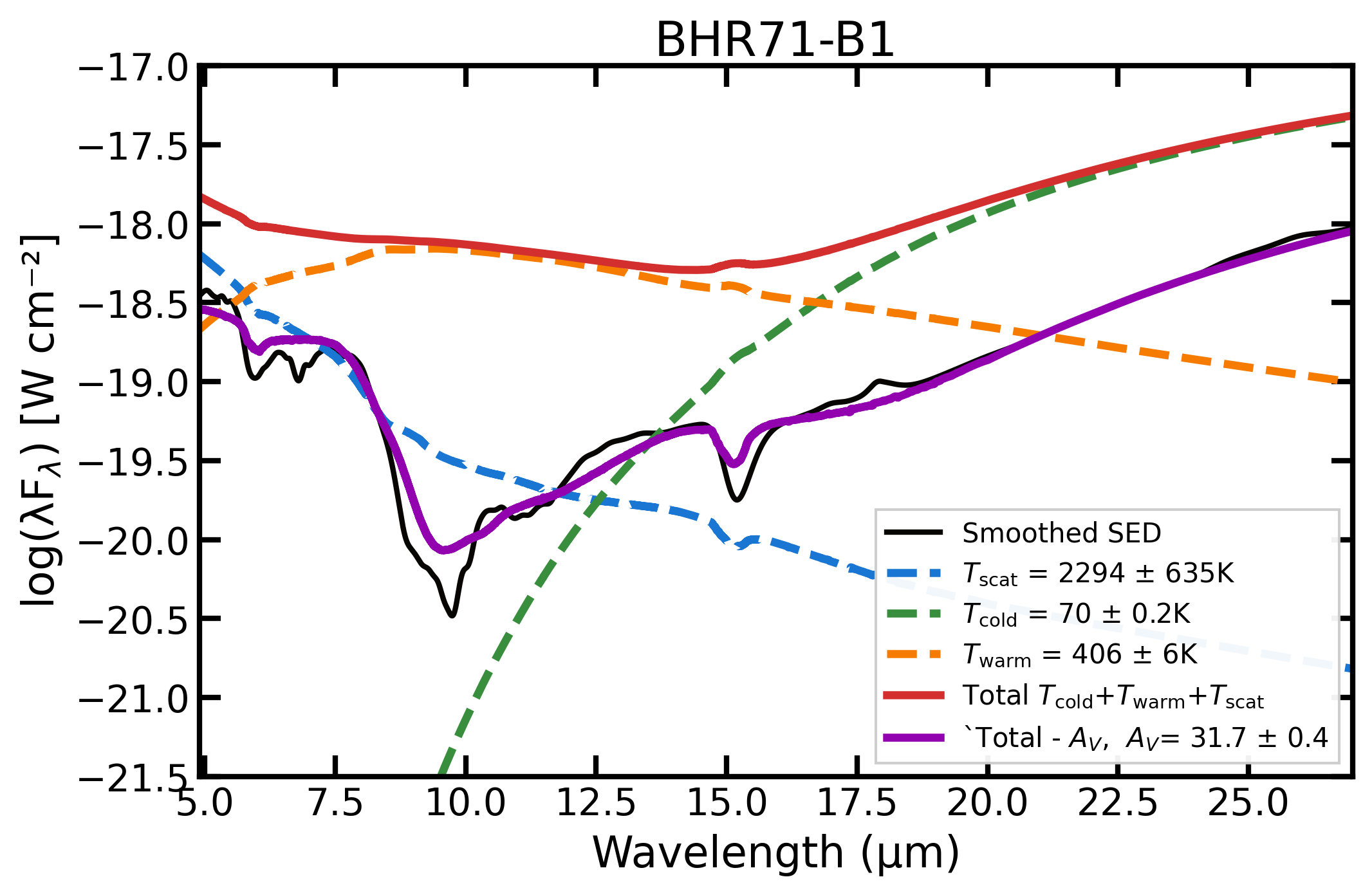}
\caption{Fits to the SED at position B1. The extracted SED was smoothed with a Gaussian kernel. The dashed lines indicate the scattered, warm, and cold dust components, shown in blue, orange, and green, respectively. The solid red line shows the sum of the three blackbodies. The dashed purple line shows the sum of the blackbody temperatures extincted with the KP5 extinction curve.}
\label{fig:sed_bhr71-irs1}
\end{center}
\end{figure}

\section{Analysis}
\label{sec:analysis}

\subsection{Characterization of the extended dust emission} 
\label{sec:extended_dust_4.1}

To characterize the extended dust emission from system BHR71-IRS1, we extracted the spectra as described in Section \ref{sec:3.2.Spectra}. We used the PSF-subtracted cubes to minimize the contribution of the central source to the extended continuum (Appendix \ref{sec:app_psf}). We modeled the dust continuum for each aperture by assuming a dusty emitting region extincted by intervening dust. We flagged line channels with the fast automatic baseline \texttt{fabc} algorithm in pybaselines \citep[$\sigma_{\rm scale} = $ 5 channels, $N_{\rm std} = 5$,][]{2022zndo...5608581E}. Then, we proceeded with uncertainty-weighted least-squares fitting with lmfit \citep{Newville.Otten.ea2025}. We assumed warm and cold emitting regions at temperatures $T_{\rm w}$ and $T_{\rm c}$, with angular areas $\Omega_{\rm w}$ and $\Omega_{\rm c}$. Both regions had dust with an absorption $\kappa_\nu^{\rm abs}$ and isotropic single-scattering $\kappa_{\nu}^{\rm sca}$ opacities (in cm$^{2}$ g$^{-1}$),
\begin{equation}
    F^{\rm dust}_{\nu} = \frac{\kappa_\nu^{\rm abs} [\Omega_{\rm c}B_\nu(T_{\rm c}) + \Omega_{\rm w}B_\nu(T_{\rm w})] + 2\kappa_\nu^{\rm sca} F^{\rm ps}_\nu}{\kappa_\nu^{\rm abs} + \kappa_\nu^{\rm sca}},
\end{equation}
where the scattered stellar radiation $F^{\rm ps}_{\nu}$ is a blackbody, subtending an angular size $\Omega_{\rm ps}$ from the perspective of the emitting region,   $F^{\rm ps}_{\nu} = \Omega_{\rm ps} B_\nu(T_{\rm ps})$.
The dust emission $F_{\nu}^{\rm dust}$ was then screened by a dust layer with surface density $\Sigma$ (g cm$^{-2}$), with an optical depth $\tau_{\nu} = \Sigma(\kappa_\nu^{\rm abs} + \kappa_{\nu}^{\rm sca})$, leading to an observed flux density of $F_\nu^{\rm obs} = F^{\rm dust}_\nu e^{-\tau_\nu}$.
We used the KP5 dust model for our opacities \citep{Pontoppidan.Evans.ea2024}. In the $5 - 27$ $\mu$m range, scattering cannot be ignored, and optically thin dust cannot necessarily be assumed. We assumed isotropic scattering, as including anisotropic scattering is a constant factor that is absorbed in $\Omega_{\rm ps}$ from an unknown inclination angle. The surface density $\Sigma$ allowed us to estimate the mass along the line of sight that caused the extinction. It does not distinguish between ISM dust in the line of sight, the envelope, or possible dust in the cavity. However, due to the large variations in $A_{\rm V}$ on the spatial scales of MIRI MRS, we considered the envelope mass to be the dominant contributor to $\Sigma$. We reserve a detailed dust modeling (composition, size distribution, and porosity) for future work.
The fit for the B1 position is presented in Fig. \ref{fig:sed_bhr71-irs1} and is shown for the remaining positions in Fig. \ref{fig:A.3_SED}.

\begin{figure}
\begin{center}
\includegraphics[width=0.48\textwidth]{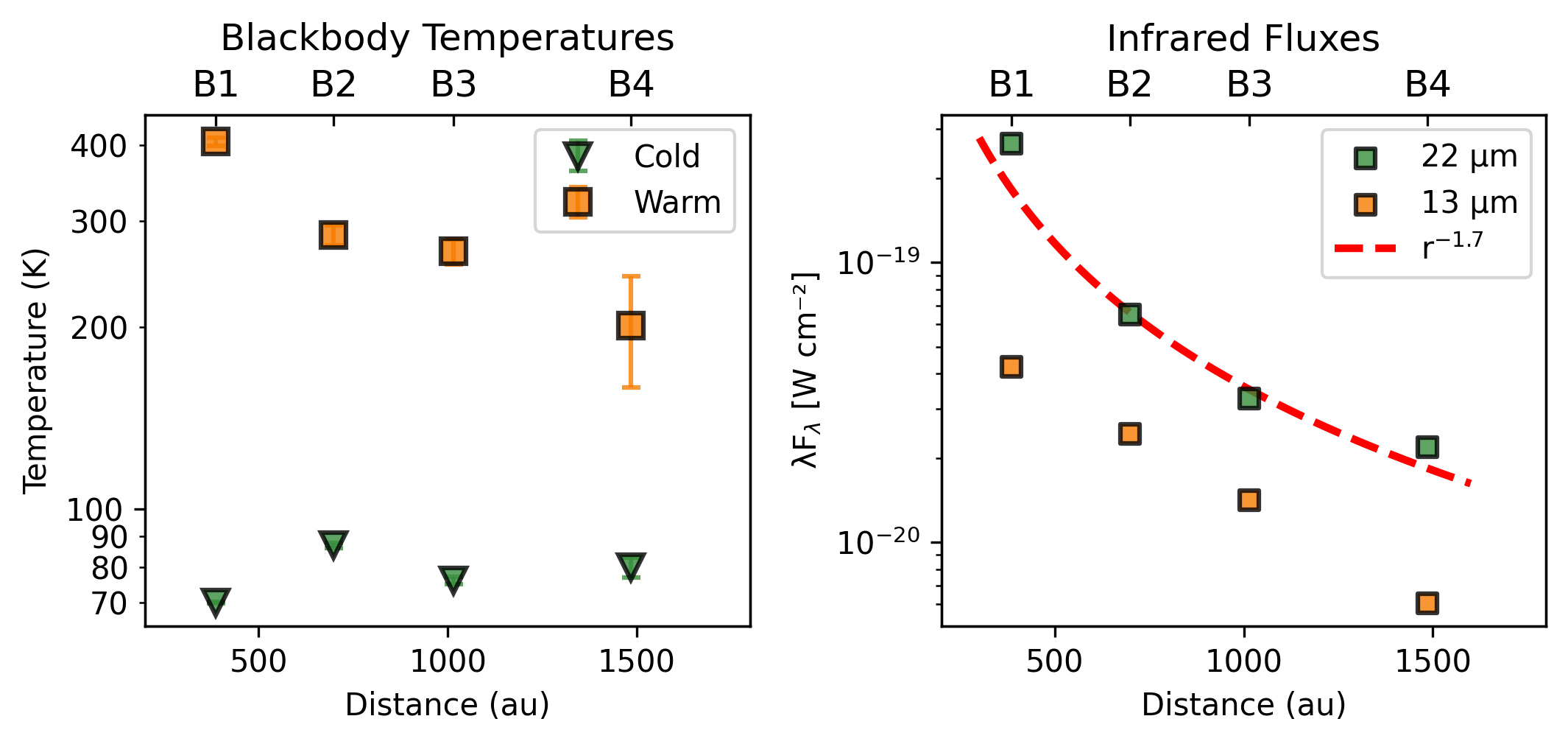}
\caption{{\it Top:} Fitted dust temperatures to warm and cold blackbody components for the SEDs in Fig. \ref{fig:sed_bhr71-irs1} and \ref{fig:A.3_SED}. {\it Middle:} IR fluxes at 13 and 22 $\mu$m as a function of distance from the protostar. The modeled envelope density as a function of radius is shown as the dashed red line \citep{Kristensen.vanDishoeck.ea2012}. {\it Bottom:} Surface density of the extincting dust as a function of distance from the protostar. The dashed red line same as in the middle panel. }
\label{fig:fig11_DustProps}

\end{center}
\end{figure}

The measured dust properties, that is, temperature, IR flux, and surface density of the extincting dust, are summarized in Fig. 
\ref{fig:fig11_DustProps} and Tab. \ref{tab:appE1_Dust}. In the top panel, we show the temperature of the warm and cold dust components as a function of distance from the protostar. The temperature of the cold component is constant with distance, at approximately 80 K. In contrast, the warm component has a clearly decreasing trend from 406 $\pm$ 6 K at B1 to 201 $\pm$ 42 at B4. In the bottom panel, we plot fluxes at 13 $\mu$m and 22 $\mu$m, where the emission is dominated by the warm and cold components, respectively. Both decrease with distance from the protostar, but the brightness of the cold component decreases more sharply. This is consistent with the power-law decrease in the envelope density modeled for this source \citep[$\rho\sim r^{-1.7}$;][]{Kristensen.vanDishoeck.ea2012}. Together with the temperature of the cold component (similar to $T_{\rm bol}$ = 66 K), this might indicate that at least part of the flux might stem from the extended envelope an not from the jet itself. 

 We next investigated whether the continuum emission observed in the jet and at the outflow cavity wall might be dominated by scattering. Based on the brightness profile as a function of wavelength, we conclude that it is dominated by thermal emission and not by scattering. Fig. \ref{fig:figure5_continuum_psfsub} shows the evolution of the dust emission as a function of wavelength. Bullets only appear as continuum emission components above 13 $\mu$m. As shown by \cite{Duchene.Menard.ea2024}, large grains show a flat scattering opacity at mid-IR wavelengths even up to 20 $\mu$m \citep[see also ][]{Kataoka.Okuzumi.ea2014}. Therefore, the bullets would have to show a similar brightness across wavelengths.  On the other hand, in the case of emission at the outflow cavity walls, where emission persists across wavelengths, the scattering can be a factor. The fitting presented here likely underestimates the scattering since we implemented large grains in the fit. Large grains ($\sim$10 $\mu$m) at the outflow cavity walls of BHR71 have been suggested before by polarization studies \citep{Hull.LeGouellec.ea2020}. In summary,  for the cavity walls, the scattering might be a significant factor and might cause us to overestimate the temperature of the grains in the envelope. However, scattering is unlikely to contribute significantly to the emission in the jet bullets.

The warm-dust component cools from 406 to 201 K on a scale of 1000 au. If the temperature decreased linearly, the dust would have a temperature above 450 K at the launch point. This estimate does not include additional heating. In the absence of a heat source, dust would cool on timescales of seconds to days \citep{Draine.Anderson1985}, and therefore, the presence of warm dust at such distances would require additional heating. Based on the average velocity of the [Fe \textsc{ii}] 5.34 $\mu$m (see Section \ref{sec:3.2.Spectra}), the travel time to B4 from the source would be 32$\pm$1 years, and it would be 23$\pm$1 years from B1 to B4. 

The surface density of the dust causing the extinction decreases with distance from the protostar. A decrease is consistent with the lower density of the envelope with radius, comparable with the model \citep[Table C.1 in ][]{Kristensen.vanDishoeck.ea2012}. The surface density can be directly converted into extinction, yielding values of 32, 15, 12, and 7 mag for the bullets B1-B4. The values are systematically lower than those measured with H$_2$. 

In summary, we identified warm (200-400 K) and cold (80~K) emitting components of the dust in the BHR71-IRS1 wind in the collimated and wide-angle components. The majority of the dust is expected to be launched directly by the wind, but detailed radiative transfer modeling is necessary to identify its physical origin.

\subsection{Shock models: Velocity, density, and elemental abundances}
\label{sec:ShockModels_main}

\begin{figure}
    \begin{center}
    \includegraphics[width=0.48\textwidth]{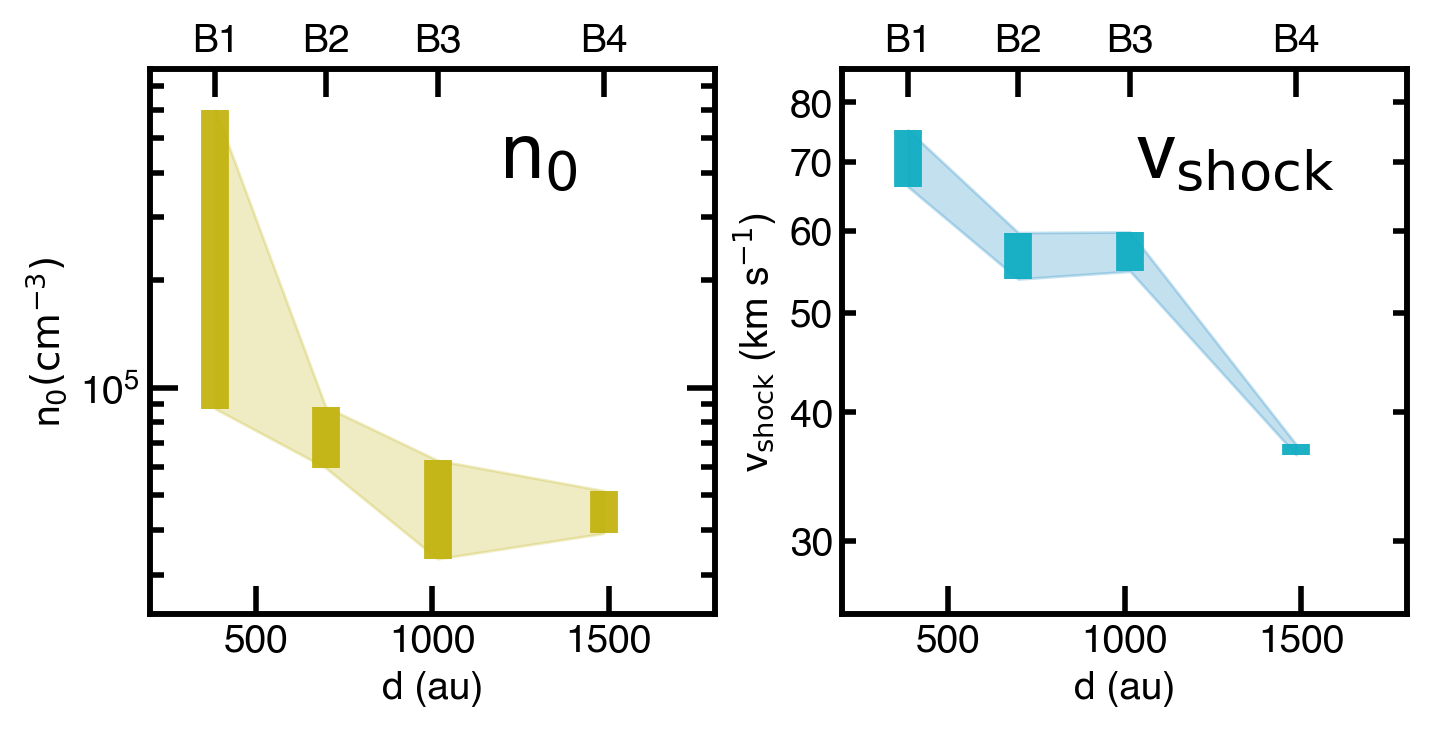}
        \includegraphics[width=0.48\textwidth]{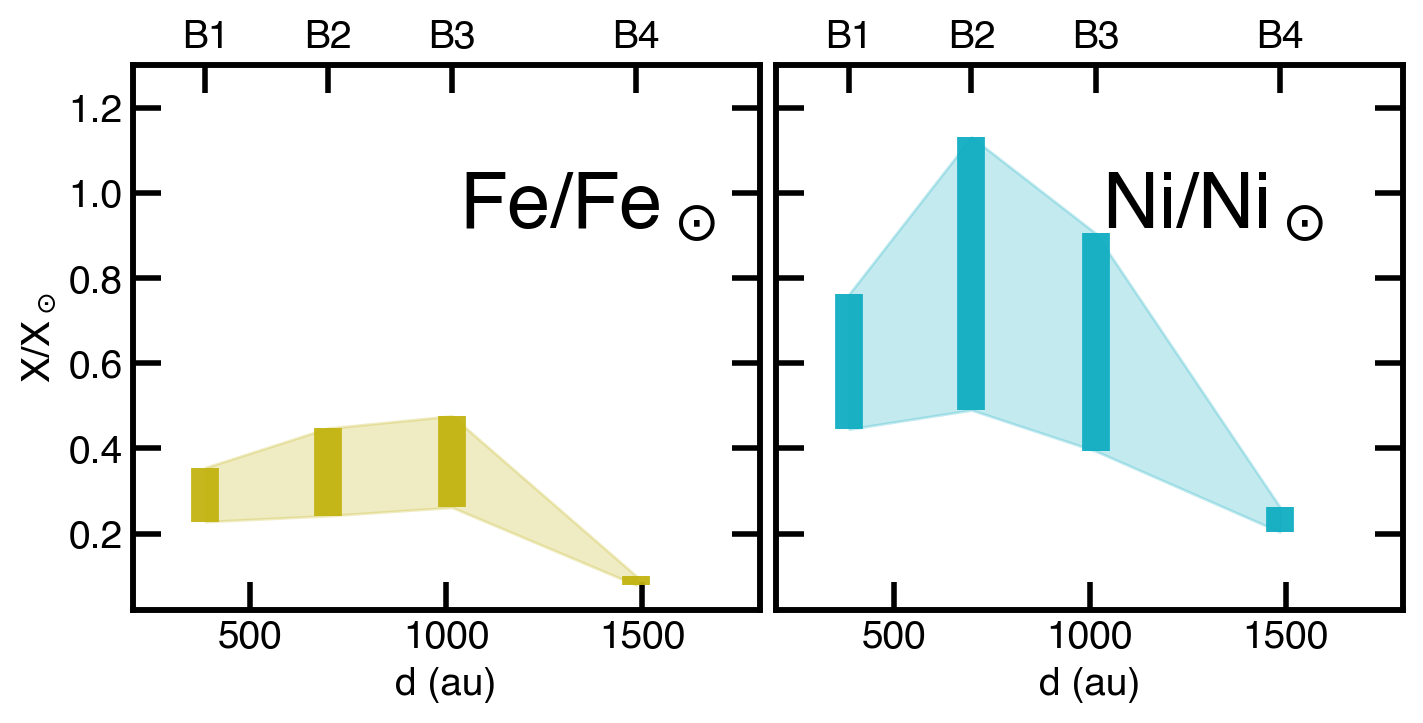}
        \caption{{\it Top:} Properties of the shock along the jet. Positions B1-B4 are indicated in Fig. \ref{fig:Figure1_overviewmaps}, and the shock modeling results are summarized in Tab. 
        \ref{tab:Hollenbach}. The range of values stems from the density estimate based on S and Cl.        {\it Bottom:} Fe and Ni abundance at the B1-B4 shock positions based on the shock model as measured in Appendix \ref{sec:appendix_hm89}. }
        \label{fig:shockprops_compared}
\end{center}
\end{figure}

We compared the extracted line intensities with the shock model of \citetalias{Hollenbach.McKee1989}. The procedure is detailed in Appendix \ref{sec:appendix_hm89}.
 The results are presented in Tab. \ref{tab:Hollenbach} and are illustrated in Fig. \ref{fig:shockprops_compared} with ranges coming from the differences between the [S \textsc{i}] and [Cl \textsc{i}] estimates. The shock velocity decreases from above 70 km s$^{-1}$ to 35 km s$^{-1}$ from bullet B1 to B4, with a constant shock velocity of 60 between bullets B2 and B3. The pre-shock density across the jet also decreases with distance from the protostar from above 10$^{5}$ cm$^{-3}$ to 4 $\times$ 10$^{4}$ cm$^{-3}$, except for bullet B4, which shows a small increase in the pre-shock density compared to a position closer to the protostar. 

We then used the derived conditions in the shock models to derive the jet mass ejection rate and also to assess the elemental depletion. The density of the pre-shock gas decreases with distance from the protostar, while at the same time, the (radial) velocity of the flow increases with distance, which results in a relatively constant mass-loss rate of $1.7-3.4\times10^{-7}$  M$_\odot$ yr$^{-1}$. When the density estimate from [Cl \textsc{i}] is used instead, a range of $1.5-23.3\times10^{-7}$  M$_\odot$ yr$^{-1}$ is measured. 

\subsection{Shock models: Elemental abundances}

Based on the estimated shock properties, we assessed the depletion of different species relative to solar values. The pre-shock density and shock velocity provide an expected line intensity for a specific tracer. The difference between the measured intensity and the value predicted by the model is assumed to be due to depletion of the species in the gas phase, as indicated by a ratio of species to atomic sulphur, which is expected to be volatile. It can therefore provide a reference of solar values as an undepleted species. We also assumed that sulphur was predominantly in atomic and not in molecular form. Higher values for Cl, especially above 1, can be attributed to the fact that some of the chlorine might be in neutral form, given its relatively high 23.81 eV ionization potential, or that Cl is less refractory than Fe, Ni, and Co. Co depletion would be of great interest, but the transitions of [Co \textsc{ii}] are not included in \citetalias{Hollenbach.McKee1989}. Overall, the significant depletion of refractories relative to solar values suggests that even under strong shocks, dust survives. This agrees with models for C-type and J-type shocks \citep{Guillet.Jones.ea2009,Gusdorf.PineaudesForets.ea2008}. If the dust we observe is lifted from the disk, its presence also means that it comes from beyond the dust-sublimation zone.

\section{Discussion}
\label{sec:discussion}

\subsection{Extended thermal dust emission}

While there is evidence of dust in protostellar winds through their refractory abundances ( \citep[e.g.,][]{Podio.Bacciotti.ea2006, Dionatos.Nisini.ea2009}; MIRI imager scattered-light observations \citep{Duchene.Menard.ea2024}, ALMA continuum observations of the low-emissivity suggestive of large grains possibly launched by the wind \citep{Sabatini.Podio.ea2024, Sabatini.Bianchi.ea2025}, and polarization studies indicating grain growth at the outflow cavity walls \citep{LeGouellec.Hull.ea2019, Hull.LeGouellec.ea2020}), 
the MIRI-MRS mode observations so far revealed extended dust continuum emission close to the source only around two protostars: HH211 \citep{CarattioGaratti.Ray.ea2024} and BHR71-IRS1 (this work). The presence of dust in the high-velocity wind (jet), spatially separated from the wide-angle wind component, and the outflow cavity walls, provides the strongest evidence so far that the grains are lifted from the disk. It remains an open question whether these grains might then fall back onto the outer disk and envelope and contribute to the reservoir of calcium-aluminium inclusions (CAI) \citep{Shu.Shang.ea2001}. In this section, we discuss the properties of the observed dust and possible scenarios for its origin.

\subsubsection{Origin of the extended continuum emission}

We detected two different spatially extended components: cone-like continuum emission along the outflow cavity walls that decreased with distance from the protostar, slightly asymmetric with a brighter western side, and emission in the jet beam, co-spatial with the ionized high-velocity gas bullets.

The cone-like emission can either be interpreted as dust that launched with the wind or was already present at the outflow cavity wall. These scenarios were discussed in detail by \cite{Sabatini.Bianchi.ea2025}, who found a dust opacity spectral index $\leq$ 1.3, which they interpreted as evidence of millimeter-sized grains. However, low spectral indices can also be caused by large temperature variations due to interior heating \citep{Ysard.Koehler.ea2019}. Further, large grains can more easily be destroyed by radiation than small grains \citep{Hoang2019}, so an investigation combining MIRI spectroscopy with ALMA continuum observations is needed. \cite{Sabatini.Bianchi.ea2025} provided strong arguments for the scenario in which the dust is launched in the wind. They rejected the scenario of grains already present in the envelope because millimeter-size grains were detected, which are difficult to form in a low-density outer envelope \citep{Silsbee.Akimkin.ea2022, Lombart.Lebreuilly.ea2026}. In the case of BHR71-IRS1, \cite{Hull.LeGouellec.ea2020} estimated the size of the largest grains in the outflow cavity walls to be 10 $\mu$m at least. 

For the dust detected in the collimated jet, we favor the scenario of dust that is directly lifted with the wind with respect to the envelope dust entrained and compressed in the jet. This latter hypothesis is disfavored for several reasons discussed below.
We detected higher dust continuum emission at the exact positions of the ionized emission from the internal working surfaces of the jet (Fig. \ref{fig:Figure1_overviewmaps}, Fig. \ref{fig:figure5_continuum_psfsub}, and Fig. \ref{fig:appE0_peaks}). At the knot positions, the continuum emission increases with respect to the background emission from the outflow cavity walls, so it is a distinct jet feature and not envelope material. In addition, if the dust were coming from the entrained envelope, it should be more strongly correlated with the H$_2$ emission or the cold CO gas (see Fig. \ref{fig:Figure3_h2maps}), which are more likely tracers of the entrained gas than the jet.

Although the jet is precessing, it is not likely that the dust in knots is entrained material from the envelope. The dust emission is seen across the bullets from 300 up to 1500 au. Dust emission is seen consistently throughout the jet and not just at the front of the jet. If this were shocked and compressed entrained dust, the limited refilling of the cavity might explain the presence of the dust at the furthermost bullet, but then, little dust would remain to be shocked in bullets closer to the source, and if anything, the dust emission would be fainter from B4 to B1, while we see an opposite trend. 

Modeling work in recent years provided substantial evidence for the feasibility of transport and launching of grains in the protostellar wind \citep[e.g.,][]{Wong.Hirashita.ea2016, Liffman.Bryan.ea2020, Vinkovic.Cemeljic2020, Booth.Clarke2021}. \cite{Giacalone.Teitler.ea2019} provided a model for dust launching in the magnetohydrodynamic (MHD) disk winds and related the $a_{\rm max}$, maximum size of the launched grains to the accretion rate, stellar mass, gas temperature in the launching zone, and the launching radius. For fiducial parameters adopted for T Tauri disks ($\macc \sim$ 10$^{-8}$, and $M_{\star} \sim$ 1 $\msun$), they obtained $a_{\rm max}=$ 0.35 $\mu$m. The mass-loss rate of our source is higher by an order of magnitude and that the stellar mass is lower by a factor of two, so that this will provide a size of 7  $\mu$m at the same launch radius and disk temperature. 
 \cite{Bhandare.Commercon.ea2024}  self-consistently explored dust transport from sub-au 
scales with the hydrodynamics code PLUTO, showing that grains can be efficiently redistributed across the disk by means of outflows for grains up to 100 $\mu$m, while \cite{Tsukamoto.Machida.ea2021} showed that even millimeter-size grains can be lifted by the winds. 

Recent direct observations of crystallization occurring in the disk during outbursts by \cite{Lee.Kim.ea2026} showed that seeds of processed grains can be formed in the inner disk. The launch of them in the wind and transport to the outer disk explains their presence in the outer Solar System. That work also predicted a variable dust composition depending on the launch radius. Detailed studies of the launched dust composition will be able to provide more information on this process.

\subsubsection{Dust survival and launch radius}
Because dust sublimates at small radii, the inner disk within the dust‑sublimation radius is expected to be essentially dust free; therefore, any dust observed in the wind must originate at disk radii larger than the sublimation radius. For BHR71‑IRS1 (\(L_{\rm bol}\approx10\,L_{\odot}\)), thia radius is \(\sim0.2\)~au under standard assumptions for the dust sublimation temperature \citep[$T_{\rm sub}\simeq 1500$ K;][]{Dullemond.Monnier2010}. The detection of dust in the ejected gas thus favors a disk‑wind origin (which launches from larger radii) over a classical X‑wind. However, under certain conditions, X‑wind models can still accommodate dust launching if the launch point shifts outward during an accretion outburst \citep{Shu.Shang.ea1997}. Therefore, the presence of dust does not conclusively distinguish between these scenarios; instead, it indicates either a launch from radii outside the sublimation zone or a temporary outward shift of the launch radius.

Accretion outbursts can increase the dust fraction in launched winds. MHD models predict that dust‑to‑gas ratios in winds can rise during outbursts \citep{Kadam.Vorobyov.ea2025}, so a recent outburst (on the flow dynamical timescale, \(\lesssim50\)~yr) might explain the dust. Outbursts also raise the IR luminosity and the grain alignment efficiency, and in turn, they increase the probability that grains are disrupted by their rotation \citep{Hoang2019}. However, according to simulations and synthetic observations, this increase in IR luminosity does not destroy all grains on typical dynamical timescales of 10–100 yr, except for the largest sizes and least cohesive grains \citep{Hull.LeGouellec.ea2020, LeGouellec.Maury.ea2023, Giang.LeGouellec.ea2025}. This leaves the possibility of the presence and survival of small and compact grains in irradiated outflow cavities. This scenario might account for the apparent truncation of the atomic jet beyond knot B4, which might mark the onset of recent outburst activity. 

\subsubsection{Dust temperature and heating source}
The observed dust temperature cannot be assumed to be simply the temperature at the launch point because grains cool very rapidly: for example, a 0.1\,$\mu$m silicate grain at 100~K radiatively cools to \(\sim10\)~K on a timescale of seconds \citep{Draine.Anderson1985}. Thus, the dust must be continuously heated by a local or external source. Maintainenance of grain temperatures \(\gtrsim100\)~K by  UV radiation alone requires a very strong field ($G_0\gtrsim10^{5}$) \citep[e.g.,][]{Hocuk2017}. The detection of prompt OH emission, a tracer of H$_2$O photodissociation, indicates an internal UV field in the jet \citep{Tabone2021}. \cite{Lehmann.Godard.ea2020} used the shock model to predict 10$^2$ $G_0$ for shock velocities of 60-75 km s$^{-1}$ , which were inferred in bullets B1-B3, and of 10$^1$ $G_0$ for bullet B4.  The lack of a clear correlation between OH flux, shock velocity, and dust temperature suggests that shock‑generated UV is not the dominant dust heater; additional heating from protostellar accretion luminosity is therefore required to explain the warm dust.
However, typical models of dust temperatures in the envelope do not predict sustained dust temperatures of several hundred K at 1000 au from the protostar \citep{Nazari.Tabone.ea2024}. While low-velocity shocks and C-type shocks cannot heat the dust above 100 K \citep{Miura.Yamamoto.ea2017, Draine.Roberge.ea1983}, the J-type shock model presented in \cite{Hollenbach.McKee1979} predicts that higher temperatures can be reached by J-type shocks ($\sim$300 K for $v_{\rm s}$= 80 km s$^{-1}$ and n$_0$= 10$^5$ cm$^{-3}$). The specific dust temperature in shocks is determined by the grain and shock properties. Qualitatively, the grain temperature increases with shock velocity and pre-shock density, and it is higher for smaller grains \citepalias{Hollenbach.McKee1989}. We note that the trend of the temperature of the warm component (Fig. \ref{fig:fig11_DustProps}, left) is similar to that of the shock velocity (Fig. \ref{fig:fig12_HM89models}, top right), which is consistent with heating of the dust grain to higher temperatures in a faster shock. The exact contributions of photospheric, accretion, and shock heating of the grains would require radiative transfer modeling.

\subsubsection{Comparison with chemical tracers}
Other species provide complementary constraints. SiO is seen in absorption toward the protostar at an offset of \(\sim40\)~km\,s$^{-1}$ \citep{vanGelder.Francis.ea2024}, which is consistent with part of the SiO gas originating from regions at or interior to the sublimation zone, where silicon is liberated from grains \citep{Gusdorf.PineaudesForets.ea2008}. ALMA does not detect extremely high‑velocity SiO in this source \citep{Gavino.Joergensen.ea2024}, which is consistent with gas mass‑loss inferred from atomic tracers of 10$^{-7}$ M$_\odot$ yr$^{-1}$, which is too low for formation of molecules \citep{Tabone.Godard.ea2020}. 

Comparisons with spectrally resolved far‑IR observations (e.g., \textit{Herschel}/HIFI) suggest that the low‑$E_{\rm up}$ H$_2$O gas identified in these data likely originated in regions more closely associated with the extended H$_2$ emission than with the high‑velocity ionic jet \citep{Kristensen.vanDishoeck.ea2012,Mottram.Kristensen.ea2014}. Faintly detected cold H$_2$O rotational transitions in the MIRI data likely trace the same component as \textit{Herschel}, but the signal-to-noise of the H$_2$O does not allow us to estimate the velocity. CO and CO$_2$ show a distinct spatial behavior: CO$_2$ coincides with warm dust at the cavity walls, whereas CO is stronger where CO$_2$ is weak near the source. This might indicate that the shocks on the western cavity wall related to jet wiggling are stronger. 

\subsection{Difference between Class 0 and Class I outflows and jets}

We compared the BHR71‑IRS1 jet to other spatially resolved jets from Class~0 and Class~I protostars, leveraging the unprecedented mid‑IR resolution of the JWST. Earlier facilities (e.g., \textit{Spitzer}) lacked the spatial resolution to separate wind layers close to the source \citep{Lahuis.vanDishoeck.ea2010}, but recent JWST studies have revealed diverse disk‑wind signatures in Class~I systems \citep{Tychoniec.vanGelder.ea2024,Delabrosse.Dougados.ea2024,Harsono.Bjerkeli.ea2023a,Assani.Harsono.ea2024}.

 In Class~I sources, the wind typically appears to broaden at lower upper‑level energies, which is consistent with a layered outflow structure, whereas in BHR71‑IRS1, the outflow boundary remains sharp, resembling that of HH46 \citep{Nisini.Navarro.ea2024}. Even high‑excitation H$_2$ lines do not show a narrow, jet‑like core, as seen in some other objects, where high‑$J$ H$_2$ traces a molecular cocoon around the jet (\citealt{CarattioGaratti.Ray.ea2024,Navarro.Nisini.ea2025, vanDishoeck.Tychoniec.ea2025,Francis.Tychoniec.ea2026}). The relative lack of molecular emission in the collimated jet, as evidenced by ALMA, might reflect evolutionary differences or distinct shock conditions \citep{Nisini.Santangelo.ea2015}.

Excitation diagnostics highlight clear contrasts with the Class~I source HH46. The line ratios indicate higher excitation in HH46: iron lines imply warmer collimated gas, [Ne\,\textsc{ii}]/[S\,\textsc{i}] is higher by some orders of magnitude in HH46 than in BHR71‑IRS1, and [Ne\,\textsc{iii}] is detected in HH46, while [Fe\,\textsc{i}] is not. These differences indicate a more highly ionized lower‑density jet in HH46 than the denser, more weakly ionized BHR71‑IRS1. By contrast, the refractory ratios with a similar excitation, for example, [Ni\,\textsc{ii}]\,6.64\,$\mu$m / [Co\,\textsc{ii}]\,10.52\,$\mu$m, are similar for BHR71‑IRS1 and HH46 (within a factor of two). This suggests that despite differences in excitation regimes, the relative abundances of these refractory species are comparable and not strongly affected by evolutionary stage.

The strong [Ne\,\textsc{ii}]\,12.8\,$\mu$m emission along the BHR71‑IRS1 jet concentrated in several bullets supports the presence of $J$‑type shocks in which H$_2$ can be dissociated and later reformed on grains \citepalias{Hollenbach.McKee1989}. In contrast to very young Class~0 sources such as HH211, where H$_2$ is clearly detected in the high‑velocity collimated wind \citep{CarattioGaratti.Ray.ea2024}, H$_2$ in BHR71‑IRS1 appears to be more extended and likely traces jet–cavity interactions and shocked cavity walls. H$_2$ centroids are also systematically slower than the ionized jet velocities, which is consistent with an origin in post‑shock or entrained gas and not in the primary high‑velocity jet.

Overall, BHR71‑IRS1 exhibits characteristics of a late-stage Class 0 source: its H$_2$ morphology resembles that of Class~0 systems (compact molecular structures near the jet and cavity), while its atomic and ionic emission indicates still lower excitation than most Class~I jets.

\begin{figure}
\centering
\includegraphics[width=0.49\textwidth]{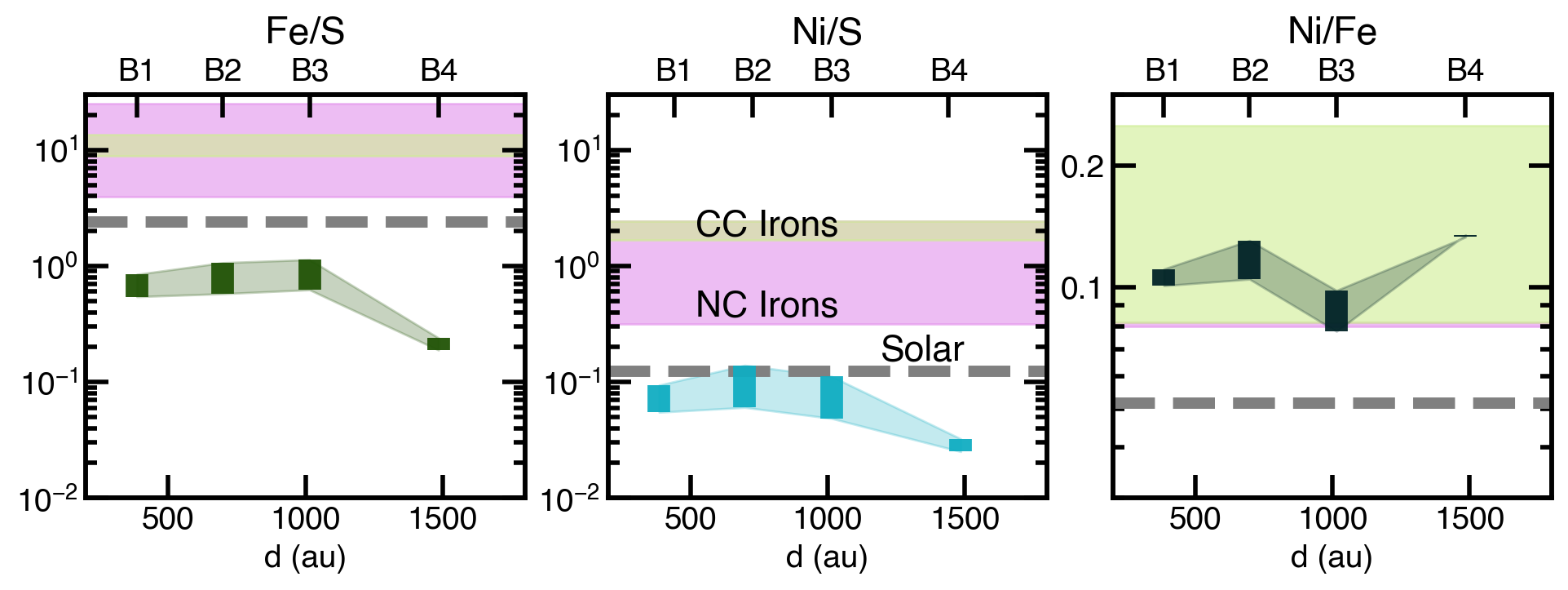}
\caption{{\it Left:} Fe to S elemental ratio compared with the solar abundance (dashed gray line) and non‑carbonaceous (NC) and carbonaceous (CC) iron meteorites \citep[green and purple shaded areas, respectively; ][]{Grewal.Nie.ea2024}; {\it Middle:} Ni to S elemental ratio. {\it Right:} Ni to Fe elemental ratio.}
\label{fig:elemental_ratios}
\end{figure}

\subsection{Refractories and planetary building blocks}

Jets sample material from the innermost disk. They can thus carry information about the composition of solids that may later contribute to planet formation. Refractory elements such as Ni, Co, and Fe have similar condensation temperatures and ionization potentials, making their relative abundances useful probes of dust processing and elemental depletion in the inner disk \citep{Wood.Smythe.ea2019,Konstantopoulou.DeCia.ea2022}. 
Figure~\ref{fig:elemental_ratios} compares the refractory ratios measured in the jet with Solar and meteoritic (non‑carbonaceous and carbonaceous iron meteorites) values \citep{Grewal.Nie.ea2024}. We also note that the abundances of primitive chondritic meteorites are usually similar to the solar values \citep{Lodders2021}.
Although Fe and Ni, the two refractories for which we can establish abundances based on the shock models, show an overall depletion relative to their solar abundances, they are compared to S, treated here as a volatile reference (Fig. \ref{fig:elemental_ratios}, left and center). However, the Ni/S ratio is closer to the solar value, especially at positions B1-B3. For Fe and Ni, this means that the values are also much lower than those of iron meteorites. 

The Ni/Fe ratio appears to be higher than the solar value and within the range of iron meteorites. The ratio is also similar to the Ikeya-Seki comet \citep[0.08;][]{Preston1967}, and at the same time, it is much lower than that of Oort cloud comets \citep[0.87;][]{Manfroid.Hutsemekers.ea2021} and pre-solar SiC \citep[$\sim$5.5;][]{Marhas.Amari.ea2008}. In comets, the ratio is explained by the lower sublimation temperature of the Ni-bearing compounds than the Fe-bearing ones \citep{Manfroid.Hutsemekers.ea2021} (see also the discussion in \cite{Samland.Henning.ea2025}). Perhaps this is the reason for the lower Ni depletion relative to solar values and the Ni/Fe excess in the BHR71-IRS1 jet, especially if it is released from NiS.

However, the interpretation of the abundance patterns requires caution. The apparent depletion of refractories relative to volatiles might reflect the retention of refractory material in large grains that survive the launch and incomplete grain destruction in the jet shocks. Our detection of warm dust in the jet supports the view that grains are not fully destroyed, implying that measured gas‑phase refractories trace only a fraction of the total refractory budget. If jet measurements reliably trace the inner‑disk composition, they can inform models of early planetesimal formation and the chemical dichotomy observed in the Solar System. However, it is difficult to assess the size distribution of ejected grains with the current data.

More systematic surveys of refractory abundances in jets across evolutionary stages are needed to establish whether jets can provide robust population‑level constraints on the initial composition of planet‑forming solids. The inclusion of cobalt in shock models, now readily detected in jets, would open a new avenue for studying the relative abundances of refractories.

\section{Summary}
\label{sec:summary}

We detected and resolved for the first time the ionized jet at the center of the outflow from the BHR71-IRS1 protostar, along with the warm dust, with JWST-MIRI MRS. The main results are summarized below.

\begin{itemize}

    \item The jet is bright in refractory gas, with species of ionized and neutral fine structure lines of Fe, Ni, Cl, and Co. The collimated emission is concentrated into four knots. In addition, molecular lines of OH, CO, H$_2$O, and CO$_2$  are detected in the outflow.
    
    \item The extended thermal dust continuum is fitted toward each bullet with three components: scattered light from the protostar, and two thermal components: warm (200-400 K) and cold ($\sim$80 K). The cold component is present throughout the envelope, with a temperature that is nearly constant; the temperature of the warm component decreases with distance. The presence of the warm component both on and off-axis of the jet (in the wide-angle wind) can be interpreted as dust entrained in the wind as well as in the collimated jet, while the origin of the cold component is more ambiguous, with possible contribution from the entrained envelope.
    
    \item The jet is surrounded by H$_2$ molecular gas, which is radially extended and peaks corresponding to the ionized jet. The H$_2$ flow extends beyond the radial distance of the jet. 

    \item Using [Ne \textsc{ii}], [S \textsc{i}], and [Cl \textsc{i}], we estimated the shock conditions. The shock velocity decreases with distance from the protostar from 70 km s$^{-1}$ to 38 km s$^{-1}$. The shock density drops from above 10$^5$ to 3-4 $\times$ 10$^{4}$ cm$^{-3}$ with distance from the protostar.
    
    \item We used the shock models to estimate the abundance of refractory species relative to volatile sulfur. We found that iron and nickel are both depleted relative to solar abundances. Together with the dust detected in the jet beam, this provides strong evidence of dust launching in the jet.    

\end{itemize}

The protostellar jet BHR71-IRS1 reveals bright refractory lines and dust thermal emission. Further studies of the dust properties and the abundance of refractory elements might help us to elucidate the composition of planetary building blocks at the earliest stages of planet formation.

\begin{acknowledgements}
We thank the referee for careful evaluation of the manuscript and comments that improved the clarity of the presented results. This work is based on observations made with the NASA/ESA/CSA James Webb Space Telescope. The data were obtained from the Mikulski Archive for Space Telescopes at the Space Telescope Science Institute, which is operated by the Association of Universities for Research in Astronomy, Inc., under NASA contract NAS 5-03127 for JWST. These observations are associated with program \#1290 \url{https://doi.org/10.17909/7eh1-8f25}.
The following National and International Funding Agencies funded and supported the MIRI development: NASA; ESA; Belgian Science Policy Office (BELSPO); Centre Nationale d’Études Spatiales (CNES); Danish National Space Centre; Deutsches Zentrum fur Luftund Raumfahrt (DLR); Enterprise Ireland; Ministerio De Economiá y Competividad; The Netherlands Research School for Astronomy (NOVA); The Netherlands Organisation for Scientific Research (NWO); Science and Technology Facilities Council; Swiss Space Office; Swedish National Space Agency; and UK Space Agency. This paper makes use of the following ALMA data: ADS/JAO.ALMA\#2019.1.00261.L ALMA is a partnership of ESO (representing its member states), NSF (USA) and NINS (Japan), together with NRC (Canada), NSTC and ASIAA (Taiwan), and KASI (Republic of Korea), in cooperation with the Republic of Chile. The Joint ALMA Observatory is operated by ESO, AUI/NRAO, and NAOJ. The National Radio Astronomy Observatory is a facility of the National Science Foundation operated under cooperative agreement by Associated Universities, Inc. This research has made use of NASA’s Astrophysics Data System Bibliographic Services.

MvG, LF, EvD, YC, KS, acknowledge support from ERC Advanced grant 101019751 MOLDISK, TOP-1 grant 614.001.751 from the Dutch Research Council (NWO), The Netherlands Research School for Astronomy (NOVA), the Danish National Research Foundation through the Center of Excellence “InterCat” (DNRF150), and DFG-grant 325594231, FOR 2634/2. T.P.R. acknowledges support from ERC grant 743029 EASY. H.B. acknowledges support from the Deutsche Forschungsgemeinschaft in the Collaborative Research Center (SFB 881) “The Milky Way System” (subproject B1). A.C.G. acknowledges support from PRIN-MUR 2022 20228JPA3A “The path to star and planet formation in the JWST era (PATH)” funded by NextGeneration EU and by INAF-GoG 2022 “NIR-dark Accretion Outbursts in Massive Young stellar objects (NAOMY)” and Large Gran INAF-2024 “Spectral Key features of Young stellar objects: Wind-Accretion LinKs Explored in the infraRed (SKYWALKER)”. KJ acknowledges the support from the Swedish National Space Agency. P.N. acknowledges support from the ESO Fellowship and IAU Gruber Foundation Fellowship programs. JMV acknowledges support from the Academy of Finland grant No 348342. D.H. is supported by the Ministry of Education of Taiwan (Center for Informatics and Computation in Astronomy grant and grant number 110J0353I9) and the National Science and Technology Council, Taiwan (Grant NSTC111-2112-M-007-014-MY3, NSTC113-2639-M-A49-002-ASP, and NSTC113-2112-M-007-027). Y.-L.Y. acknowledges support from Grant-in-Aid from the Ministry of Education, Culture, Sports, Science, and Technology of Japan (20H05844 and 25H00676). The project was supported by RIKEN Incentive Research Project (Emergence of pre-biotic extraterrestrial world: Using machine-learning techniques to uncover the link of organic matter at the dawn of planetary systems).
V.J.M.L.G. acknowledges support by the Spanish program Unidad de Excelencia María de Maeztu CEX2020-001058-M, financed by MCIN/AEI/10.13039/501100011033, and by the MaX-CSIC Excellence Award MaX4-SOMMA-ICE. V.J.M.L.G. acknowledges support by the European Research Council (ERC) 
under the European Union’s Horizon 2020 research and innovation program (grant agreement No. 101098309 - PEBBLES).
This research made use of NumPy \citep[][]{Harris2020}; {\tt scipy} \citep{SciPy},  Astropy, a community-developed core Python package for Astronomy \citep[][]{AstropyCollaboration2022}; Matplotlib \citep[][]{Hunter2007}, lmfit \citep{Newville.Otten.ea2025}, {\tt stpsf} \citep{Perrin.Long.ea2025},  jwst \citep{Bushouse.Eisenhamer.ea2023}, {\tt pandas} \citep{pandas} . This research made use of Photutils, an Astropy package for
detection and photometry of astronomical sources \citep{Bradley.Sipoecz.ea2025}. We utilized GitHub CoPilot
to assist in writing parts of the code for the analysis, accessed through the Visual Studio Code extension GitHub
Copilot Chat. We used Grammarly for grammar and spelling corrections.

\end{acknowledgements}

\bibliography{main.bib}

@Article{Vioque.Booth.ea2026,
  author        = {{Vioque}, Miguel and {Booth}, Richard A. and {Ragusa}, Enrico and {Ribas}, {\'A}lvaro and {Kurtovic}, Nicol{\'a}s T. and {Rosotti}, Giovanni P. and {Penoyre}, Zephyr and {Facchini}, Stefano and {Garufi}, Antonio and {Manara}, Carlo F. and {Hu{\'e}lamo}, Nuria and {Winter}, Andrew and {P{\'e}rez}, Sebasti{\'a}n and {Benisty}, Myriam and {Mendigut{\'\i}a}, Ignacio and {Cuello}, Nicol{\'a}s and {Penzlin}, Anna B.~T. and {Castro-Ginard}, Alfred and {Teague}, Richard},
  journal       = {\aap},
  title         = {{Astrometric view of companions in the inner dust cavities of protoplanetary discs}},
  year          = {2026},
  month         = jan,
  pages         = {A238},
  volume        = {705},
  adsurl        = {https://ui.adsabs.harvard.edu/abs/2026A&A...705A.238V},
  archiveprefix = {arXiv},
  doi           = {10.1051/0004-6361/202557086},
  eid           = {A238},
  eprint        = {2512.00157},
  primaryclass  = {astro-ph.EP},
}

@Article{Federman.Megeath.ea2026,
  author        = {{Federman}, Samuel A. and {Megeath}, S. Thomas and {Caratti o Garatti}, Alessio and {Narang}, Mayank and {Tyagi}, Himanshu and {Evans}, II, Neal J. and {Kimmig}, Carolin N. and {Tychoniec}, {\L}ukasz and {Beuther}, Henrik and {Stutz}, Amelia and {Manoj}, P. and {Gutermuth}, Robert and {Bourke}, Tyler L. and {Green}, Joel and {Hartmann}, Lee and {Klaassen}, Pamela and {Kuiper}, Rolf and {Looney}, Leslie W. and {Nazari}, Pooneh and {Stanke}, Thomas and {Watson}, Dan M. and {Yang}, Yao-Lun and {Zakri}, Wafa},
  journal       = {\apj},
  title         = {{The Structure and Kinematics of Three Class 0 Protostellar Jets from JWST}},
  year          = {2026},
  month         = feb,
  number        = {2},
  pages         = {282},
  volume        = {998},
  adsurl        = {https://ui.adsabs.harvard.edu/abs/2026ApJ...998..282F},
  archiveprefix = {arXiv},
  doi           = {10.3847/1538-4357/ae34b2},
  eid           = {282},
  eprint        = {2601.09587},
  primaryclass  = {astro-ph.SR},
}

@Article{Giang.LeGouellec.ea2025,
  author    = {Giang, Nguyen Chau and Le Gouellec, V. J. M. and Hoang, Thiem and Maury, A. J. and Hennebelle, P.},
  journal   = {\apj},
  title     = {Synthetic Modeling of Polarized Dust Emission in Intermediate-mass Young Stellar Objects. II. Effects of Radiative Torque Disruption on Dust Grains in Protostellar Jets/Outflows},
  year      = {2025},
  month     = {oct},
  number    = {1},
  pages     = {131},
  volume    = {993},
  doi       = {10.3847/1538-4357/adff4f},
  publisher = {The American Astronomical Society},
  url       = {https://doi.org/10.3847/1538-4357/adff4f},
}

@Article{Francis.Tychoniec.ea2026,
  author        = {{Francis}, L. and {Tychoniec}, {\L}. and {van Dishoeck}, E.~F. and {Sellek}, A.~D. and {Caratti o Garatti}, A. and {Le Gouellec}, V.~J.~M. and {Gieser}, C. and {Beuther}, H. and {Vorster}, J.~M. and {Ressler}, M.~E. and {Nazari}, P. and {Tabone}, B. and {Assani}, K. and {Devaraj}, R. and {Tobin}, J.~J. and {Navarro}, M.~G. and {Cort{\'e}s}, P.~C. and {Girart}, J.~M. and {G{\"u}del}, M. and {Henning}, Th. and {{\"O}stlin}, G. and {Wright}, G. and {Ray}, T.},
  journal       = {\aap},
  title         = {{JOYS+: A JWST/MIRI survey of the evolution of H$_{2}$ winds and jets from low-mass protostars}},
  year          = {2026},
  month         = jul,
  pages         = {A258},
  volume        = {711},
  adsurl        = {https://ui.adsabs.harvard.edu/abs/2026A&A...711A.258F},
  archiveprefix = {arXiv},
  doi           = {10.1051/0004-6361/202558772},
  eid           = {A258},
  eprint        = {2604.13773},
  primaryclass  = {astro-ph.SR},
}

@Article{Vinkovic.Cemeljic2020,
  author   = {Vinković, Dejan and Čemeljić, Miljenko},
  journal  = {\mnras},
  title    = {Inner dusty regions of protoplanetary discs – II. Dust dynamics driven by radiation pressure and disc winds},
  year     = {2020},
  issn     = {0035-8711},
  month    = {10},
  number   = {1},
  pages    = {506-519},
  volume   = {500},
  doi      = {10.1093/mnras/staa3272},
  eprint   = {https://academic.oup.com/mnras/article-pdf/500/1/506/34260895/staa3272.pdf},
  url      = {https://doi.org/10.1093/mnras/staa3272},
}

@Article{Liffman.Bryan.ea2020,
  author        = {{Liffman}, Kurt and {Bryan}, Geoffrey and {Hutchison}, Mark and {Maddison}, Sarah T.},
  journal       = {\mnras},
  title         = {{Infrared variability due to magnetic pressure-driven jets, dust ejection and quasi-puffed-up inner rims}},
  year          = {2020},
  month         = apr,
  number        = {3},
  pages         = {4022-4038},
  volume        = {493},
  adsurl        = {https://ui.adsabs.harvard.edu/abs/2020MNRAS.493.4022L},
  archiveprefix = {arXiv},
  doi           = {10.1093/mnras/staa402},
  eprint        = {2002.08432},
  primaryclass  = {astro-ph.SR},
}

@Article{Booth.Clarke2021,
  author        = {{Booth}, Richard A. and {Clarke}, Cathie J.},
  journal       = {\mnras},
  title         = {{Modelling the delivery of dust from discs to ionized winds}},
  year          = {2021},
  month         = apr,
  number        = {2},
  pages         = {1569-1578},
  volume        = {502},
  adsurl        = {https://ui.adsabs.harvard.edu/abs/2021MNRAS.502.1569B},
  archiveprefix = {arXiv},
  doi           = {10.1093/mnras/stab090},
  eprint        = {2101.04121},
  primaryclass  = {astro-ph.EP},
}

@Article{Wong.Hirashita.ea2016,
  author        = {{Wong}, Yi Hang Valerie and {Hirashita}, Hiroyuki and {Li}, Zhi-Yun},
  journal       = {\pasj},
  title         = {{Millimeter-sized grains in the protostellar envelopes: Where do they come from?}},
  year          = {2016},
  month         = aug,
  number        = {4},
  pages         = {67},
  volume        = {68},
  adsurl        = {https://ui.adsabs.harvard.edu/abs/2016PASJ...68...67W},
  archiveprefix = {arXiv},
  doi           = {10.1093/pasj/psw066},
  eid           = {67},
  eprint        = {1606.03277},
  primaryclass  = {astro-ph.SR},
}

@Article{LeGouellec.Hull.ea2019,
  author        = {{Le Gouellec}, Valentin J.~M. and {Hull}, Charles L.~H. and {Maury}, Ana{\"e}lle J. and {Girart}, Josep M. and {Tychoniec}, {\L}ukasz and {Kristensen}, Lars E. and {Li}, Zhi-Yun and {Louvet}, Fabien and {Cortes}, Paulo C. and {Rao}, Ramprasad},
  journal       = {\apj},
  title         = {{Characterizing Magnetic Field Morphologies in Three Serpens Protostellar Cores with ALMA}},
  year          = {2019},
  month         = nov,
  number        = {2},
  pages         = {106},
  volume        = {885},
  adsurl        = {https://ui.adsabs.harvard.edu/abs/2019ApJ...885..106L},
  archiveprefix = {arXiv},
  doi           = {10.3847/1538-4357/ab43c2},
  eid           = {106},
  eprint        = {1909.00046},
  primaryclass  = {astro-ph.GA},
}

@Article{AgraAmboage.Dougados.ea2011,
  author        = {{Agra-Amboage}, V. and {Dougados}, C. and {Cabrit}, S. and {Reunanen}, J.},
  journal       = {\aap},
  title         = {{Sub-arcsecond [Fe ii] spectro-imaging of the DG Tauri jet. Periodic bubbles and a dusty disk wind?}},
  year          = {2011},
  month         = aug,
  pages         = {A59},
  volume        = {532},
  adsurl        = {https://ui.adsabs.harvard.edu/abs/2011A&A...532A..59A},
  archiveprefix = {arXiv},
  doi           = {10.1051/0004-6361/201015886},
  eid           = {A59},
  eprint        = {1106.2690},
  primaryclass  = {astro-ph.SR},
}

@Article{Guillet.Jones.ea2009,
  author   = {{Guillet}, V. and {Jones}, A.~P. and {Pineau Des For{\^e}ts}, G.},
  journal  = {\aap},
  title    = {{Shocks in dense clouds. II. Dust destruction and SiO formation in J shocks}},
  year     = {2009},
  month    = apr,
  number   = {1},
  pages    = {145-153},
  volume   = {497},
  adsurl   = {https://ui.adsabs.harvard.edu/abs/2009A&A...497..145G},
  doi      = {10.1051/0004-6361/200811115},
}

@Article{Lombart.Lebreuilly.ea2026,
  author        = {{Lombart}, M. and {Lebreuilly}, U. and {Maury}, A.},
  journal       = {\aap},
  title         = {{The 3D time evolution of the dust size distribution in protostellar envelopes}},
  year          = {2026},
  month         = mar,
  pages         = {A358},
  volume        = {707},
  adsurl        = {https://ui.adsabs.harvard.edu/abs/2026A&A...707A.358L},
  archiveprefix = {arXiv},
  doi           = {10.1051/0004-6361/202557277},
  eid           = {A358},
  eprint        = {2601.07088},
  primaryclass  = {astro-ph.SR},
}

@Article{Silsbee.Akimkin.ea2022,
  author        = {{Silsbee}, Kedron and {Akimkin}, Vitaly and {Ivlev}, Alexei V. and {Testi}, Leonardo and {Gong}, Munan and {Caselli}, Paola},
  journal       = {\apj},
  title         = {{Dust Grains Cannot Grow to Millimeter Sizes in Protostellar Envelopes}},
  year          = {2022},
  month         = dec,
  number        = {2},
  pages         = {188},
  volume        = {940},
  adsurl        = {https://ui.adsabs.harvard.edu/abs/2022ApJ...940..188S},
  archiveprefix = {arXiv},
  doi           = {10.3847/1538-4357/ac978b},
  eid           = {188},
  eprint        = {2210.01832},
  primaryclass  = {astro-ph.GA},
}

@Article{Greenfield.Miller2016,
  author   = {{Greenfield}, P. and {Miller}, T.},
  journal  = {Astronomy and Computing},
  title    = {{The Calibration Reference Data System}},
  year     = {2016},
  month    = jul,
  pages    = {41-53},
  volume   = {16},
  adsurl   = {https://ui.adsabs.harvard.edu/abs/2016A&C....16...41G},
  doi      = {10.1016/j.ascom.2016.04.001},
}

@Article{Hoang2019,
  author        = {{Hoang}, Thiem},
  journal       = {\apj},
  title         = {{A Dynamical Constraint on Interstellar Dust Models from Radiative Torque Disruption}},
  year          = {2019},
  month         = may,
  number        = {1},
  pages         = {13},
  volume        = {876},
  adsurl        = {https://ui.adsabs.harvard.edu/abs/2019ApJ...876...13H},
  archiveprefix = {arXiv},
  doi           = {10.3847/1538-4357/ab1075},
  eid           = {13},
  eprint        = {1812.08391},
  primaryclass  = {astro-ph.GA},
}

@Article{Ysard.Koehler.ea2019,
  author        = {{Ysard}, N. and {Koehler}, M. and {Jimenez-Serra}, I. and {Jones}, A.~P. and {Verstraete}, L.},
  journal       = {\aap},
  title         = {{From grains to pebbles: the influence of size distribution and chemical composition on dust emission properties}},
  year          = {2019},
  month         = {Nov},
  pages         = {A88},
  volume        = {631},
  adsurl        = {https://ui.adsabs.harvard.edu/abs/2019A&A...631A..88Y},
  archiveprefix = {arXiv},
  doi           = {10.1051/0004-6361/201936089},
  eid           = {A88},
  eprint        = {1909.05015},
  primaryclass  = {astro-ph.GA},
}

@Article{LeGouellec.Maury.ea2023,
  author        = {{Le Gouellec}, V.~J.~M. and {Maury}, A.~J. and {Hull}, C.~L.~H. and {Verliat}, A. and {Hennebelle}, P. and {Valdivia}, V.},
  journal       = {\aap},
  title         = {{Physical conditions for dust grain alignment in Class 0 protostellar cores. II. The role of the radiation field in models that align and disrupt dust grains}},
  year          = {2023},
  month         = jul,
  pages         = {A133},
  volume        = {675},
  adsurl        = {https://ui.adsabs.harvard.edu/abs/2023A&A...675A.133L},
  archiveprefix = {arXiv},
  doi           = {10.1051/0004-6361/202245346},
  eid           = {A133},
  eprint        = {2303.12275},
  primaryclass  = {astro-ph.GA},
}

@Article{Lee.Kim.ea2026,
  author  = {{Lee}, Jeong-Eun and {Kim}, Chul-Hwan and {Kim}, Jaeyeong and {Lee}, Seokho and {Kim}, Young-Jun and {Lee}, Seonjae and {Baek}, Giseon and {Green}, Joel D. and {Herczeg}, Gregory J. and {Johnstone}, Doug and {Pontoppidan}, Klaus M. and {Aikawa}, Yuri and {Yang}, Yao-Lun and {Francis}, Logan and {Jin}, Mihwa and {Jang}, Hyerin},
  journal = {\nat},
  title   = {{Accretion bursts crystallize silicates in a planet-forming disk}},
  year    = {2026},
  month   = jan,
  number  = {8098},
  pages   = {853-858},
  volume  = {649},
  adsurl  = {https://ui.adsabs.harvard.edu/abs/2026Natur.649..853L},
  doi     = {10.1038/s41586-025-09939-3},
}

@Article{Bhandare.Commercon.ea2024,
  author        = {{Bhandare}, Asmita and {Commer{\c{c}}on}, Beno{\^\i}t and {Laibe}, Guillaume and {Flock}, Mario and {Kuiper}, Rolf and {Henning}, Thomas and {Mignone}, Andrea and {Marleau}, Gabriel-Dominique},
  journal       = {\aap},
  title         = {{Mixing is easy: New insights for cosmochemical evolution from pre-stellar core collapse}},
  year          = {2024},
  month         = jul,
  pages         = {A158},
  volume        = {687},
  adsurl        = {https://ui.adsabs.harvard.edu/abs/2024A&A...687A.158B},
  archiveprefix = {arXiv},
  doi           = {10.1051/0004-6361/202449594},
  eid           = {A158},
  eprint        = {2404.09257},
  primaryclass  = {astro-ph.SR},
}

@Article{Draine.Roberge.ea1983,
  author   = {Draine, B. T. and Roberge, W. G. and Dalgarno, A.},
  journal  = {\apj},
  title    = {Magnetohydrodynamic shock waves in molecular clouds},
  year     = {1983},
  month    = jan,
  pages    = {485-507},
  volume   = {264},
  doi      = {10.1086/160617},
}

@Article{Herzberg.Monfils1961,
  author   = {G Herzberg and A Monfils},
  journal  = {JoMS},
  title    = {The dissociation energies of the H2, HD, and D2 molecules},
  year     = {1961},
  issn     = {0022-2852},
  number   = {1},
  pages    = {482-498},
  volume   = {5},
  doi      = {https://doi.org/10.1016/0022-2852(61)90111-4},
  url      = {https://www.sciencedirect.com/science/article/pii/0022285261901114},
}

@Article{Giacalone.Teitler.ea2019,
  author    = {Giacalone, Steven and Teitler, Seth and Königl, Arieh and Krijt, Sebastiaan and Ciesla, Fred J.},
  journal   = {\apj},
  title     = {Dust Transport and Processing in Centrifugally Driven Protoplanetary Disk Winds},
  year      = {2019},
  month     = {aug},
  number    = {1},
  pages     = {33},
  volume    = {882},
  doi       = {10.3847/1538-4357/ab311a},
  publisher = {The American Astronomical Society},
  url       = {https://doi.org/10.3847/1538-4357/ab311a},
}

@Article{Nazari.Tabone.ea2024,
  author        = {{Nazari}, P. and {Tabone}, B. and {Rosotti}, G.~P. and {van Dishoeck}, E.~F.},
  journal       = {\aap},
  title         = {{Correlations among complex organic molecules around protostars: Effects of physical structure}},
  year          = {2024},
  month         = jul,
  pages         = {A263},
  volume        = {687},
  adsurl        = {https://ui.adsabs.harvard.edu/abs/2024A&A...687A.263N},
  archiveprefix = {arXiv},
  doi           = {10.1051/0004-6361/202347033},
  eid           = {A263},
  eprint        = {2404.10045},
  primaryclass  = {astro-ph.GA},
}

@Article{Lehmann.Godard.ea2020,
  author        = {{Lehmann}, A. and {Godard}, B. and {Pineau des For{\^e}ts}, G. and {Falgarone}, E.},
  journal       = {\aap},
  title         = {{Self-generated ultraviolet radiation in molecular shock waves. I. Effects of Lyman {\ensuremath{\alpha}}, Lyman {\ensuremath{\beta}}, and two-photon continuum}},
  year          = {2020},
  month         = nov,
  pages         = {A101},
  volume        = {643},
  adsurl        = {https://ui.adsabs.harvard.edu/abs/2020A&A...643A.101L},
  archiveprefix = {arXiv},
  doi           = {10.1051/0004-6361/202038644},
  eid           = {A101},
  eprint        = {2010.01042},
  primaryclass  = {astro-ph.GA},
}

@Article{Marhas.Amari.ea2008,
  author   = {{Marhas}, Kuljeet K. and {Amari}, Sachiko and {Gyngard}, Frank and {Zinner}, Ernst and {Gallino}, Roberto},
  journal  = {\apj},
  title    = {{Iron and Nickel Isotopic Ratios in Presolar SiC Grains}},
  year     = {2008},
  month    = dec,
  number   = {1},
  pages    = {622-645},
  volume   = {689},
  adsurl   = {https://ui.adsabs.harvard.edu/abs/2008ApJ...689..622M},
  doi      = {10.1086/592599},
}

@Article{Samland.Henning.ea2025,
  author        = {{Samland}, Matthias and {Henning}, Thomas and {Caratti o Garatti}, Alessio and {Giannini}, Teresa and {Bouwman}, Jeroen and {Tabone}, Beno{\^\i}t and {Arabhavi}, Aditya M. and {Olofsson}, G{\"o}ran and {G{\"u}del}, Manuel and {Pawellek}, Nicole and {Kamp}, Inga and {Waters}, L.~B.~F.~M. and {Semenov}, Dmitry and {van Dishoeck}, Ewine F. and {Absil}, Olivier and {Barrado}, David and {Boccaletti}, Anthony and {Christiaens}, Valentin and {Gasman}, Danny and {Grant}, Sierra L. and {Jang}, Hyerin and {Kaeufer}, Till and {Kanwar}, Jayatee and {Perotti}, Giulia and {Schwarz}, Kamber and {Temmink}, Milou},
  journal       = {\apj},
  title         = {{MINDS: Detection of an Inner Gas Disk Caused by Evaporating Bodies around HD 172555}},
  year          = {2025},
  month         = aug,
  number        = {2},
  pages         = {132},
  volume        = {989},
  adsurl        = {https://ui.adsabs.harvard.edu/abs/2025ApJ...989..132S},
  archiveprefix = {arXiv},
  doi           = {10.3847/1538-4357/ade2db},
  eid           = {132},
  eprint        = {2506.09976},
  primaryclass  = {astro-ph.EP},
}

@Article{Manfroid.Hutsemekers.ea2021,
  author  = {{Manfroid}, J. and {Hutsem{\'e}kers}, D. and {Jehin}, E.},
  journal = {\nat},
  title   = {{Iron and nickel atoms in cometary atmospheres even far from the Sun}},
  year    = {2021},
  month   = may,
  number  = {7859},
  pages   = {372-374},
  volume  = {593},
  adsurl  = {https://ui.adsabs.harvard.edu/abs/2021Natur.593..372M},
  doi     = {10.1038/s41586-021-03435-0},
}

@Article{Preston1967,
  author  = {{Preston}, G.~W.},
  journal = {\apj},
  title   = {{The spectrum of Ikeya-Seki (1965f)}},
  year    = {1967},
  month   = feb,
  pages   = {718-742},
  volume  = {147},
  adsurl  = {https://ui.adsabs.harvard.edu/abs/1967ApJ...147..718P},
  doi     = {10.1086/149049},
}

@Article{Kataoka.Okuzumi.ea2014,
  author        = {{Kataoka}, Akimasa and {Okuzumi}, Satoshi and {Tanaka}, Hidekazu and {Nomura}, Hideko},
  journal       = {\aap},
  title         = {{Opacity of fluffy dust aggregates}},
  year          = {2014},
  month         = {Aug},
  pages         = {A42},
  volume        = {568},
  adsurl        = {https://ui.adsabs.harvard.edu/abs/2014A&A...568A..42K},
  archiveprefix = {arXiv},
  doi           = {10.1051/0004-6361/201323199},
  eid           = {A42},
  eprint        = {1312.1459},
  primaryclass  = {astro-ph.EP},
}

@Article{Kristensen.Godard.ea2023,
  author        = {{Kristensen}, L.~E. and {Godard}, B. and {Guillard}, P. and {Gusdorf}, A. and {Pineau des For{\^e}ts}, G.},
  journal       = {\aap},
  title         = {{Shock excitation of H$_{2}$ in the James Webb Space Telescope era}},
  year          = {2023},
  month         = jul,
  pages         = {A86},
  volume        = {675},
  adsurl        = {https://ui.adsabs.harvard.edu/abs/2023A&A...675A..86K},
  archiveprefix = {arXiv},
  doi           = {10.1051/0004-6361/202346254},
  eid           = {A86},
  eprint        = {2307.04178},
  primaryclass  = {astro-ph.GA},
}

@Article{Lodders2021,
  author   = {{Lodders}, Katharina},
  journal  = {\ssr},
  title    = {{Relative Atomic Solar System Abundances, Mass Fractions, and Atomic Masses of the Elements and Their Isotopes, Composition of the Solar Photosphere, and Compositions of the Major Chondritic Meteorite Groups}},
  year     = {2021},
  month    = apr,
  number   = {3},
  pages    = {44},
  volume   = {217},
  adsurl   = {https://ui.adsabs.harvard.edu/abs/2021SSRv..217...44L},
  doi      = {10.1007/s11214-021-00825-8},
  eid      = {44},
}

@Article{Miura.Yamamoto.ea2017,
  author   = {{Miura}, Hitoshi and {Yamamoto}, Tetsuo and {Nomura}, Hideko and {Nakamoto}, Taishi and {Tanaka}, Kyoko K. and {Tanaka}, Hidekazu and {Nagasawa}, Makiko},
  journal  = {\apj},
  title    = {{Comprehensive Study of Thermal Desorption of Grain-surface Species by Accretion Shocks around Protostars}},
  year     = {2017},
  month    = apr,
  number   = {1},
  pages    = {47},
  volume   = {839},
  adsurl   = {https://ui.adsabs.harvard.edu/abs/2017ApJ...839...47M},
  doi      = {10.3847/1538-4357/aa67df},
  eid      = {47},
}

@Software{Bradley.Sipoecz.ea2025,
  author    = {Larry Bradley and Brigitta Sip{\H o}cz and Thomas Robitaille and Erik Tollerud and Z\`e Vin{\'{\i}}cius and Christoph Deil and Kyle Barbary and Tom J Wilson and Ivo Busko and Axel Donath and Hans Moritz G{\"u}nther and Mihai Cara and P. L. Lim and Sebastian Me{\ss}linger and Zach Burnett and Simon Conseil and Michael Droettboom and Azalee Bostroem and E. M. Bray and Lars Andersen Bratholm and William Jamieson and Adam Ginsburg and Geert Barentsen and Matt Craig and Sergio Pascual and Shivangee Rathi and Marshall Perrin and Brett M. Morris},
  doi       = {10.5281/zenodo.14889440},
  month     = feb,
  publisher = {Zenodo},
  swhid     = {swh:1:dir:11159107f27a28985192ed1118b1f2055709d093 ;origin=https://doi.org/10.5281/zenodo.596036;visi t=swh:1:snp:ae8c4a55d349d43e53cfe9ce92e678fcfe840f 3b;anchor=swh:1:rel:0117f67e8888adcdfc85308287dd9c 854b466389;path=astropy-photutils-ffb96c5},
  title     = {astropy/photutils: 2.2.0},
  url       = {https://doi.org/10.5281/zenodo.14889440},
  version   = {2.2.0},
  year      = {2025},
}

@Article{Shu.Shang.ea2001,
  author   = {{Shu}, Frank H. and {Shang}, Hsien and {Gounelle}, Matthieu and {Glassgold}, Alfred E. and {Lee}, Typhoon},
  journal  = {\apj},
  title    = {{The Origin of Chondrules and Refractory Inclusions in Chondritic Meteorites}},
  year     = {2001},
  month    = feb,
  number   = {2},
  pages    = {1029-1050},
  volume   = {548},
  adsurl   = {https://ui.adsabs.harvard.edu/abs/2001ApJ...548.1029S},
  doi      = {10.1086/319018},
}

@ARTICLE{Hunter2007,
       author = {{Hunter}, John D.},
        title = "{Matplotlib: A 2D Graphics Environment}",
      journal = {Computing in Science and Engineering},
         year = 2007,
        month = may,
       volume = {9},
       number = {3},
        pages = {90-95},
          doi = {10.1109/MCSE.2007.55},
       adsurl = {https://ui.adsabs.harvard.edu/abs/2007CSE.....9...90H}
}

@ARTICLE{AstropyCollaboration2022,
       author = {{Astropy Collaboration} and {Price-Whelan}, Adrian M. and {Lim}, Pey Lian and {Earl}, Nicholas and {Starkman}, Nathaniel and {Bradley}, Larry and {Shupe}, David L. and {Patil}, Aarya A. and {Corrales}, Lia and {Brasseur}, C.~E. and {N{\"o}the}, Maximilian and {Donath}, Axel and {Tollerud}, Erik and {Morris}, Brett M. and {Ginsburg}, Adam and {Vaher}, Eero and {Weaver}, Benjamin A. and {Tocknell}, James and {Jamieson}, William and {van Kerkwijk}, Marten H. and {Robitaille}, Thomas P. and {Merry}, Bruce and {Bachetti}, Matteo and {G{\"u}nther}, H. Moritz and {Aldcroft}, Thomas L. and {Alvarado-Montes}, Jaime A. and {Archibald}, Anne M. and {B{\'o}di}, Attila and {Bapat}, Shreyas and {Barentsen}, Geert and {Baz{\'a}n}, Juanjo and {Biswas}, Manish and {Boquien}, M{\'e}d{\'e}ric and {Burke}, D.~J. and {Cara}, Daria and {Cara}, Mihai and {Conroy}, Kyle E. and {Conseil}, Simon and {Craig}, Matthew W. and {Cross}, Robert M. and {Cruz}, Kelle L. and {D'Eugenio}, Francesco and {Dencheva}, Nadia and {Devillepoix}, Hadrien A.~R. and {Dietrich}, J{\"o}rg P. and {Eigenbrot}, Arthur Davis and {Erben}, Thomas and {Ferreira}, Leonardo and {Foreman-Mackey}, Daniel and {Fox}, Ryan and {Freij}, Nabil and {Garg}, Suyog and {Geda}, Robel and {Glattly}, Lauren and {Gondhalekar}, Yash and {Gordon}, Karl D. and {Grant}, David and {Greenfield}, Perry and {Groener}, Austen M. and {Guest}, Steve and {Gurovich}, Sebastian and {Handberg}, Rasmus and {Hart}, Akeem and {Hatfield-Dodds}, Zac and {Homeier}, Derek and {Hosseinzadeh}, Griffin and {Jenness}, Tim and {Jones}, Craig K. and {Joseph}, Prajwel and {Kalmbach}, J. Bryce and {Karamehmetoglu}, Emir and {Ka{\l}uszy{\'n}ski}, Miko{\l}aj and {Kelley}, Michael S.~P. and {Kern}, Nicholas and {Kerzendorf}, Wolfgang E. and {Koch}, Eric W. and {Kulumani}, Shankar and {Lee}, Antony and {Ly}, Chun and {Ma}, Zhiyuan and {MacBride}, Conor and {Maljaars}, Jakob M. and {Muna}, Demitri and {Murphy}, N.~A. and {Norman}, Henrik and {O'Steen}, Richard and {Oman}, Kyle A. and {Pacifici}, Camilla and {Pascual}, Sergio and {Pascual-Granado}, J. and {Patil}, Rohit R. and {Perren}, Gabriel I. and {Pickering}, Timothy E. and {Rastogi}, Tanuj and {Roulston}, Benjamin R. and {Ryan}, Daniel F. and {Rykoff}, Eli S. and {Sabater}, Jose and {Sakurikar}, Parikshit and {Salgado}, Jes{\'u}s and {Sanghi}, Aniket and {Saunders}, Nicholas and {Savchenko}, Volodymyr and {Schwardt}, Ludwig and {Seifert-Eckert}, Michael and {Shih}, Albert Y. and {Jain}, Anany Shrey and {Shukla}, Gyanendra and {Sick}, Jonathan and {Simpson}, Chris and {Singanamalla}, Sudheesh and {Singer}, Leo P. and {Singhal}, Jaladh and {Sinha}, Manodeep and {Sip{\H{o}}cz}, Brigitta M. and {Spitler}, Lee R. and {Stansby}, David and {Streicher}, Ole and {{\v{S}}umak}, Jani and {Swinbank}, John D. and {Taranu}, Dan S. and {Tewary}, Nikita and {Tremblay}, Grant R. and {de Val-Borro}, Miguel and {Van Kooten}, Samuel J. and {Vasovi{\'c}}, Zlatan and {Verma}, Shresth and {de Miranda Cardoso}, Jos{\'e} Vin{\'\i}cius and {Williams}, Peter K.~G. and {Wilson}, Tom J. and {Winkel}, Benjamin and {Wood-Vasey}, W.~M. and {Xue}, Rui and {Yoachim}, Peter and {Zhang}, Chen and {Zonca}, Andrea and {Astropy Project Contributors}},
        title = "{The Astropy Project: Sustaining and Growing a Community-oriented Open-source Project and the Latest Major Release (v5.0) of the Core Package}",
      journal = {\apj},
         year = 2022,
        month = aug,
       volume = {935},
       number = {2},
          eid = {167},
        pages = {167},
          doi = {10.3847/1538-4357/ac7c74},
archivePrefix = {arXiv},
       eprint = {2206.14220},
 primaryClass = {astro-ph.IM},
       adsurl = {https://ui.adsabs.harvard.edu/abs/2022ApJ...935..167A}
}

@ARTICLE{Harris2020,
       author = {{Harris}, Charles R. and {Millman}, K. Jarrod and {van der Walt}, St{\'e}fan J. and {Gommers}, Ralf and {Virtanen}, Pauli and {Cournapeau}, David and {Wieser}, Eric and {Taylor}, Julian and {Berg}, Sebastian and {Smith}, Nathaniel J. and {Kern}, Robert and {Picus}, Matti and {Hoyer}, Stephan and {van Kerkwijk}, Marten H. and {Brett}, Matthew and {Haldane}, Allan and {del R{\'\i}o}, Jaime Fern{\'a}ndez and {Wiebe}, Mark and {Peterson}, Pearu and {G{\'e}rard-Marchant}, Pierre and {Sheppard}, Kevin and {Reddy}, Tyler and {Weckesser}, Warren and {Abbasi}, Hameer and {Gohlke}, Christoph and {Oliphant}, Travis E.},
        title = "{Array programming with NumPy}",
      journal = {\nat},
         year = 2020,
        month = sep,
       volume = {585},
       number = {7825},
        pages = {357-362},
          doi = {10.1038/s41586-020-2649-2},
archivePrefix = {arXiv},
       eprint = {2006.10256},
 primaryClass = {cs.MS},
       adsurl = {https://ui.adsabs.harvard.edu/abs/2020Natur.585..357H}
}

@ARTICLE{SciPy,
  author  = {Virtanen, Pauli and Gommers, Ralf and Oliphant, Travis E. and
            Haberland, Matt and Reddy, Tyler and Cournapeau, David and
            Burovski, Evgeni and Peterson, Pearu and Weckesser, Warren and
            Bright, Jonathan and {van der Walt}, St{\'e}fan J. and
            Brett, Matthew and Wilson, Joshua and Millman, K. Jarrod and
            Mayorov, Nikolay and Nelson, Andrew R. J. and Jones, Eric and
            Kern, Robert and Larson, Eric and Carey, C J and
            Polat, {\.I}lhan and Feng, Yu and Moore, Eric W. and
            {VanderPlas}, Jake and Laxalde, Denis and Perktold, Josef and
            Cimrman, Robert and Henriksen, Ian and Quintero, E. A. and
            Harris, Charles R. and Archibald, Anne M. and
            Ribeiro, Ant{\^o}nio H. and Pedregosa, Fabian and
            {van Mulbregt}, Paul and {SciPy 1.0 Contributors}},
  title   = {{{SciPy} 1.0: Fundamental Algorithms for Scientific
            Computing in Python}},
  journal = {Nature Methods},
  year    = {2020},
  volume  = {17},
  pages   = {261--272},
  adsurl  = {https://rdcu.be/b08Wh},
  doi     = {10.1038/s41592-019-0686-2},
}

@software{pandas,
    author       = {The pandas development team},
    title        = {pandas-dev/pandas: Pandas},
    month        = feb,
    year         = 2020,
    publisher    = {Zenodo},
    version      = {latest},
    doi          = {10.5281/zenodo.3509134},
    url          = {https://doi.org/10.5281/zenodo.3509134}
}

@ARTICLE{Tabone2021,
       author = {{Tabone}, Beno{\^\i}t and {van Hemert}, Marc C. and {van Dishoeck}, Ewine F. and {Black}, John H.},
        title = "{OH mid-infrared emission as a diagnostic of H$_{2}$O UV photodissociation. I. Model and application to the HH 211 shock}",
      journal = {\aap},
         year = 2021,
        month = jun,
       volume = {650},
          eid = {A192},
        pages = {A192},
          doi = {10.1051/0004-6361/202039549},
archivePrefix = {arXiv},
       eprint = {2101.01989},
 primaryClass = {astro-ph.GA},
       adsurl = {https://ui.adsabs.harvard.edu/abs/2021A&A...650A.192T}
}

@ARTICLE{Hocuk2017,
       author = {{Hocuk}, S. and {Sz{\H{u}}cs}, L. and {Caselli}, P. and {Cazaux}, S. and {Spaans}, M. and {Esplugues}, G.~B.},
        title = "{Parameterizing the interstellar dust temperature}",
      journal = {\aap},
         year = 2017,
        month = aug,
       volume = {604},
          eid = {A58},
        pages = {A58},
          doi = {10.1051/0004-6361/201629944},
archivePrefix = {arXiv},
       eprint = {1704.02763},
 primaryClass = {astro-ph.GA},
       adsurl = {https://ui.adsabs.harvard.edu/abs/2017A&A...604A..58H}
}

@BOOK{Allen1973,
       author = {{Allen}, C.~W.},
        title = "{Astrophysical quantities}",
         year = 1973,
       adsurl = {https://ui.adsabs.harvard.edu/abs/1973asqu.book.....A}
}

@Article{Draine.Anderson1985,
  author   = {{Draine}, B.~T. and {Anderson}, N.},
  journal  = {\apj},
  title    = {{Temperature fluctuations and infrared emission from interstellar grains.}},
  year     = {1985},
  month    = may,
  pages    = {494-499},
  volume   = {292},
  adsurl   = {https://ui.adsabs.harvard.edu/abs/1985ApJ...292..494D},
  doi      = {10.1086/163181},
}

@Article{Schoenrich.Binney.ea2010,
  author        = {{Sch{\"o}nrich}, Ralph and {Binney}, James and {Dehnen}, Walter},
  journal       = {\mnras},
  title         = {{Local kinematics and the local standard of rest}},
  year          = {2010},
  month         = apr,
  number        = {4},
  pages         = {1829-1833},
  volume        = {403},
  adsurl        = {https://ui.adsabs.harvard.edu/abs/2010MNRAS.403.1829S},
  archiveprefix = {arXiv},
  doi           = {10.1111/j.1365-2966.2010.16253.x},
  eprint        = {0912.3693},
  primaryclass  = {astro-ph.GA},
}

@Article{Law.Morrison.ea2023,
  author        = {{Law}, David D. and {Morrison}, Jane E. and {Argyriou}, Ioannis and {Patapis}, Polychronis and {{\'A}lvarez-M{\'a}rquez}, J. and {Labiano}, Alvaro and {Vandenbussche}, Bart},
  journal       = {\aj},
  title         = {{A 3D Drizzle Algorithm for JWST and Practical Application to the MIRI Medium Resolution Spectrometer}},
  year          = {2023},
  month         = aug,
  number        = {2},
  pages         = {45},
  volume        = {166},
  adsurl        = {https://ui.adsabs.harvard.edu/abs/2023AJ....166...45L},
  archiveprefix = {arXiv},
  doi           = {10.3847/1538-3881/acdddc},
  eid           = {45},
  eprint        = {2306.05520},
  primaryclass  = {astro-ph.IM},
}

@Article{Podio.Medves.ea2009,
  author        = {{Podio}, L. and {Medves}, S. and {Bacciotti}, F. and {Eisl{\"o}ffel}, J. and {Ray}, T.},
  journal       = {\aap},
  title         = {{Physical structure and dust reprocessing in a sample of HH jets}},
  year          = {2009},
  month         = nov,
  number        = {2},
  pages         = {779-788},
  volume        = {506},
  adsurl        = {https://ui.adsabs.harvard.edu/abs/2009A&A...506..779P},
  archiveprefix = {arXiv},
  doi           = {10.1051/0004-6361/200912408},
  eprint        = {0907.3842},
  primaryclass  = {astro-ph.SR},
}

@Article{Navarro.Nisini.ea2025,
  author        = {{Navarro}, Maria Gabriela and {Nisini}, Brunella and {Giannini}, Teresa and {Kavanagh}, Patrick J. and {Caratti o Garatti}, Alessio and {Antoniucci}, Simone and {Arce}, Hector G. and {Bacciotti}, Francesca and {Cabrit}, Sylvie and {Coffey}, Deirdre and {Dougados}, Catherine and {Eisl{\"o}ffel}, Jochen and {Hartigan}, Patrick and {Crespo}, Alberto Noriega- and {Podio}, Linda and {van Dishoeck}, Ewine F. and {Whelan}, Emma T.},
  journal       = {\apj},
  title         = {{PROJECT-J: The Shocking H$_{2}$ Outflow from HH 46}},
  year          = {2025},
  month         = dec,
  number        = {2},
  pages         = {199},
  volume        = {995},
  adsurl        = {https://ui.adsabs.harvard.edu/abs/2025ApJ...995..199N},
  archiveprefix = {arXiv},
  doi           = {10.3847/1538-4357/ae1f8f},
  eid           = {199},
  eprint        = {2511.17712},
  primaryclass  = {astro-ph.GA},
}

@Article{Ouyed.Pudritz1997,
  author   = {{Ouyed}, Rachid and {Pudritz}, Ralph E.},
  journal  = {\apj},
  title    = {{Numerical Simulations of Astrophysical Jets from Keplerian Disks. II. Episodic Outflows}},
  year     = {1997},
  month    = jul,
  number   = {2},
  pages    = {794-809},
  volume   = {484},
  adsurl   = {https://ui.adsabs.harvard.edu/abs/1997ApJ...484..794O},
  doi      = {10.1086/304355},
}

@Article{Kratter.Lodato2016,
  author        = {{Kratter}, Kaitlin and {Lodato}, Giuseppe},
  journal       = {\araa},
  title         = {{Gravitational Instabilities in Circumstellar Disks}},
  year          = {2016},
  month         = {Sep},
  pages         = {271-311},
  volume        = {54},
  adsurl        = {https://ui.adsabs.harvard.edu/abs/2016ARA&A..54..271K},
  archiveprefix = {arXiv},
  doi           = {10.1146/annurev-astro-081915-023307},
  eprint        = {1603.01280},
  primaryclass  = {astro-ph.SR},
}

@Article{Speedie.Dong.ea2024,
  author        = {{Speedie}, Jessica and {Dong}, Ruobing and {Hall}, Cassandra and {Longarini}, Cristiano and {Veronesi}, Benedetta and {Paneque-Carre{\~n}o}, Teresa and {Lodato}, Giuseppe and {Tang}, Ya-Wen and {Teague}, Richard and {Hashimoto}, Jun},
  journal       = {\nat},
  title         = {{Gravitational instability in a planet-forming disk}},
  year          = {2024},
  month         = sep,
  number        = {8028},
  pages         = {58-62},
  volume        = {633},
  adsurl        = {https://ui.adsabs.harvard.edu/abs/2024Natur.633...58S},
  archiveprefix = {arXiv},
  doi           = {10.1038/s41586-024-07877-0},
  eprint        = {2409.02196},
  primaryclass  = {astro-ph.EP},
}

@Article{Harris.Gry.ea1984,
  author   = {{Harris}, A.~W. and {Gry}, C. and {Bromage}, G.~E.},
  journal  = {\apj},
  title    = {{The correlation of interstellar elememt depletions with mean gas density.}},
  year     = {1984},
  month    = sep,
  pages    = {157-160},
  volume   = {284},
  adsurl   = {https://ui.adsabs.harvard.edu/abs/1984ApJ...284..157H},
  doi      = {10.1086/162395},
}

@Article{Giannini.Nisini.ea2013,
  author        = {{Giannini}, T. and {Nisini}, B. and {Antoniucci}, S. and {Alcal{\'a}}, J.~M. and {Bacciotti}, F. and {Bonito}, R. and {Podio}, L. and {Stelzer}, B. and {Whelan}, E.~T.},
  journal       = {\apj},
  title         = {{The Diagnostic Potential of Fe Lines Applied to Protostellar Jets}},
  year          = {2013},
  month         = nov,
  number        = {1},
  pages         = {71},
  volume        = {778},
  adsurl        = {https://ui.adsabs.harvard.edu/abs/2013ApJ...778...71G},
  archiveprefix = {arXiv},
  doi           = {10.1088/0004-637X/778/1/71},
  eid           = {71},
  eprint        = {1309.5827},
  primaryclass  = {astro-ph.GA},
}

@Misc{NIST_ASD,
author = {A.~Kramida and {Yu.~Ralchenko} and
J.~Reader and {and NIST ASD Team}},
HOWPUBLISHED = {{NIST Atomic Spectra Database
(ver. 5.12), [Online]. Available:
{\tt{https://physics.nist.gov/asd}} [2025, November 28].
National Institute of Standards and Technology,
Gaithersburg, MD.}},
year = {2024},
}

@Article{Kristensen.Ravkilde.ea2007,
  author   = {{Kristensen}, L.~E. and {Ravkilde}, T.~L. and {Field}, D. and {Lemaire}, J.~L. and {Pineau Des For{\^e}ts}, G.},
  journal  = {\aap},
  title    = {{Excitation conditions in the Orion molecular cloud obtained from observations of ortho- and para-lines of H \{2\}}},
  year     = {2007},
  month    = jul,
  number   = {2},
  pages    = {561-574},
  volume   = {469},
  adsurl   = {https://ui.adsabs.harvard.edu/abs/2007A&A...469..561K},
  doi      = {10.1051/0004-6361:20065786},
}

@Article{LeBourlot.PineaudesForets.ea1999,
  author   = {{Le Bourlot}, J. and {Pineau des For{\^e}ts}, G. and {Flower}, D.~R.},
  journal  = {\mnras},
  title    = {{The cooling of astrophysical media by H\_2}},
  year     = {1999},
  month    = may,
  number   = {4},
  pages    = {802-810},
  volume   = {305},
  adsurl   = {https://ui.adsabs.harvard.edu/abs/1999MNRAS.305..802L},
  doi      = {10.1046/j.1365-8711.1999.02497.x},
}

@Article{Wilgenbus.Cabrit.ea2000,
  author   = {{Wilgenbus}, D. and {Cabrit}, S. and {Pineau des For{\^e}ts}, G. and {Flower}, D.~R.},
  journal  = {\aap},
  title    = {{The ortho:para-H\_2 ratio in C- and J-type shocks}},
  year     = {2000},
  month    = apr,
  pages    = {1010-1022},
  volume   = {356},
  adsurl   = {https://ui.adsabs.harvard.edu/abs/2000A&A...356.1010W},
}

@Article{Sternberg.Neufeld1999,
  author        = {{Sternberg}, Amiel and {Neufeld}, David A.},
  journal       = {\apj},
  title         = {{The Ratio of Ortho- to Para-H$_{2}$ in Photodissociation Regions}},
  year          = {1999},
  month         = may,
  number        = {1},
  pages         = {371-380},
  volume        = {516},
  adsurl        = {https://ui.adsabs.harvard.edu/abs/1999ApJ...516..371S},
  archiveprefix = {arXiv},
  doi           = {10.1086/307115},
  eprint        = {astro-ph/9812049},
  primaryclass  = {astro-ph},
}

@Article{Tappe.Lada.ea2008,
  author        = {{Tappe}, A. and {Lada}, C.~J. and {Black}, J.~H. and {Muench}, A.~A.},
  journal       = {\apjl},
  title         = {{Discovery of Superthermal Hydroxyl (OH) in the HH 211 Outflow}},
  year          = {2008},
  month         = jun,
  number        = {2},
  pages         = {L117},
  volume        = {680},
  adsurl        = {https://ui.adsabs.harvard.edu/abs/2008ApJ...680L.117T},
  archiveprefix = {arXiv},
  doi           = {10.1086/589998},
  eprint        = {0805.1737},
  primaryclass  = {astro-ph},
}

@Article{Hollenbach.McKee1979,
  author   = {{Hollenbach}, D. and {McKee}, C.~F.},
  journal  = {\apjs},
  title    = {{Molecule formation and infrared emission in fast interstellar shocks. I. Physical processes.}},
  year     = {1979},
  month    = nov,
  pages    = {555-592},
  volume   = {41},
  adsurl   = {https://ui.adsabs.harvard.edu/abs/1979ApJS...41..555H},
  doi      = {10.1086/190631},
}

@Article{Shu.Najita.ea1994a,
  author   = {{Shu}, Frank and {Najita}, Joan and {Ostriker}, Eve and {Wilkin}, Frank and {Ruden}, Steven and {Lizano}, Susana},
  journal  = {\apj},
  title    = {{Magnetocentrifugally Driven Flows from Young Stars and Disks. I. A Generalized Model}},
  year     = {1994},
  month    = jul,
  pages    = {781},
  volume   = {429},
  adsurl   = {https://ui.adsabs.harvard.edu/abs/1994ApJ...429..781S},
  doi      = {10.1086/174363},
}

@Article{Pudritz.Norman1983a,
  author   = {{Pudritz}, R.~E. and {Norman}, C.~A.},
  journal  = {\apj},
  title    = {{Centrifugally driven winds from contracting molecular disks}},
  year     = {1983},
  month    = nov,
  pages    = {677-697},
  volume   = {274},
  adsurl   = {https://ui.adsabs.harvard.edu/abs/1983ApJ...274..677P},
  doi      = {10.1086/161481},
}

@InProceedings{Pudritz.Ouyed.ea2007,
  author        = {{Pudritz}, R.~E. and {Ouyed}, R. and {Fendt}, Ch. and {Brandenburg}, A.},
  booktitle     = {PP V},
  title         = {{Disk Winds, Jets, and Outflows: Theoretical and Computational Foundations}},
  year          = {2007},
  editor        = {{Reipurth}, Bo and {Jewitt}, David and {Keil}, Klaus},
  month         = jan,
  pages         = {277},
  adsurl        = {https://ui.adsabs.harvard.edu/abs/2007prpl.conf..277P},
  archiveprefix = {arXiv},
  doi           = {10.48550/arXiv.astro-ph/0603592},
  eprint        = {astro-ph/0603592},
  primaryclass  = {astro-ph},
}

@InProceedings{Frank.Ray.ea2014,
  author        = {{Frank}, A. and {Ray}, T.~P. and {Cabrit}, S. and {Hartigan}, P. and {Arce}, H.~G. and {Bacciotti}, F. and {Bally}, J. and {Benisty}, M. and {Eisl{\"o}ffel}, J. and {G{\"u}del}, M. and {Lebedev}, S. and {Nisini}, B. and {Raga}, A.},
  booktitle     = {PP VI},
  title         = {{Jets and Outflows from Star to Cloud: Observations Confront Theory}},
  year          = {2014},
  editor        = {{Beuther}, H. and {Klessen}, Ralf S. and {Dullemond}, Cornelis P. and {Henning}, Thomas},
  pages         = {451-474},
  publisher     = {Tucson, AZ: Univ. Arizona Press},
  adsurl        = {http://adsabs.harvard.edu/abs/2014prpl.conf..451F},
  archiveprefix = {arXiv},
  doi           = {10.2458/azu_uapress_9780816531240-ch020},
  eprint        = {1402.3553},
  primaryclass  = {astro-ph.SR},
}

@Article{Skretas.Karska.ea2025,
  author        = {{Skretas}, I.~M. and {Karska}, A. and {Francis}, L. and {Rocha}, W.~R.~M. and {van Gelder}, M.~L. and {Tychoniec}, {\L}. and {Figueira}, M. and {Sewi{\l}o}, M. and {Wyrowski}, F. and {Schilke}, P.},
  journal       = {\aap},
  title         = {{UV-irradiated outflows from low-mass protostars in Ophiuchus with JWST/MIRI}},
  year          = {2025},
  month         = nov,
  pages         = {A139},
  volume        = {703},
  adsurl        = {https://ui.adsabs.harvard.edu/abs/2025A&A...703A.139S},
  archiveprefix = {arXiv},
  doi           = {10.1051/0004-6361/202554977},
  eid           = {A139},
  eprint        = {2509.10256},
  primaryclass  = {astro-ph.GA},
}

@Article{Raga.Canto.ea1990,
  author   = {{Raga}, A.~C. and {Canto}, J. and {Binette}, L. and {Calvet}, N.},
  journal  = {\apj},
  title    = {{Stellar Jets with Intrinsically Variable Sources}},
  year     = {1990},
  month    = dec,
  pages    = {601},
  volume   = {364},
  adsurl   = {https://ui.adsabs.harvard.edu/abs/1990ApJ...364..601R},
  doi      = {10.1086/169443},
}

@Article{Chapman.Mundy.ea2008,
  author    = {Chapman, Nicholas L. and Mundy, Lee G. and Lai, Shih-Ping and Evans, Neal J.},
  journal   = {\apj},
  title     = {THE MID-INFRARED EXTINCTION LAW IN THE OPHIUCHUS, PERSEUS, AND SERPENS MOLECULAR CLOUDS},
  year      = {2008},
  month     = {dec},
  number    = {1},
  pages     = {496},
  volume    = {690},
  doi       = {10.1088/0004-637X/690/1/496},
  publisher = {The American Astronomical Society},
  url       = {https://doi.org/10.1088/0004-637X/690/1/496},
}

@Article{Shu.Shang.ea1997,
  author   = {{Shu}, F.~H. and {Shang}, H. and {Glassgold}, A.~E. and {Lee}, T.},
  journal  = {Science},
  title    = {{X-rays and fluctuating X-winds from protostars.}},
  year     = {1997},
  month    = jan,
  pages    = {1475-1479},
  volume   = {277},
  adsurl   = {https://ui.adsabs.harvard.edu/abs/1997Sci...277.1475S},
  doi      = {10.1126/science.277.5331.1475},
}

@Article{Grewal.Nie.ea2024,
  author        = {{Grewal}, Damanveer S. and {Nie}, Nicole X. and {Zhang}, Bidong and {Izidoro}, Andre and {Asimow}, Paul D.},
  journal       = {Nature Astronomy},
  title         = {{Accretion of the earliest inner Solar System planetesimals beyond the water snowline}},
  year          = {2024},
  month         = mar,
  pages         = {290-297},
  volume        = {8},
  adsurl        = {https://ui.adsabs.harvard.edu/abs/2024NatAs...8..290G},
  archiveprefix = {arXiv},
  doi           = {10.1038/s41550-023-02172-w},
  eprint        = {2408.17032},
  primaryclass  = {astro-ph.EP},
}

@Article{Nisini.Santangelo.ea2015,
  author        = {{Nisini}, B. and {Santangelo}, G. and {Giannini}, T. and {Antoniucci}, S. and {Cabrit}, S. and {Codella}, C. and {Davis}, C.~J. and {Eisl{\"o}ffel}, J. and {Kristensen}, L. and {Herczeg}, G. and {Neufeld}, D. and {van Dishoeck}, E.~F.},
  journal       = {\apj},
  title         = {{[O I] 63 {$\mu$}m Jets in Class 0 Sources Detected By Herschel}},
  year          = {2015},
  month         = mar,
  pages         = {121},
  volume        = {801},
  adsurl        = {http://adsabs.harvard.edu/abs/2015ApJ...801..121N},
  archiveprefix = {arXiv},
  doi           = {10.1088/0004-637X/801/2/121},
  eid           = {121},
  eprint        = {1501.03681},
  primaryclass  = {astro-ph.SR},
}

@Article{Dullemond.Monnier2010,
  author        = {{Dullemond}, C.~P. and {Monnier}, J.~D.},
  journal       = {\araa},
  title         = {{The Inner Regions of Protoplanetary Disks}},
  year          = {2010},
  month         = sep,
  pages         = {205-239},
  volume        = {48},
  adsurl        = {https://ui.adsabs.harvard.edu/abs/2010ARA&A..48..205D},
  archiveprefix = {arXiv},
  doi           = {10.1146/annurev-astro-081309-130932},
  eprint        = {1006.3485},
  primaryclass  = {astro-ph.SR},
}

@Article{vanDishoeck.Tychoniec.ea2025,
  author        = {{van Dishoeck}, E.~F. and {Tychoniec}, {\L}. and {Rocha}, W.~R.~M. and {Slavicinska}, K. and {Francis}, L. and {van Gelder}, M.~L. and {Ray}, T.~P. and {Beuther}, H. and {Caratti o Garatti}, A. and {Brunken}, N.~G.~C. and {Chen}, Y. and {Devaraj}, R. and {Geers}, V.~C. and {Gieser}, C. and {Greene}, T.~P. and {Justtanont}, K. and {Le Gouellec}, V.~J.~M. and {Kavanagh}, P.~J. and {Klaassen}, P.~D. and {Janssen}, A.~G.~M. and {Navarro}, M.~G. and {Nazari}, P. and {Notsu}, S. and {Perotti}, G. and {Ressler}, M.~E. and {Reyes}, S.~D. and {Sellek}, A.~D. and {Tabone}, B. and {Tap}, C. and {Theijssen}, N.~C.~M.~A. and {Colina}, L. and {G{\"u}del}, M. and {Henning}, Th. and {Lagage}, P.-O. and {{\"O}stlin}, G. and {Vandenbussche}, B. and {Wright}, G.~S.},
  journal       = {\aap},
  title         = {{JWST Observations of Young protoStars (JOYS): Overview of program and early results}},
  year          = {2025},
  month         = jul,
  pages         = {A361},
  volume        = {699},
  adsurl        = {https://ui.adsabs.harvard.edu/abs/2025A&A...699A.361V},
  archiveprefix = {arXiv},
  doi           = {10.1051/0004-6361/202554444},
  eid           = {A361},
  eprint        = {2505.08002},
  primaryclass  = {astro-ph.GA},
}

@Article{Gusdorf.Riquelme.ea2015,
  author        = {{Gusdorf}, A. and {Riquelme}, D. and {Anderl}, S. and {Eisl{\"o}ffel}, J. and {Codella}, C. and {G{\'o}mez-Ruiz}, A.~I. and {Graf}, U.~U. and {Kristensen}, L.~E. and {Leurini}, S. and {Parise}, B. and {Requena-Torres}, M.~A. and {Ricken}, O. and {G{\"u}sten}, R.},
  journal       = {\aap},
  title         = {{Impacts of pure shocks in the BHR71 bipolar outflow}},
  year          = {2015},
  month         = mar,
  pages         = {A98},
  volume        = {575},
  adsurl        = {https://ui.adsabs.harvard.edu/abs/2015A&A...575A..98G},
  archiveprefix = {arXiv},
  doi           = {10.1051/0004-6361/201425142},
  eid           = {A98},
  eprint        = {1502.00488},
  primaryclass  = {astro-ph.GA},
}

@Article{Micolta.Calvet.ea2024,
  author        = {{Micolta}, Marbely and {Calvet}, Nuria and {Thanathibodee}, Thanawuth and {Magris C.}, Gladis and {Manara}, Carlo F. and {Venuti}, Laura and {Alcal{\'a}}, Juan Manuel and {Herczeg}, Gregory J.},
  journal       = {\apj},
  title         = {{Using the Ca II Lines in T Tauri Stars to Infer the Abundance of Refractory Elements in the Innermost Disk Region}},
  year          = {2024},
  month         = dec,
  number        = {2},
  pages         = {251},
  volume        = {976},
  adsurl        = {https://ui.adsabs.harvard.edu/abs/2024ApJ...976..251M},
  archiveprefix = {arXiv},
  doi           = {10.3847/1538-4357/ad8884},
  eid           = {251},
  eprint        = {2410.17327},
  primaryclass  = {astro-ph.SR},
}

@Article{McClure.vantHoff.ea2025,
  author   = {{McClure}, M.~K. and {van't Hoff}, Merel and {Francis}, Logan and {Bergin}, Edwin and {Rocha}, Will R.~M. and {Sturm}, J.~A. and {Harsono}, Daniel and {van Dishoeck}, Ewine F. and {Black}, John H. and {Noble}, J.~A. and {Qasim}, D. and {Dartois}, E.},
  journal  = {\nat},
  title    = {{Refractory solid condensation detected in an embedded protoplanetary disk}},
  year     = {2025},
  month    = jul,
  number   = {8072},
  pages    = {649-653},
  volume   = {643},
  adsurl   = {https://ui.adsabs.harvard.edu/abs/2025Natur.643..649M},
  doi      = {10.1038/s41586-025-09163-z},
}

@InProceedings{Davis.Alexander.ea2014,
  author    = {{Davis}, A.~M. and {Alexander}, C.~M.~O. 'D. and {Ciesla}, F.~J. and {Gounelle}, M. and {Krot}, A.~N. and {Petaev}, M.~I. and {Stephan}, T.},
  booktitle = {PP VI},
  title     = {{Samples of the Solar System: Recent Developments}},
  year      = {2014},
  editor    = {{Beuther}, Henrik and {Klessen}, Ralf S. and {Dullemond}, Cornelis P. and {Henning}, Thomas},
  month     = jan,
  pages     = {809-831},
  adsurl    = {https://ui.adsabs.harvard.edu/abs/2014prpl.conf..809D},
  doi       = {10.2458/azu_uapress_9780816531240-ch035},
}

@Article{Phuong.Lee.ea2025,
  author        = {{Phuong}, Nguyen Thi and {Lee}, Chang Won and {Tobin}, John J. and {Ohashi}, Nagayoshi and {J{\o}rgensen}, Jes K. and {Takakuwa}, Shigehisa and {Aikawa}, Yuri and {Aso}, Yusuke and {Li}, Zhi-Yun and {Koch}, Patrick M. and {Williams}, Jonathan P. and {Gavino}, Sacha and {Lin}, Zhe-Yu Daniel and {Tomida}, Kengo and {Kwon}, Woojin and {Looney}, Leslie W. and {Han}, Ilseung and {Santamar{\'\i}a-Miranda}, Alejandro and {Lai}, Shih-Ping and {Yen}, Hsi-Wei and {Thieme}, Travis J. and {Sai}, Jinshi and {Flores}, Christian},
  journal       = {\apj},
  title         = {{Early Planet Formation in Embedded Disks (eDisk). XXII. Keplerian Disk, Disk Structures, and Jets/Outflows in the Class 0 Protostar IRAS 04166+2706}},
  year          = {2025},
  month         = oct,
  number        = {1},
  pages         = {18},
  volume        = {992},
  adsurl        = {https://ui.adsabs.harvard.edu/abs/2025ApJ...992...18P},
  archiveprefix = {arXiv},
  doi           = {10.3847/1538-4357/adfa92},
  eid           = {18},
  eprint        = {2508.07212},
  primaryclass  = {astro-ph.SR},
}

@Article{Giannini.Nisini.ea2011,
  author        = {{Giannini}, Teresa and {Nisini}, Brunella and {Neufeld}, David and {Yuan}, Yuan and {Antoniucci}, Simone and {Gusdorf}, Antoine},
  journal       = {\apj},
  title         = {{Spitzer Spectral Line Mapping of Protostellar Outflows. III. H$_{2}$ Emission in L1448, BHR71, and NGC2071}},
  year          = {2011},
  month         = sep,
  number        = {1},
  pages         = {80},
  volume        = {738},
  adsurl        = {https://ui.adsabs.harvard.edu/abs/2011ApJ...738...80G},
  archiveprefix = {arXiv},
  doi           = {10.1088/0004-637X/738/1/80},
  eid           = {80},
  eprint        = {1106.2722},
  primaryclass  = {astro-ph.SR},
}

@Article{Neufeld.Nisini.ea2009,
  author        = {{Neufeld}, David A. and {Nisini}, Brunella and {Giannini}, Teresa and {Melnick}, Gary J. and {Bergin}, Edwin A. and {Yuan}, Yuan and {Maret}, S{\'e}bastien and {Tolls}, Volker and {G{\"u}sten}, Rolf and {Kaufman}, Michael J.},
  journal       = {\apj},
  title         = {{Spitzer Spectral Line Mapping of Protostellar Outflows. I. Basic Data and Outflow Energetics}},
  year          = {2009},
  month         = nov,
  number        = {1},
  pages         = {170-183},
  volume        = {706},
  adsurl        = {https://ui.adsabs.harvard.edu/abs/2009ApJ...706..170N},
  archiveprefix = {arXiv},
  doi           = {10.1088/0004-637X/706/1/170},
  eprint        = {0910.1107},
  primaryclass  = {astro-ph.GA},
}

@Article{Sabatini.Podio.ea2024,
  author        = {{Sabatini}, G. and {Podio}, L. and {Codella}, C. and {Watanabe}, Y. and {De Simone}, M. and {Bianchi}, E. and {Ceccarelli}, C. and {Chandler}, C.~J. and {Sakai}, N. and {Svoboda}, B. and {Testi}, L. and {Aikawa}, Y. and {Balucani}, N. and {Bouvier}, M. and {Caselli}, P. and {Caux}, E. and {Chahine}, L. and {Charnley}, S. and {Cuello}, N. and {Dulieu}, F. and {Evans}, L. and {Fedele}, D. and {Feng}, S. and {Fontani}, F. and {Hama}, T. and {Hanawa}, T. and {Herbst}, E. and {Hirota}, T. and {Isella}, A. and {J{\'\i}menez-Serra}, I. and {Johnstone}, D. and {Lefloch}, B. and {Le Gal}, R. and {Loinard}, L. and {Liu}, H.~B. and {L{\'o}pez-Sepulcre}, A. and {Maud}, L.~T. and {Maureira}, M.~J. and {Menard}, F. and {Miotello}, A. and {Moellenbrock}, G. and {Nomura}, H. and {Oba}, Y. and {Ohashi}, S. and {Okoda}, Y. and {Oya}, Y. and {Pineda}, J. and {Rimola}, A. and {Sakai}, T. and {Segura-Cox}, D. and {Shirley}, Y. and {Vastel}, C. and {Viti}, S. and {Watanabe}, N. and {Zhang}, Y. and {Zhang}, Z.~E. and {Yamamoto}, S.},
  journal       = {\aap},
  title         = {{FAUST. XIII. Dusty cavity and molecular shock driven by IRS7B in the Corona Australis cluster}},
  year          = {2024},
  month         = apr,
  pages         = {L12},
  volume        = {684},
  adsurl        = {https://ui.adsabs.harvard.edu/abs/2024A&A...684L..12S},
  archiveprefix = {arXiv},
  doi           = {10.1051/0004-6361/202449616},
  eid           = {L12},
  eprint        = {2403.18108},
  primaryclass  = {astro-ph.GA},
}

@Article{Sabatini.Bianchi.ea2025,
  author        = {{Sabatini}, G. and {Bianchi}, E. and {Chandler}, C.~J. and {Cacciapuoti}, L. and {Podio}, L. and {Maureira}, M.~J. and {Codella}, C. and {Ceccarelli}, C. and {Sakai}, N. and {Testi}, L. and {Toci}, C. and {Svoboda}, B. and {Sakai}, T. and {Bouvier}, M. and {Caselli}, P. and {Cuello}, N. and {De Simone}, M. and {J{\'\i}menez-Serra}, I. and {Johnstone}, D. and {Loinard}, L. and {Zhang}, Z.~E. and {Yamamoto}, S.},
  journal       = {\aap},
  title         = {{FAUST: XXIV. Large dust grains in the protostellar outflow cavity walls of the Class I binary L1551 IRS5}},
  year          = {2025},
  month         = jun,
  pages         = {L16},
  volume        = {698},
  adsurl        = {https://ui.adsabs.harvard.edu/abs/2025A&A...698L..16S},
  archiveprefix = {arXiv},
  doi           = {10.1051/0004-6361/202554750},
  eid           = {L16},
  eprint        = {2505.13596},
  primaryclass  = {astro-ph.GA},
}

@Article{Hull.LeGouellec.ea2020,
  author        = {{Hull}, Charles L.~H. and {Le Gouellec}, Valentin J.~M. and {Girart}, Josep M. and {Tobin}, John J. and {Bourke}, Tyler L.},
  journal       = {\apj},
  title         = {{Understanding the Origin of the Magnetic Field Morphology in the Wide-binary Protostellar System BHR 71}},
  year          = {2020},
  month         = apr,
  number        = {2},
  pages         = {152},
  volume        = {892},
  adsurl        = {https://ui.adsabs.harvard.edu/abs/2020ApJ...892..152H},
  archiveprefix = {arXiv},
  doi           = {10.3847/1538-4357/ab5809},
  eid           = {152},
  eprint        = {1910.07290},
  primaryclass  = {astro-ph.SR},
}

@Article{Smith.Bally.ea2005,
  author   = {Smith, Nathan and Bally, John and Shuping, Ralph Y. and Morris, Mark and Kassis, Marc},
  journal  = {\aj},
  title    = {Thermal Dust Emission from Proplyds, Unresolved Disks, and Shocks in the Orion Nebula*},
  year     = {2005},
  month    = {oct},
  number   = {4},
  pages    = {1763},
  volume   = {130},
  doi      = {10.1086/432912},
  url      = {https://dx.doi.org/10.1086/432912},
}

@Article{Konstantopoulou.DeCia.ea2022,
  author        = {{Konstantopoulou}, Christina and {De Cia}, Annalisa and {Krogager}, Jens-Kristian and {Ledoux}, C{\'e}dric and {Noterdaeme}, Pasquier and {Fynbo}, Johan P.~U. and {Heintz}, Kasper E. and {Watson}, Darach and {Andersen}, Anja C. and {Ramburuth-Hurt}, Tanita and {Jermann}, Iris},
  journal       = {\aap},
  title         = {{Dust depletion of metals from local to distant galaxies. I. Peculiar nucleosynthesis effects and grain growth in the ISM}},
  year          = {2022},
  month         = oct,
  pages         = {A12},
  volume        = {666},
  adsurl        = {https://ui.adsabs.harvard.edu/abs/2022A&A...666A..12K},
  archiveprefix = {arXiv},
  doi           = {10.1051/0004-6361/202243994},
  eid           = {A12},
  eprint        = {2207.08804},
  primaryclass  = {astro-ph.GA},
}

@Article{Black.vanDishoeck1987,
  author   = {{Black}, John H. and {van Dishoeck}, Ewine F.},
  journal  = {\apj},
  title    = {{Fluorescent Excitation of Interstellar H 2}},
  year     = {1987},
  month    = nov,
  pages    = {412},
  volume   = {322},
  adsurl   = {https://ui.adsabs.harvard.edu/abs/1987ApJ...322..412B},
  doi      = {10.1086/165740},
}

@Article{Harsono.Bjerkeli.ea2023a,
  author        = {{Harsono}, D. and {Bjerkeli}, P. and {Ramsey}, J.~P. and {Pontoppidan}, K.~M. and {Kristensen}, L.~E. and {J{\o}rgensen}, J.~K. and {Calcutt}, H. and {Li}, Z. -Y. and {Plunkett}, A.},
  journal       = {\apjl},
  title         = {{JWST Peers into the Class I Protostar TMC1A: Atomic Jet and Spatially Resolved Dissociative Shock Region}},
  year          = {2023},
  month         = jul,
  number        = {2},
  pages         = {L32},
  volume        = {951},
  adsurl        = {https://ui.adsabs.harvard.edu/abs/2023ApJ...951L..32H},
  archiveprefix = {arXiv},
  doi           = {10.3847/2041-8213/acdfca},
  eid           = {L32},
  eprint        = {2306.08380},
  primaryclass  = {astro-ph.SR},
}

@Article{Anglada.Lopez.ea2007,
  author        = {{Anglada}, Guillem and {L{\'o}pez}, Rosario and {Estalella}, Robert and {Masegosa}, Josefa and {Riera}, Angels and {Raga}, Alejandro C.},
  journal       = {\aj},
  title         = {{Proper Motions of the Jets in the Region of HH 30 and HL/XZ Tau: Evidence for a Binary Exciting Source of the HH 30 Jet}},
  year          = {2007},
  month         = jun,
  number        = {6},
  pages         = {2799-2814},
  volume        = {133},
  adsurl        = {https://ui.adsabs.harvard.edu/abs/2007AJ....133.2799A},
  archiveprefix = {arXiv},
  doi           = {10.1086/517493},
  eprint        = {astro-ph/0703155},
  primaryclass  = {astro-ph},
}

@software{2022zndo...5608581E,
       author = {{Erb}, Donald},
        title = "{pybaselines: A Python library of algorithms for the baseline correction of experimental data}",
         year = 2025,
        month = mar,
          eid = {10.5281/zenodo.5608581},
          doi = {10.5281/zenodo.5608581},
      version = {1.2.0},
    publisher = {Zenodo},
       adsurl = {https://ui.adsabs.harvard.edu/abs/2022zndo...5608581E}
}

@Article{Louvet.Dougados.ea2018,
  author        = {{Louvet}, F. and {Dougados}, C. and {Cabrit}, S. and {Mardones}, D. and {M{\'e}nard}, F. and {Tabone}, B. and {Pinte}, C. and {Dent}, W.~R.~F.},
  journal       = {\aap},
  title         = {{The HH30 edge-on T Tauri star. A rotating and precessing monopolar outflow scrutinized by ALMA}},
  year          = {2018},
  month         = oct,
  pages         = {A120},
  volume        = {618},
  adsurl        = {https://ui.adsabs.harvard.edu/abs/2018A&A...618A.120L},
  archiveprefix = {arXiv},
  doi           = {10.1051/0004-6361/201731733},
  eid           = {A120},
  eprint        = {1808.03285},
  primaryclass  = {astro-ph.GA},
}

@Article{Lee.Hasegawa.ea2010,
  author        = {{Lee}, Chin-Fei and {Hasegawa}, Tatsuhiko I. and {Hirano}, Naomi and {Palau}, Aina and {Shang}, Hsien and {Ho}, Paul T.~P. and {Zhang}, Qizhou},
  journal       = {\apj},
  title         = {{The Reflection-Symmetric Wiggle of the Young Protostellar Jet HH 211}},
  year          = {2010},
  month         = apr,
  number        = {2},
  pages         = {731-737},
  volume        = {713},
  adsurl        = {https://ui.adsabs.harvard.edu/abs/2010ApJ...713..731L},
  archiveprefix = {arXiv},
  doi           = {10.1088/0004-637X/713/2/731},
  eprint        = {1003.1355},
  primaryclass  = {astro-ph.GA},
}

@Article{Maret.Bergin.ea2009,
  author        = {{Maret}, S{\'e}bastien and {Bergin}, Edwin A. and {Neufeld}, David A. and {Green}, Joel D. and {Watson}, Dan M. and {Harwit}, Martin O. and {Kristensen}, Lars E. and {Melnick}, Gary J. and {Sonnentrucker}, Paule and {Tolls}, Volker and {Werner}, Michael W. and {Willacy}, Karen and {Yuan}, Yuan},
  journal       = {\apj},
  title         = {{Spitzer Mapping of Molecular Hydrogen Pure Rotational Lines in NGC 1333: A Detailed Study of Feedback in Star Formation}},
  year          = {2009},
  month         = jun,
  number        = {2},
  pages         = {1244-1260},
  volume        = {698},
  adsurl        = {https://ui.adsabs.harvard.edu/abs/2009ApJ...698.1244M},
  archiveprefix = {arXiv},
  doi           = {10.1088/0004-637X/698/2/1244},
  eprint        = {0904.0603},
  primaryclass  = {astro-ph.SR},
}

@Article{Rosenthal.Bertoldi.ea2000,
  author        = {{Rosenthal}, D. and {Bertoldi}, F. and {Drapatz}, S.},
  journal       = {\aap},
  title         = {{ISO-SWS observations of OMC-1: H\_2 and fine structure lines}},
  year          = {2000},
  month         = apr,
  pages         = {705-723},
  volume        = {356},
  adsurl        = {https://ui.adsabs.harvard.edu/abs/2000A&A...356..705R},
  archiveprefix = {arXiv},
  doi           = {10.48550/arXiv.astro-ph/0002456},
  eprint        = {astro-ph/0002456},
  primaryclass  = {astro-ph},
}

@InProceedings{vanderBliek.Norman.ea2004,
  author    = {{van der Bliek}, Nicole S. and {Norman}, Dara and {Blum}, Robert D. and {Probst}, Ronald G. and {Montane}, Andres and {Galvez}, Ramon and {Warner}, Michael and {Tighe}, Roberto and {Delgado}, Francisco and {Martinez}, Manuel},
  booktitle = {Ground-based Instrumentation for Astronomy},
  title     = {{ISPI: a wide-field NIR imager for the CTIO Blanco 4-m telescope}},
  year      = {2004},
  editor    = {{Moorwood}, Alan F.~M. and {Iye}, Masanori},
  month     = sep,
  pages     = {1582-1589},
  series    = {SPIE},
  volume    = {5492},
  adsurl    = {https://ui.adsabs.harvard.edu/abs/2004SPIE.5492.1582V},
  doi       = {10.1117/12.550973},
}

@InProceedings{Martini.Persson.ea2004,
  author        = {{Martini}, Paul and {Persson}, S.~E. and {Murphy}, David C. and {Birk}, Christoph and {Shectman}, Stephen A. and {Gunnels}, Steve M. and {Koch}, Erich},
  booktitle     = {Ground-based Instrumentation for Astronomy},
  title         = {{PANIC: a near-infrared camera for the Magellan telescopes}},
  year          = {2004},
  editor        = {{Moorwood}, Alan F.~M. and {Iye}, Masanori},
  month         = sep,
  pages         = {1653-1660},
  series        = {SPIE},
  volume        = {5492},
  adsurl        = {https://ui.adsabs.harvard.edu/abs/2004SPIE.5492.1653M},
  archiveprefix = {arXiv},
  doi           = {10.1117/12.551828},
  eprint        = {astro-ph/0406666},
  primaryclass  = {astro-ph},
}

@Article{Tobin.Hartmann.ea2010,
  author        = {{Tobin}, John J. and {Hartmann}, Lee and {Looney}, Leslie W. and {Chiang}, Hsin-Fang},
  journal       = {\apj},
  title         = {{Complex Structure in Class 0 Protostellar Envelopes}},
  year          = {2010},
  month         = apr,
  number        = {2},
  pages         = {1010-1028},
  volume        = {712},
  adsurl        = {https://ui.adsabs.harvard.edu/abs/2010ApJ...712.1010T},
  archiveprefix = {arXiv},
  doi           = {10.1088/0004-637X/712/2/1010},
  eprint        = {1002.2362},
  primaryclass  = {astro-ph.SR},
}

@Article{Crouzet.Mueller.ea2025,
  author        = {{Crouzet}, N. and {Mueller}, M. and {Sargent}, B. and {Lahuis}, F. and {Kester}, D. and {Yang}, G. and {Argyriou}, I. and {Gasman}, D. and {Kavanagh}, P.~J. and {Labiano}, A. and {Larson}, K. and {Law}, D.~R. and {{\'A}lvarez-M{\'a}rquez}, J. and {Brandl}, B.~R. and {Glasse}, A. and {Patapis}, P. and {Roelfsema}, P.~R. and {Tychoniec}, {\L}. and {van Dishoeck}, E.~F. and {Wright}, G.~S.},
  journal       = {\aap},
  title         = {{Extended source fringe flats for the JWST MIRI Medium Resolution Spectrometer}},
  year          = {2025},
  month         = may,
  pages         = {A77},
  volume        = {698},
  adsurl        = {https://ui.adsabs.harvard.edu/abs/2025A&A...698A..77C},
  archiveprefix = {arXiv},
  doi           = {10.1051/0004-6361/202452903},
  eid           = {A77},
  eprint        = {2504.11328},
  primaryclass  = {astro-ph.IM},
}

@Misc{Perrin.Long.ea2025,
  author    = {Perrin, Marshall and Long, Joseph and Osborne, Shannon and Geda, Robel and Sappington, Bradley and Meléndez, Marcio and Lajoie, Charles-Philippe and Leisenring, Jarron and Zimmerman, Neil and Brooks, Keira and Otor, O. Justin and Kulp, Trey and Chambers, Lauren and Jurling, Alden},
  title     = {STPSF},
  year      = {2025},
  copyright = {BSD 3-Clause "New" or "Revised" License},
  doi       = {10.5281/ZENODO.15747364},
  publisher = {Zenodo},
}

@Article{Mottram.Kristensen.ea2014,
  author        = {{Mottram}, J.~C. and {Kristensen}, L.~E. and {van Dishoeck}, E.~F. and {Bruderer}, S. and {San Jos{\'e}-Garc{\'{\i}}a}, I. and {Karska}, A. and {Visser}, R. and {Santangelo}, G. and {Benz}, A.~O. and {Bergin}, E.~A. and {Caselli}, P. and {Herpin}, F. and {Hogerheijde}, M.~R. and {Johnstone}, D. and {van Kempen}, T.~A. and {Liseau}, R. and {Nisini}, B. and {Tafalla}, M. and {van der Tak}, F.~F.~S. and {Wyrowski}, F.},
  journal       = {\aap},
  title         = {{Water in star-forming regions with Herschel (WISH). V. The physical conditions in low-mass protostellar outflows revealed by multi-transition water observations}},
  year          = {2014},
  month         = dec,
  pages         = {A21},
  volume        = {572},
  adsurl        = {http://adsabs.harvard.edu/abs/2014A%26A...572A..21M},
  archiveprefix = {arXiv},
  doi           = {10.1051/0004-6361/201424267},
  eid           = {A21},
  eprint        = {1409.5704},
  primaryclass  = {astro-ph.SR},
}

@Article{Kadam.Vorobyov.ea2025,
  author        = {{Kadam}, Kundan and {Vorobyov}, Eduard and {Woitke}, Peter and {G{\"u}del}, Manuel},
  journal       = {\aap},
  title         = {{Dust in the wind of outbursting young stars}},
  year          = {2025},
  month         = may,
  pages         = {A43},
  volume        = {697},
  adsurl        = {https://ui.adsabs.harvard.edu/abs/2025A&A...697A..43K},
  archiveprefix = {arXiv},
  doi           = {10.1051/0004-6361/202554021},
  eid           = {A43},
  eprint        = {2503.24077},
  primaryclass  = {astro-ph.SR},
}

@Article{Labdon.Kraus.ea2023,
  author        = {{Labdon}, Aaron and {Kraus}, Stefan and {Davies}, Claire L. and {Kreplin}, Alexander and {Zarrilli}, Sebastian and {Monnier}, John D. and {Le Bouquin}, Jean-Baptiste and {Anugu}, Narsireddy and {Setterholm}, Benjamin and {Gardner}, Tyler and {Ennis}, Jacob and {Lanthermann}, Cyprien and {ten Brummelaar}, Theo and {Schaefer}, Gail and {Harries}, Tim J.},
  journal       = {\aap},
  title         = {{Imaging the warped dusty disk wind environment of SU Aurigae with MIRC-X}},
  year          = {2023},
  month         = oct,
  pages         = {A6},
  volume        = {678},
  adsurl        = {https://ui.adsabs.harvard.edu/abs/2023A&A...678A...6L},
  archiveprefix = {arXiv},
  doi           = {10.1051/0004-6361/202245813},
  eid           = {A6},
  eprint        = {2306.06240},
  primaryclass  = {astro-ph.SR},
}

@Article{Dartois.Noble.ea2025,
  author        = {{Dartois}, E. and {Noble}, J.~A. and {McClure}, M.~K. and {Sturm}, J.~A. and {Beck}, T.~L. and {Arulanantham}, N. and {Drozdovskaya}, M.~N. and {Espaillat}, C.~C. and {Harsono}, D. and {Palumbo}, M.-E. and {Pendleton}, Y.~J. and {Pontoppidan}, K.~M.},
  journal       = {\aap},
  title         = {{The edge-on disc Tau 042021: Icy grains at high altitudes and a wind containing astronomical polycyclic aromatic hydrocarbons}},
  year          = {2025},
  month         = jun,
  pages         = {A8},
  volume        = {698},
  adsurl        = {https://ui.adsabs.harvard.edu/abs/2025A&A...698A...8D},
  archiveprefix = {arXiv},
  doi           = {10.1051/0004-6361/202452966},
  eid           = {A8},
  eprint        = {2503.24309},
  primaryclass  = {astro-ph.EP},
}

@Article{Delabrosse.Dougados.ea2024,
  author        = {{Delabrosse}, V. and {Dougados}, C. and {Cabrit}, S. and {Tabone}, B. and {Tychoniec}, L. and {Ray}, T. and {Podio}, L. and {McClure}, M.},
  journal       = {\aap},
  title         = {{JWST study of the DG Tau B disk-wind candidate. I. Overview and nested H$_{2}$-CO outflows}},
  year          = {2024},
  month         = aug,
  pages         = {A173},
  volume        = {688},
  adsurl        = {https://ui.adsabs.harvard.edu/abs/2024A&A...688A.173D},
  archiveprefix = {arXiv},
  doi           = {10.1051/0004-6361/202449176},
  eid           = {A173},
  eprint        = {2403.19400},
  primaryclass  = {astro-ph.SR},
}

@Article{Nisini.Giannini.ea2010,
  author        = {{Nisini}, Brunella and {Giannini}, Teresa and {Neufeld}, David A. and {Yuan}, Yuan and {Antoniucci}, Simone and {Bergin}, Edwin A. and {Melnick}, Gary J.},
  journal       = {\apj},
  title         = {{Spitzer Spectral Line Mapping of Protostellar Outflows. II. H$_{2}$ Emission in L1157}},
  year          = {2010},
  month         = nov,
  number        = {1},
  pages         = {69-79},
  volume        = {724},
  adsurl        = {https://ui.adsabs.harvard.edu/abs/2010ApJ...724...69N},
  archiveprefix = {arXiv},
  doi           = {10.1088/0004-637X/724/1/69},
  eprint        = {1009.3834},
  primaryclass  = {astro-ph.SR},
}

@Article{Schwarz.Henning.ea2024,
  author        = {{Schwarz}, Kamber R. and {Henning}, Thomas and {Christiaens}, Valentin and {Gasman}, Danny and {Samland}, Matthias and {Perotti}, Giulia and {Jang}, Hyerin and {Grant}, Sierra L. and {Tabone}, Beno{\^\i}t and {Morales-Calder{\'o}n}, Maria and {Kamp}, Inga and {van Dishoeck}, Ewine F. and {G{\"u}del}, Manuel and {Lagage}, Pierre-Olivier and {Barrado}, David and {Caratti o Garatti}, Alessio and {Glauser}, Adrian M. and {Ray}, Tom P. and {Vandenbussche}, Bart and {Waters}, L.~B.~F.~M. and {Arabhavi}, Aditya M. and {Kanwar}, Jayatee and {Olofsson}, G{\"o}ran and {Rodgers-Lee}, Donna and {Schreiber}, J{\"u}rgen and {Temmink}, Milou},
  journal       = {\apj},
  title         = {{MINDS. JWST/MIRI Reveals a Dynamic Gas-rich Inner Disk inside the Cavity of SY Cha}},
  year          = {2024},
  month         = feb,
  number        = {1},
  pages         = {8},
  volume        = {962},
  adsurl        = {https://ui.adsabs.harvard.edu/abs/2024ApJ...962....8S},
  archiveprefix = {arXiv},
  doi           = {10.3847/1538-4357/ad1393},
  eid           = {8},
  eprint        = {2312.07135},
  primaryclass  = {astro-ph.EP},
}

@Article{Gieser.Beuther.ea2024,
  author   = {{Gieser}, C. and {Beuther}, H. and {van Dishoeck}, E.~F. and {Francis}, L. and {van Gelder}, M.~L. and {Tychoniec}, L. and {Kavanagh}, P.~J. and {Perotti}, G. and {Caratti o Garatti}, A. and {Ray}, T.~P. and {Klaassen}, P. and {Justtanont}, K. and {Linnartz}, H. and {Rocha}, W.~R.~M. and {Slavicinska}, K. and {Colina}, L. and {G{\"u}del}, M. and {Henning}, Th. and {Lagage}, P. -O. and {{\"O}stlin}, G. and {Vandenbussche}, B. and {Waelkens}, C. and {Wright}, G.},
  journal  = {\aap},
  title    = {{JOYS: Disentangling the warm and cold material in the high-mass IRAS 23385+6053 cluster (Corrigendum)}},
  year     = {2024},
  month    = may,
  pages    = {C5},
  volume   = {685},
  adsurl   = {https://ui.adsabs.harvard.edu/abs/2024A&A...685C...5G},
  doi      = {10.1051/0004-6361/202450520e},
  eid      = {C5},
}

@Article{Zapata.FernandezLopez.ea2018,
  author        = {{Zapata}, Luis A. and {Fern{\'a}ndez-L{\'o}pez}, Manuel and {Rodr{\'\i}guez}, Luis F. and {Garay}, Guido and {Takahashi}, Satoko and {Lee}, Chin-Fei and {Hern{\'a}ndez-G{\'o}mez}, Antonio},
  journal       = {\aj},
  title         = {{ALMA Reveals a Collision between Protostellar Outflows in BHR 71}},
  year          = {2018},
  month         = nov,
  number        = {5},
  pages         = {239},
  volume        = {156},
  adsurl        = {https://ui.adsabs.harvard.edu/abs/2018AJ....156..239Z},
  archiveprefix = {arXiv},
  doi           = {10.3847/1538-3881/aae51e},
  eid           = {239},
  eprint        = {1804.00625},
  primaryclass  = {astro-ph.SR},
}

@Article{Lahuis.vanDishoeck.ea2010,
  author        = {{Lahuis}, F. and {van Dishoeck}, E.~F. and {J{\o}rgensen}, J.~K. and {Blake}, G.~A. and {Evans}, N.~J.},
  journal       = {\aap},
  title         = {{c2d Spitzer IRS spectra of embedded low-mass young stars: gas-phase emission lines}},
  year          = {2010},
  month         = sep,
  pages         = {A3},
  volume        = {519},
  adsurl        = {http://adsabs.harvard.edu/abs/2010A%26A...519A...3L},
  archiveprefix = {arXiv},
  doi           = {10.1051/0004-6361/200913957},
  eid           = {A3},
  eprint        = {1005.2867},
}

@Article{Kristensen.vanDishoeck.ea2012,
  author        = {{Kristensen}, L.~E. and {van Dishoeck}, E.~F. and {Bergin}, E.~A. and {Visser}, R. and {Y{\i}ld{\i}z}, U.~A. and {San Jose-Garcia}, I. and {J{\o}rgensen}, J.~K. and {Herczeg}, G.~J. and {Johnstone}, D. and {Wampfler}, S.~F. and {Benz}, A.~O. and {Bruderer}, S. and {Cabrit}, S. and {Caselli}, P. and {Doty}, S.~D. and {Harsono}, D. and {Herpin}, F. and {Hogerheijde}, M.~R. and {Karska}, A. and {van Kempen}, T.~A. and {Liseau}, R. and {Nisini}, B. and {Tafalla}, M. and {van der Tak}, F. and {Wyrowski}, F.},
  journal       = {\aap},
  title         = {{Water in star-forming regions with Herschel (WISH). II. Evolution of 557 GHz 1$_{10}$-1$_{01}$ emission in low-mass protostars}},
  year          = {2012},
  month         = jun,
  pages         = {A8},
  volume        = {542},
  adsurl        = {http://adsabs.harvard.edu/abs/2012A%26A...542A...8K},
  archiveprefix = {arXiv},
  doi           = {10.1051/0004-6361/201118146},
  eid           = {A8},
  eprint        = {1204.0009},
  primaryclass  = {astro-ph.SR},
}

@Article{Bertoldi.Timmermann.ea1999,
  author        = {{Bertoldi}, Frank and {Timmermann}, Ralf and {Rosenthal}, Dirk and {Drapatz}, Siegfried and {Wright}, Christopher M.},
  journal       = {\aap},
  title         = {{Detection of HD in the Orion molecular outflow}},
  year          = {1999},
  month         = jun,
  pages         = {267-277},
  volume        = {346},
  adsurl        = {https://ui.adsabs.harvard.edu/abs/1999A&A...346..267B},
  archiveprefix = {arXiv},
  doi           = {10.48550/arXiv.astro-ph/9904261},
  eprint        = {astro-ph/9904261},
  primaryclass  = {astro-ph},
}

@Article{Barsony.WolfChase.ea2010,
  author   = {{Barsony}, Mary and {Wolf-Chase}, Grace A. and {Ciardi}, David R. and {O'Linger}, JoAnn},
  journal  = {\apj},
  title    = {{IRS Scan-mapping of the Wasp-waist Nebula (IRAS 16253-2429). I. Derivation of Shock Conditions from H$_{2}$ Emission and Discovery of 11.3 {\ensuremath{\mu}}m PAH Absorption}},
  year     = {2010},
  month    = sep,
  number   = {1},
  pages    = {64-86},
  volume   = {720},
  adsurl   = {https://ui.adsabs.harvard.edu/abs/2010ApJ...720...64B},
  doi      = {10.1088/0004-637X/720/1/64},
}

@Article{vanGelder.Francis.ea2024,
  author        = {{van Gelder}, M.~L. and {Francis}, L. and {van Dishoeck}, E.~F. and {Tychoniec}, {\L}. and {Ray}, T.~P. and {Beuther}, H. and {Caratti o Garatti}, A. and {Chen}, Y. and {Devaraj}, R. and {Gieser}, C. and {Justtanont}, K. and {Kavanagh}, P.~J. and {Nazari}, P. and {Reyes}, S. and {Rocha}, W.~R.~M. and {Slavicinska}, K. and {G{\"u}del}, M. and {Henning}, Th. and {Lagage}, P. -O. and {Wright}, G.},
  journal       = {\aap},
  title         = {{JWST Observations of Young protoStars (JOYS): Overview of gaseous molecular emission and absorption in low-mass protostars}},
  year          = {2024},
  month         = dec,
  pages         = {A197},
  volume        = {692},
  adsurl        = {https://ui.adsabs.harvard.edu/abs/2024A&A...692A.197V},
  archiveprefix = {arXiv},
  doi           = {10.1051/0004-6361/202451967},
  eid           = {A197},
  eprint        = {2410.01636},
  primaryclass  = {astro-ph.SR},
}

@Article{Francis.vanDishoeck.ea2025,
  author        = {{Francis}, L. and {van Dishoeck}, E.~F. and {Caratti o Garatti}, A. and {van Gelder}, M.~L. and {Gieser}, C. and {Beuther}, H. and {Ray}, T.~P. and {Tychoniec}, L. and {Nazari}, P. and {Reyes}, S. and {Kavanagh}, P.~J. and {Klaassen}, P. and {G{\"u}del}, M. and {Henning}, T.},
  journal       = {\aap},
  title         = {{JOYS: The [D/H] abundance derived from protostellar outflows across the Galactic disk measured with JWST}},
  year          = {2025},
  month         = feb,
  pages         = {A174},
  volume        = {694},
  adsurl        = {https://ui.adsabs.harvard.edu/abs/2025A&A...694A.174F},
  archiveprefix = {arXiv},
  doi           = {10.1051/0004-6361/202451629},
  eid           = {A174},
  eprint        = {2501.02085},
  primaryclass  = {astro-ph.GA},
}

@Article{Neufeld.Melnick.ea1998,
  author   = {{Neufeld}, David A. and {Melnick}, Gary J. and {Harwit}, Martin},
  journal  = {\apjl},
  title    = {{Infrared Space Observatory Observations of Molecular Hydrogen in HH 54:Measurement of a Nonequilibrium Ratio of Ortho- to Para-H$_{2}$}},
  year     = {1998},
  month    = oct,
  number   = {1},
  pages    = {L75-L78},
  volume   = {506},
  adsurl   = {https://ui.adsabs.harvard.edu/abs/1998ApJ...506L..75N},
  doi      = {10.1086/311636},
}

@Article{Neufeld.Melnick.ea2006,
  author        = {{Neufeld}, David A. and {Melnick}, Gary J. and {Sonnentrucker}, Paule and {Bergin}, Edwin A. and {Green}, Joel D. and {Kim}, Kyoung Hee and {Watson}, Dan M. and {Forrest}, William J. and {Pipher}, Judith L.},
  journal       = {\apj},
  title         = {{Spitzer Observations of HH 54 and HH 7-11: Mapping the H$_{2}$ Ortho-to-Para Ratio in Shocked Molecular Gas}},
  year          = {2006},
  month         = oct,
  number        = {2},
  pages         = {816-835},
  volume        = {649},
  adsurl        = {https://ui.adsabs.harvard.edu/abs/2006ApJ...649..816N},
  archiveprefix = {arXiv},
  doi           = {10.1086/506604},
  eprint        = {astro-ph/0606232},
  primaryclass  = {astro-ph},
}

@Article{Okoda.Yang.ea2025,
  author        = {{Okoda}, Yuki and {Yang}, Yao-Lun and {Evans}, II, Neal J. and {Kim}, Jaeyeong and {Jin}, Mihwa and {Garrod}, Robin T. and {Francis}, Logan and {Johnstone}, Doug and {Ceccarelli}, Cecilia and {Codella}, Claudio and {Chandler}, Claire J. and {Yamamoto}, Satoshi and {Sakai}, Nami},
  journal       = {\apj},
  title         = {{CORINOS. III. Outflow Shocked Regions of the Low-mass Protostellar Source IRAS 15398{\textendash}3359 with JWST and ALMA}},
  year          = {2025},
  month         = apr,
  number        = {2},
  pages         = {149},
  volume        = {982},
  adsurl        = {https://ui.adsabs.harvard.edu/abs/2025ApJ...982..149O},
  archiveprefix = {arXiv},
  doi           = {10.3847/1538-4357/adb83f},
  eid           = {149},
  eprint        = {2503.03050},
  primaryclass  = {astro-ph.SR},
}

@article{Christiaens.Gonzalez.ea2023,
  author   = {{Christiaens}, Valentin and {Gonzalez}, Carlos and {Farkas}, Ralf and {Dahlqvist}, Carl-Henrik and {Nasedkin}, Evert and {Milli}, Julien and {Absil}, Olivier and {Ngo}, Henry and {Cantero}, Carles and {Rainot}, Alan and {Hammond}, Iain and {Bonse}, Markus and {Cantalloube}, Faustine and {Vigan}, Arthur and {Kompella}, Vijay and {Hancock}, Paul},
  journal  = {JOSS},
  title    = {{VIP: A Python package for high-contrast imaging}},
  year     = {2023},
  month    = jan,
  number   = {81},
  pages    = {4774},
  volume   = {8},
  adsurl   = {https://ui.adsabs.harvard.edu/abs/2023JOSS....8.4774C},
  doi      = {10.21105/joss.04774},
  eid      = {4774},
}

@Misc{Bushouse.Eisenhamer.ea2023,
  author       = {{Bushouse}, Howard and {Eisenhamer}, Jonathan and {Dencheva}, Nadia and {Davies}, James and {Greenfield}, Perry and {Morrison}, Jane and {Hodge}, Phil and {Simon}, Bernie and {Grumm}, David and {Droettboom}, Michael and {Slavich}, Edward and {Sosey}, Megan and {Pauly}, Tyler and {Miller}, Todd and {Jedrzejewski}, Robert and {Hack}, Warren and {Davis}, David and {Crawford}, Steven and {Law}, David and {Gordon}, Karl and {Regan}, Michael and {Cara}, Mihai and {MacDonald}, Ken and {Bradley}, Larry and {Shanahan}, Clare and {Jamieson}, William and {Teodoro}, Mairan and {Williams}, Thomas},
  howpublished = {Zenodo},
  month        = mar,
  title        = {{JWST Calibration Pipeline}},
  year         = {2023},
  adsurl       = {https://ui.adsabs.harvard.edu/abs/2023zndo...7714020B},
  doi          = {10.5281/zenodo.7714020},
  eid          = {10.5281/zenodo.7714020},
  publisher    = {Zenodo},
  version      = {1.9.6},
}

@Article{Wright.Wright.ea2015,
  author        = {{Wright}, G.~S. and {Wright}, David and {Goodson}, G.~B. and {Rieke}, G.~H. and {Aitink-Kroes}, Gabby and {Amiaux}, J. and {Aricha-Yanguas}, Ana and {Azzollini}, Ruym{\'a}n and {Banks}, Kimberly and {Barrado-Navascues}, D. and {Belenguer-Davila}, T. and {Bloemmart}, J.~A.~D.~L. and {Bouchet}, Patrice and {Brandl}, B.~R. and {Colina}, L. and {Detre}, {\"O}rs and {Diaz-Catala}, Eva and {Eccleston}, Paul and {Friedman}, Scott D. and {Garc{\'\i}a-Mar{\'\i}n}, Macarena and {G{\"u}del}, Manuel and {Glasse}, Alistair and {Glauser}, Adrian M. and {Greene}, T.~P. and {Groezinger}, Uli and {Grundy}, Tim and {Hastings}, Peter and {Henning}, Th. and {Hofferbert}, Ralph and {Hunter}, Faye and {Jessen}, N.~C. and {Justtanont}, K. and {Karnik}, Avinash R. and {Khorrami}, Mori A. and {Krause}, Oliver and {Labiano}, Alvaro and {Lagage}, P. -O. and {Langer}, Ulrich and {Lemke}, Dietrich and {Lim}, Tanya and {Lorenzo-Alvarez}, Jose and {Mazy}, Emmanuel and {McGowan}, Norman and {Meixner}, M.~E. and {Morris}, Nigel and {Morrison}, Jane E. and {M{\"u}ller}, Friedrich and {rgaard-Nielson}, H. -U. N{\o} and {Olofsson}, G{\"o}ran and {O'Sullivan}, Brian and {Pel}, J. -W. and {Penanen}, Konstantin and {Petach}, M.~B. and {Pye}, J.~P. and {Ray}, T.~P. and {Renotte}, Etienne and {Renouf}, Ian and {Ressler}, M.~E. and {Samara-Ratna}, Piyal and {Scheithauer}, Silvia and {Schneider}, Analyn and {Shaughnessy}, Bryan and {Stevenson}, Tim and {Sukhatme}, Kalyani and {Swinyard}, Bruce and {Sykes}, Jon and {Thatcher}, John and {Tikkanen}, Tuomo and {van Dishoeck}, E.~F. and {Waelkens}, C. and {Walker}, Helen and {Wells}, Martyn and {Zhender}, Alex},
  journal       = {\pasp},
  title         = {{The Mid-Infrared Instrument for the James Webb Space Telescope, II: Design and Build}},
  year          = {2015},
  month         = jul,
  number        = {953},
  pages         = {595},
  volume        = {127},
  adsurl        = {https://ui.adsabs.harvard.edu/abs/2015PASP..127..595W},
  archiveprefix = {arXiv},
  doi           = {10.1086/682253},
  eprint        = {1508.02333},
  primaryclass  = {astro-ph.IM},
}

@Article{Rigby.Perrin.ea2023,
  author        = {{Rigby}, Jane and {Perrin}, Marshall and {McElwain}, Michael and {Kimble}, Randy and {Friedman}, Scott and {Lallo}, Matt and {Doyon}, Ren{\'e} and {Feinberg}, Lee and {Ferruit}, Pierre and {Glasse}, Alistair and {Rieke}, Marcia and {Rieke}, George and {Wright}, Gillian and {Willott}, Chris and {Colon}, Knicole and {Milam}, Stefanie and {Neff}, Susan and {Stark}, Christopher and {Valenti}, Jeff and {Abell}, Jim and {Abney}, Faith and {Abul-Huda}, Yasin and {Acton}, D. Scott and {Adams}, Evan and {Adler}, David and {Aguilar}, Jonathan and {Ahmed}, Nasif and {Albert}, Lo{\"\i}c and {Alberts}, Stacey and {Aldridge}, David and {Allen}, Marsha and {Altenburg}, Martin and {{\'A}lvarez-M{\'a}rquez}, Javier and {Alves de Oliveira}, Catarina and {Andersen}, Greg and {Anderson}, Harry and {Anderson}, Sara and {Argyriou}, Ioannis and {Armstrong}, Amber and {Arribas}, Santiago and {Artigau}, Etienne and {Arvai}, Amanda and {Atkinson}, Charles and {Bacon}, Gregory and {Bair}, Thomas and {Banks}, Kimberly and {Barrientes}, Jaclyn and {Barringer}, Bruce and {Bartosik}, Peter and {Bast}, William and {Baudoz}, Pierre and {Beatty}, Thomas and {Bechtold}, Katie and {Beck}, Tracy and {Bergeron}, Eddie and {Bergkoetter}, Matthew and {Bhatawdekar}, Rachana and {Birkmann}, Stephan and {Blazek}, Ronald and {Blome}, Claire and {Boccaletti}, Anthony and {B{\"o}ker}, Torsten and {Boia}, John and {Bonaventura}, Nina and {Bond}, Nicholas and {Bosley}, Kari and {Boucarut}, Ray and {Bourque}, Matthew and {Bouwman}, Jeroen and {Bower}, Gary and {Bowers}, Charles and {Boyer}, Martha and {Bradley}, Larry and {Brady}, Greg and {Braun}, Hannah and {Breda}, David and {Bresnahan}, Pamela and {Bright}, Stacey and {Britt}, Christopher and {Bromenschenkel}, Asa and {Brooks}, Brian and {Brooks}, Keira and {Brown}, Bob and {Brown}, Matthew and {Brown}, Patricia and {Bunker}, Andy and {Burger}, Matthew and {Bushouse}, Howard and {Cale}, Steven and {Cameron}, Alex and {Cameron}, Peter and {Canipe}, Alicia and {Caplinger}, James and {Caputo}, Francis and {Cara}, Mihai and {Carey}, Larkin and {Carniani}, Stefano and {Carrasquilla}, Maria and {Carruthers}, Margaret and {Case}, Michael and {Catherine}, Riggs and {Chance}, Don and {Chapman}, George and {Charlot}, St{\'e}phane and {Charlow}, Brian and {Chayer}, Pierre and {Chen}, Bin and {Cherinka}, Brian and {Chichester}, Sarah and {Chilton}, Zack and {Chonis}, Taylor and {Clampin}, Mark and {Clark}, Charles and {Clark}, Kerry and {Coe}, Dan and {Coleman}, Benee and {Comber}, Brian and {Comeau}, Tom and {Connolly}, Dennis and {Cooper}, James and {Cooper}, Rachel and {Coppock}, Eric and {Correnti}, Matteo and {Cossou}, Christophe and {Coulais}, Alain and {Coyle}, Laura and {Cracraft}, Misty and {Curti}, Mirko and {Cuturic}, Steven and {Davis}, Katherine and {Davis}, Michael and {Dean}, Bruce and {DeLisa}, Amy and {deMeester}, Wim and {Dencheva}, Nadia and {Dencheva}, Nadezhda and {DePasquale}, Joseph and {Deschenes}, Jeremy and {Hunor Detre}, {\"O}rs and {Diaz}, Rosa and {Dicken}, Dan and {DiFelice}, Audrey and {Dillman}, Matthew and {Dixon}, William and {Doggett}, Jesse and {Donaldson}, Tom and {Douglas}, Rob and {DuPrie}, Kimberly and {Dupuis}, Jean and {Durning}, John and {Easmin}, Nilufar and {Eck}, Weston and {Edeani}, Chinwe and {Egami}, Eiichi and {Ehrenwinkler}, Ralf and {Eisenhamer}, Jonathan and {Eisenhower}, Michael and {Elie}, Michelle and {Elliott}, James and {Elliott}, Kyle and {Ellis}, Tracy and {Engesser}, Michael and {Espinoza}, Nestor and {Etienne}, Odessa and {Etxaluze}, Mireya and {Falini}, Patrick and {Feeney}, Matthew and {Ferry}, Malcolm and {Filippazzo}, Joseph and {Fincham}, Brian and {Fix}, Mees and {Flagey}, Nicolas and {Florian}, Michael and {Flynn}, Jim and {Fontanella}, Erin and {Ford}, Terrance and {Forshay}, Peter and {Fox}, Ori and {Franz}, David and {Fu}, Henry and {Fullerton}, Alexander and {Galkin}, Sergey and {Galyer}, Anthony and {Garc{\'\i}a Mar{\'\i}n}, Macarena and {Gardner}, Jonathan P. and {Gardner}, Lisa and {Garland}, Dennis and {Garrett}, Bruce and {Gasman}, Danny and {Gaspar}, Andras and {Gaudreau}, Daniel and {Gauthier}, Peter and {Geers}, Vincent and {Geithner}, Paul and {Gennaro}, Mario and {Giardino}, Giovanna and {Girard}, Julien and {Giuliano}, Mark and {Glassmire}, Kirk and {Glauser}, Adrian and {Glazer}, Stuart and {Godfrey}, John and {Golimowski}, David and {Gollnitz}, David and {Gong}, Fan and {Gonzaga}, Shireen and {Gordon}, Michael and {Gordon}, Karl and {Goudfrooij}, Paul and {Greene}, Thomas and {Greenhouse}, Matthew and {Grimaldi}, Stefano and {Groebner}, Andrew and {Grundy}, Timothy and {Guillard}, Pierre and {Gutman}, Irvin and {Ha}, Kong Q. and {Haderlein}, Peter and {Hagedorn}, Andria and {Hainline}, Kevin and {Haley}, Craig and {Hami}, Maryam and {Hamilton}, Forrest and {Hammel}, Heidi and {Hansen}, Carl and {Harkins}, Tom and {Harr}, Michael and {Hart}, Jessica and {Hart}, Quyen and {Hartig}, George and {Hashimoto}, Ryan and {Haskins}, Sujee and {Hathaway}, William and {Havey}, Keith and {Hayden}, Brian and {Hecht}, Karen and {Heller-Boyer}, Chris and {Henriques}, Caroline and {Henry}, Alaina and {Hermann}, Karl and {Hernandez}, Scarlin and {Hesman}, Brigette and {Hicks}, Brian and {Hilbert}, Bryan and {Hines}, Dean and {Hoffman}, Melissa and {Holfeltz}, Sherie and {Holler}, Bryan J. and {Hoppa}, Jennifer and {Hott}, Kyle and {Howard}, Joseph M. and {Howard}, Rick and {Hunter}, Alexander and {Hunter}, David and {Hurst}, Brendan and {Husemann}, Bernd and {Hustak}, Leah and {Ilinca Ignat}, Luminita and {Illingworth}, Garth and {Irish}, Sandra and {Jackson}, Wallace and {Jahromi}, Amir and {Jakobsen}, Peter and {James}, LeAndrea and {James}, Bryan and {Januszewski}, William and {Jenkins}, Ann and {Jirdeh}, Hussein and {Johnson}, Phillip and {Johnson}, Timothy and {Jones}, Vicki and {Jones}, Ron and {Jones}, Danny and {Jones}, Olivia and {Jordan}, Ian and {Jordan}, Margaret and {Jurczyk}, Sarah and {Jurling}, Alden and {Kaleida}, Catherine and {Kalmanson}, Phillip and {Kammerer}, Jens and {Kang}, Huijo and {Kao}, Shaw-Hong and {Karakla}, Diane and {Kavanagh}, Patrick and {Kelly}, Doug and {Kendrew}, Sarah and {Kennedy}, Herbert and {Kenny}, Deborah and {Keski-kuha}, Ritva and {Keyes}, Charles and {Kidwell}, Richard and {Kinzel}, Wayne and {Kirk}, Jeff and {Kirkpatrick}, Mark and {Kirshenblat}, Danielle and {Klaassen}, Pamela and {Knapp}, Bryan and {Knight}, J. Scott and {Knollenberg}, Perry and {Koehler}, Robert and {Koekemoer}, Anton and {Kovacs}, Aiden and {Kulp}, Trey and {Kumari}, Nimisha and {Kyprianou}, Mark and {La Massa}, Stephanie and {Labador}, Aurora and {Labiano}, Alvaro and {Lagage}, Pierre-Olivier and {Lajoie}, Charles-Philippe and {Lallo}, Matthew and {Lam}, May and {Lamb}, Tracy and {Lambros}, Scott and {Lampenfield}, Richard and {Langston}, James and {Larson}, Kirsten and {Law}, David and {Lawrence}, Jon and {Lee}, David and {Leisenring}, Jarron and {Lepo}, Kelly and {Leveille}, Michael and {Levenson}, Nancy and {Levine}, Marie and {Levy}, Zena and {Lewis}, Dan and {Lewis}, Hannah and {Libralato}, Mattia and {Lightsey}, Paul and {Link}, Miranda and {Liu}, Lily and {Lo}, Amy and {Lockwood}, Alexandra and {Logue}, Ryan and {Long}, Chris and {Long}, Douglas and {Loomis}, Charles and {Lopez-Caniego}, Marcos and {Lorenzo Alvarez}, Jose and {Love-Pruitt}, Jennifer and {Lucy}, Adrian and {Luetzgendorf}, Nora and {Maghami}, Peiman and {Maiolino}, Roberto and {Major}, Melissa and {Malla}, Sunita and {Malumuth}, Eliot and {Manjavacas}, Elena and {Mannfolk}, Crystal and {Marrione}, Amanda and {Marston}, Anthony and {Martel}, Andr{\'e} and {Maschmann}, Marc and {Masci}, Gregory and {Masciarelli}, Michaela and {Maszkiewicz}, Michael and {Mather}, John and {McKenzie}, Kenny and {McLean}, Brian and {McMaster}, Matthew and {Melbourne}, Katie and {Mel{\'e}ndez}, Marcio and {Menzel}, Michael and {Merz}, Kaiya and {Meyett}, Michele and {Meza}, Luis and {Miskey}, Cherie and {Misselt}, Karl and {Moller}, Christopher and {Morrison}, Jane and {Morse}, Ernie and {Moseley}, Harvey and {Mosier}, Gary and {Mountain}, Matt and {Mueckay}, Julio and {Mueller}, Michael and {Mullally}, Susan and {Murphy}, Jess and {Murray}, Katherine and {Murray}, Claire and {Mustelier}, David and {Muzerolle}, James and {Mycroft}, Matthew and {Myers}, Richard and {Myrick}, Kaila and {Nanavati}, Shashvat and {Nance}, Elizabeth and {Nayak}, Omnarayani and {Naylor}, Bret and {Nelan}, Edmund and {Nickson}, Bryony and {Nielson}, Alethea and {Nieto-Santisteban}, Maria and {Nikolov}, Nikolay and {Noriega-Crespo}, Alberto and {O'Shaughnessy}, Brian and {O'Sullivan}, Brian and {Ochs}, William and {Ogle}, Patrick and {Oleszczuk}, Brenda and {Olmsted}, Joseph and {Osborne}, Shannon and {Ottens}, Richard and {Owens}, Beverly and {Pacifici}, Camilla and {Pagan}, Alyssa and {Page}, James and {Park}, Sang and {Parrish}, Keith and {Patapis}, Polychronis and {Paul}, Lee and {Pauly}, Tyler and {Pavlovsky}, Cheryl and {Pedder}, Andrew and {Peek}, Matthew and {Pena-Guerrero}, Maria and {Penanen}, Konstantin and {Perez}, Yesenia and {Perna}, Michele and {Perriello}, Beth and {Phillips}, Kevin and {Pietraszkiewicz}, Martin and {Pinaud}, Jean-Paul and {Pirzkal}, Norbert and {Pitman}, Joseph and {Piwowar}, Aidan and {Platais}, Vera and {Player}, Danielle and {Plesha}, Rachel and {Pollizi}, Joe and {Polster}, Ethan and {Pontoppidan}, Klaus and {Porterfield}, Blair and {Proffitt}, Charles and {Pueyo}, Laurent and {Pulliam}, Christine and {Quirt}, Brian and {Quispe Neira}, Irma and {Ramos Alarcon}, Rafael and {Ramsay}, Leah and {Rapp}, Greg and {Rapp}, Robert and {Rauscher}, Bernard and {Ravindranath}, Swara and {Rawle}, Timothy and {Regan}, Michael and {Reichard}, Timothy A. and {Reis}, Carl and {Ressler}, Michael E. and {Rest}, Armin and {Reynolds}, Paul and {Rhue}, Timothy and {Richon}, Karen and {Rickman}, Emily and {Ridgaway}, Michael and {Ritchie}, Christine and {Rix}, Hans-Walter and {Robberto}, Massimo and {Robinson}, Gregory and {Robinson}, Michael and {Robinson}, Orion and {Rock}, Frank and {Rodriguez}, David and {Rodriguez Del Pino}, Bruno and {Roellig}, Thomas and {Rohrbach}, Scott and {Roman}, Anthony and {Romelfanger}, Fred and {Rose}, Perry and {Roteliuk}, Anthony and {Roth}, Marc and {Rothwell}, Braden and {Rowlands}, Neil and {Roy}, Arpita and {Royer}, Pierre and {Royle}, Patricia and {Rui}, Chunlei and {Rumler}, Peter and {Runnels}, Joel and {Russ}, Melissa and {Rustamkulov}, Zafar and {Ryden}, Grant and {Ryer}, Holly and {Sabata}, Modhumita and {Sabatke}, Derek and {Sabbi}, Elena and {Samuelson}, Bridget and {Sapp}, Benjamin and {Sappington}, Bradley and {Sargent}, B. and {Sauer}, Arne and {Scheithauer}, Silvia and {Schlawin}, Everett and {Schlitz}, Joseph and {Schmitz}, Tyler and {Schneider}, Analyn and {Schreiber}, J{\"u}rgen and {Schulze}, Vonessa and {Schwab}, Ryan and {Scott}, John and {Sembach}, Kenneth and {Shanahan}, Clare and {Shaughnessy}, Bryan and {Shaw}, Richard and {Shawger}, Nanci and {Shay}, Christopher and {Sheehan}, Evan and {Shen}, Sharon and {Sherman}, Allan and {Shiao}, Bernard and {Shih}, Hsin-Yi and {Shivaei}, Irene and {Sienkiewicz}, Matthew and {Sing}, David and {Sirianni}, Marco and {Sivaramakrishnan}, Anand and {Skipper}, Joy and {Sloan}, G.~C. and {Slocum}, Christine and {Slowinski}, Steven and {Smith}, Erin and {Smith}, Eric and {Smith}, Denise and {Smith}, Corbett and {Snyder}, Gregory and {Soh}, Warren and {Sohn}, Sangmo Tony and {Soto}, Christian and {Spencer}, Richard and {Stallcup}, Scott and {Stansberry}, John and {Starr}, Carl and {Starr}, Elysia and {Stewart}, Alphonso and {Stiavelli}, Massimo and {Straughn}, Amber and {Strickland}, David and {Stys}, Jeff and {Summers}, Francis and {Sun}, Fengwu and {Sunnquist}, Ben and {Swade}, Daryl and {Swam}, Michael and {Swaters}, Robert and {Swoish}, Robby and {Taylor}, Joanna M. and {Taylor}, Rolanda and {Te Plate}, Maurice and {Tea}, Mason and {Teague}, Kelly and {Telfer}, Randal and {Temim}, Tea and {Thatte}, Deepashri and {Thompson}, Christopher and {Thompson}, Linda and {Thomson}, Shaun and {Tikkanen}, Tuomo and {Tippet}, William and {Todd}, Connor and {Toolan}, Sharon and {Tran}, Hien and {Trejo}, Edwin and {Truong}, Justin and {Tsukamoto}, Chris and {Tustain}, Samuel and {Tyra}, Harrison and {Ubeda}, Leonardo and {Underwood}, Kelli and {Uzzo}, Michael and {Van Campen}, Julie and {Vandal}, Thomas and {Vandenbussche}, Bart and {Vila}, Bego{\~n}a and {Volk}, Kevin and {Wahlgren}, Glenn and {Waldman}, Mark and {Walker}, Chanda and {Wander}, Michel and {Warfield}, Christine and {Warner}, Gerald and {Wasiak}, Matthew and {Watkins}, Mitchell and {Weaver}, Andrew and {Weilert}, Mark and {Weiser}, Nick and {Weiss}, Ben and {Weissman}, Sarah and {Welty}, Alan and {West}, Garrett and {Wheate}, Lauren and {Wheatley}, Elizabeth and {Wheeler}, Thomas and {White}, Rick and {Whiteaker}, Kevin and {Whitehouse}, Paul and {Whiteleather}, Jennifer and {Whitman}, William and {Williams}, Christina and {Willmer}, Christopher and {Willoughby}, Scott and {Wilson}, Andrew and {Wirth}, Gregory and {Wislowski}, Emily and {Wolf}, Erin and {Wolfe}, David and {Wolff}, Schuyler and {Workman}, Bill and {Wright}, Ray and {Wu}, Carl and {Wu}, Rai and {Wymer}, Kristen and {Yates}, Kayla and {Yeager}, Christopher and {Yeates}, Jared and {Yerger}, Ethan and {Yoon}, Jinmi and {Young}, Alice and {Yu}, Susan and {Zak}, Dean and {Zeidler}, Peter and {Zhou}, Julia and {Zielinski}, Thomas and {Zincke}, Cristian and {Zonak}, Stephanie},
  journal       = {\pasp},
  title         = {{The Science Performance of JWST as Characterized in Commissioning}},
  year          = {2023},
  month         = apr,
  number        = {1046},
  pages         = {048001},
  volume        = {135},
  adsurl        = {https://ui.adsabs.harvard.edu/abs/2023PASP..135d8001R},
  archiveprefix = {arXiv},
  doi           = {10.1088/1538-3873/acb293},
  eid           = {048001},
  eprint        = {2207.05632},
  primaryclass  = {astro-ph.IM},
}

@Article{Rieke.Ressler.ea2015,
  author        = {{Rieke}, G.~H. and {Ressler}, M.~E. and {Morrison}, Jane E. and {Bergeron}, L. and {Bouchet}, Patrice and {Garc{\'\i}a-Mar{\'\i}n}, Macarena and {Greene}, T.~P. and {Regan}, M.~W. and {Sukhatme}, K.~G. and {Walker}, Helen},
  journal       = {\pasp},
  title         = {{The Mid-Infrared Instrument for the James Webb Space Telescope, VII: The MIRI Detectors}},
  year          = {2015},
  month         = jul,
  number        = {953},
  pages         = {665},
  volume        = {127},
  adsurl        = {https://ui.adsabs.harvard.edu/abs/2015PASP..127..665R},
  archiveprefix = {arXiv},
  doi           = {10.1086/682257},
  eprint        = {1508.02362},
  primaryclass  = {astro-ph.IM},
}

@Article{Wright.Rieke.ea2023,
  author   = {{Wright}, Gillian S. and {Rieke}, George H. and {Glasse}, Alistair and {Ressler}, Michael and {Garc{\'\i}a Mar{\'\i}n}, Macarena and {Aguilar}, Jonathan and {Alberts}, Stacey and {{\'A}lvarez-M{\'a}rquez}, Javier and {Argyriou}, Ioannis and {Banks}, Kimberly and {Baudoz}, Pierre and {Boccaletti}, Anthony and {Bouchet}, Patrice and {Bouwman}, Jeroen and {Brandl}, Bernard R. and {Breda}, David and {Bright}, Stacey and {Cale}, Steven and {Colina}, Luis and {Cossou}, Christophe and {Coulais}, Alain and {Cracraft}, Misty and {De Meester}, Wim and {Dicken}, Daniel and {Engesser}, Michael and {Etxaluze}, Mireya and {Fox}, Ori D. and {Friedman}, Scott and {Fu}, Henry and {Gasman}, Danny and {G{\'a}sp{\'a}r}, Andr{\'a}s and {Gastaud}, Ren{\'e} and {Geers}, Vincent and {Glauser}, Adrian Michael and {Gordon}, Karl D. and {Greene}, Thomas and {Greve}, Thomas R. and {Grundy}, Timothy and {G{\"u}del}, Manuel and {Guillard}, Pierre and {Haderlein}, Peter and {Hashimoto}, Ryan and {Henning}, Thomas and {Hines}, Dean and {Holler}, Bryan and {Detre}, {\"O}rs Hunor and {Jahromi}, Amir and {James}, Bryan and {Jones}, Olivia C. and {Justtanont}, Kay and {Kavanagh}, Patrick and {Kendrew}, Sarah and {Klaassen}, Pamela and {Krause}, Oliver and {Labiano}, Alvaro and {Lagage}, Pierre-Olivier and {Lambros}, Scott and {Larson}, Kirsten and {Law}, David and {Lee}, David and {Libralato}, Mattia and {Lorenzo Alverez}, Jose and {Meixner}, Margaret and {Morrison}, Jane and {Mueller}, Migo and {Murray}, Katherine and {Mycroft}, Matthew and {Myers}, Richard and {Nayak}, Omnarayani and {Naylor}, Bret and {Nickson}, Bryony and {Noriega-Crespo}, Alberto and {{\"O}stlin}, G{\"o}ran and {O'Sullivan}, Brian and {Ottens}, Richard and {Patapis}, Polychronis and {Penanen}, Konstantin and {Pietraszkiewicz}, Martin and {Ray}, Tom and {Regan}, Michael and {Roteliuk}, Anthony and {Royer}, Pierre and {Samara-Ratna}, Piyal and {Samuelson}, Bridget and {Sargent}, Beth A. and {Scheithauer}, Silvia and {Schneider}, Analyn and {Schreiber}, J{\"u}rgen and {Shaughnessy}, Bryan and {Sheehan}, Even and {Shivaei}, Irene and {Sloan}, G.~C. and {Tamas}, Laszlo and {Teague}, Kelly and {Temim}, Tea and {Tikkanen}, Tuomo and {Tustain}, Samuel and {van Dishoeck}, Ewine F. and {Vandenbussche}, Bart and {Weilert}, Mark and {Whitehouse}, Paul and {Wolff}, Schuyler},
  journal  = {\pasp},
  title    = {{The Mid-infrared Instrument for JWST and Its In-flight Performance}},
  year     = {2023},
  month    = apr,
  number   = {1046},
  pages    = {048003},
  volume   = {135},
  adsurl   = {https://ui.adsabs.harvard.edu/abs/2023PASP..135d8003W},
  doi      = {10.1088/1538-3873/acbe66},
  eid      = {048003},
}

@Article{Argyriou.Glasse.ea2023,
  author        = {{Argyriou}, Ioannis and {Glasse}, Alistair and {Law}, David R. and {Labiano}, Alvaro and {{\'A}lvarez-M{\'a}rquez}, Javier and {Patapis}, Polychronis and {Kavanagh}, Patrick J. and {Gasman}, Danny and {Mueller}, Michael and {Larson}, Kirsten and {Vandenbussche}, Bart and {Glauser}, Adrian M. and {Royer}, Pierre and {Dicken}, Daniel and {Harkett}, Jake and {Sargent}, Beth A. and {Engesser}, Michael and {Jones}, Olivia C. and {Kendrew}, Sarah and {Noriega-Crespo}, Alberto and {Brandl}, Bernhard and {Rieke}, George H. and {Wright}, Gillian S. and {Lee}, David and {Wells}, Martyn},
  journal       = {\aap},
  title         = {{JWST MIRI flight performance: The Medium-Resolution Spectrometer}},
  year          = {2023},
  month         = jul,
  pages         = {A111},
  volume        = {675},
  adsurl        = {https://ui.adsabs.harvard.edu/abs/2023A&A...675A.111A},
  archiveprefix = {arXiv},
  doi           = {10.1051/0004-6361/202346489},
  eid           = {A111},
  eprint        = {2303.13469},
  primaryclass  = {astro-ph.IM},
}

@Article{Wells.Pel.ea2015,
  author        = {{Wells}, Martyn and {Pel}, J. -W. and {Glasse}, Alistair and {Wright}, G.~S. and {Aitink-Kroes}, Gabby and {Azzollini}, Ruym{\'a}n and {Beard}, Steven and {Brandl}, B.~R. and {Gallie}, Angus and {Geers}, V.~C. and {Glauser}, A.~M. and {Hastings}, Peter and {Henning}, Th. and {Jager}, Rieks and {Justtanont}, K. and {Kruizinga}, Bob and {Lahuis}, Fred and {Lee}, David and {Martinez-Delgado}, I. and {Mart{\'\i}nez-Galarza}, J.~R. and {Meijers}, M. and {Morrison}, Jane E. and {M{\"u}ller}, Friedrich and {Nakos}, Thodori and {O'Sullivan}, Brian and {Oudenhuysen}, Ad and {Parr-Burman}, P. and {Pauwels}, Evert and {Rohloff}, R. -R. and {Schmalzl}, Eva and {Sykes}, Jon and {Thelen}, M.~P. and {van Dishoeck}, E.~F. and {Vandenbussche}, Bart and {Venema}, Lars B. and {Visser}, Huib and {Waters}, L.~B.~F.~M. and {Wright}, David},
  journal       = {\pasp},
  title         = {{The Mid-Infrared Instrument for the James Webb Space Telescope, VI: The Medium Resolution Spectrometer}},
  year          = {2015},
  month         = jul,
  number        = {953},
  pages         = {646},
  volume        = {127},
  adsurl        = {https://ui.adsabs.harvard.edu/abs/2015PASP..127..646W},
  archiveprefix = {arXiv},
  doi           = {10.1086/682281},
  eprint        = {1508.03070},
  primaryclass  = {astro-ph.IM},
}

@Article{Yang.Evans.ea2020,
  author        = {{Yang}, Yao-Lun and {Evans}, Neal J., II and {Smith}, Aaron and {Lee}, Jeong-Eun and {Tobin}, John J. and {Terebey}, Susan and {Calcutt}, Hannah and {J{\o}rgensen}, Jes K. and {Green}, Joel D. and {Bourke}, Tyler L.},
  journal       = {\apj},
  title         = {{Constraining the Infalling Envelope Models of Embedded Protostars: BHR 71 and Its Hot Corino}},
  year          = {2020},
  month         = mar,
  number        = {1},
  pages         = {61},
  volume        = {891},
  adsurl        = {https://ui.adsabs.harvard.edu/abs/2020ApJ...891...61Y},
  archiveprefix = {arXiv},
  doi           = {10.3847/1538-4357/ab7201},
  eid           = {61},
  eprint        = {2002.01478},
  primaryclass  = {astro-ph.SR},
}

@Article{Bourke2001,
  author        = {{Bourke}, Tyler L.},
  journal       = {\apjl},
  title         = {{IRAS 11590-6452 in BHR 71: A Binary Protostellar System?}},
  year          = {2001},
  month         = jun,
  number        = {1},
  pages         = {L91-L94},
  volume        = {554},
  adsurl        = {https://ui.adsabs.harvard.edu/abs/2001ApJ...554L..91B},
  archiveprefix = {arXiv},
  doi           = {10.1086/320921},
  eprint        = {astro-ph/0105204},
  primaryclass  = {astro-ph},
}

@Article{Tobin.Bourke.ea2019,
  author        = {{Tobin}, John J. and {Bourke}, Tyler L. and {Mader}, Stacy and {Kristensen}, Lars and {Arce}, Hector and {Gueth}, Fr{\'e}d{\'e}ric and {Gusdorf}, Antoine and {Codella}, Claudio and {Leurini}, Silvia and {Chen}, Xuepeng},
  journal       = {\apj},
  title         = {{The Formation Conditions of the Wide Binary Class 0 Protostars within BHR 71}},
  year          = {2019},
  month         = jan,
  number        = {2},
  pages         = {81},
  volume        = {870},
  adsurl        = {https://ui.adsabs.harvard.edu/abs/2019ApJ...870...81T},
  archiveprefix = {arXiv},
  doi           = {10.3847/1538-4357/aaef87},
  eid           = {81},
  eprint        = {1811.03059},
  primaryclass  = {astro-ph.GA},
}

@Article{Ohashi.Tobin.ea2023,
  author        = {{Ohashi}, Nagayoshi and {Tobin}, John J. and {J{\o}rgensen}, Jes K. and {Takakuwa}, Shigehisa and {Sheehan}, Patrick and {Aikawa}, Yuri and {Li}, Zhi-Yun and {Looney}, Leslie W. and {Williams}, Jonathan P. and {Aso}, Yusuke and {Sharma}, Rajeeb and {Sai}, Jinshi (Insa Choi) and {Yamato}, Yoshihide and {Lee}, Jeong-Eun and {Tomida}, Kengo and {Yen}, Hsi-Wei and {Encalada}, Frankie J. and {Flores}, Christian and {Gavino}, Sacha and {Kido}, Miyu and {Han}, Ilseung and {Lin}, Zhe-Yu Daniel and {Narayanan}, Suchitra and {Phuong}, Nguyen Thi and {Santamar{\'\i}a-Miranda}, Alejandro and {Thieme}, Travis J. and {van't Hoff}, Merel L.~R. and {de Gregorio-Monsalvo}, Itziar and {Koch}, Patrick M. and {Kwon}, Woojin and {Lai}, Shih-Ping and {Lee}, Chang Won and {Plunkett}, Adele and {Saigo}, Kazuya and {Hirano}, Shingo and {Lam}, Ka Ho and {Mori}, Shoji},
  journal       = {\apj},
  title         = {{Early Planet Formation in Embedded Disks (eDisk). I. Overview of the Program and First Results}},
  year          = {2023},
  month         = jul,
  number        = {1},
  pages         = {8},
  volume        = {951},
  adsurl        = {https://ui.adsabs.harvard.edu/abs/2023ApJ...951....8O},
  archiveprefix = {arXiv},
  doi           = {10.3847/1538-4357/acd384},
  eid           = {8},
  eprint        = {2306.15406},
  primaryclass  = {astro-ph.EP},
}

@Article{Voirin.Manara.ea2018,
  author        = {{Voirin}, Jordan and {Manara}, Carlo F. and {Prusti}, Timo},
  journal       = {\aap},
  title         = {{A revised estimate of the distance to the clouds in the Chamaeleon complex using the Tycho-Gaia Astrometric Solution}},
  year          = {2018},
  month         = mar,
  pages         = {A64},
  volume        = {610},
  adsurl        = {https://ui.adsabs.harvard.edu/abs/2018A&A...610A..64V},
  archiveprefix = {arXiv},
  doi           = {10.1051/0004-6361/201731153},
  eid           = {A64},
  eprint        = {1710.04528},
  primaryclass  = {astro-ph.SR},
}

@Article{Gavino.Joergensen.ea2024,
  author        = {{Gavino}, Sacha and {J{\o}rgensen}, Jes K. and {Sharma}, Rajeeb and {Yang}, Yao-Lun and {Li}, Zhi-Yun and {Tobin}, John J. and {Ohashi}, Nagayoshi and {Takakuwa}, Shigehisa and {Plunkett}, Adele L. and {Kwon}, Woojin and {de Gregorio-Monsalvo}, Itziar and {Lin}, Zhe-Yu Daniel and {Santamar{\'\i}a-Miranda}, Alejandro and {Aso}, Yusuke and {Sai}, Jinshi and {Aikawa}, Yuri and {Tomida}, Kengo and {Koch}, Patrick M. and {Lee}, Jeong-Eun and {Lee}, Chang Won and {Lai}, Shih-Ping and {Looney}, Leslie W. and {Narayanan}, Suchitra and {Phuong}, Nguyen Thi and {Thieme}, Travis J. and {van't Hoff}, Merel L.~R. and {Williams}, Jonathan P. and {Yen}, Hsi-Wei},
  journal       = {\apj},
  title         = {{Early Planet Formation in Embedded Disks. XI. A High-resolution View Toward the BHR 71 Class 0 Protostellar Wide Binary}},
  year          = {2024},
  month         = oct,
  number        = {1},
  pages         = {21},
  volume        = {974},
  adsurl        = {https://ui.adsabs.harvard.edu/abs/2024ApJ...974...21G},
  archiveprefix = {arXiv},
  doi           = {10.3847/1538-4357/ad655e},
  eid           = {21},
  eprint        = {2407.17249},
  primaryclass  = {astro-ph.EP},
}

@Article{Narang.Manoj.ea2024,
  author        = {{Narang}, Mayank and {Manoj}, P. and {Tyagi}, Himanshu and {Watson}, Dan M. and {Megeath}, S. Thomas and {Federman}, Samuel and {Rubinstein}, Adam E. and {Gutermuth}, Robert and {Caratti o Garatti}, Alessio and {Beuther}, Henrik and {Bourke}, Tyler L. and {Van Dishoeck}, Ewine F. and {Evans}, Neal J. and {Anglada}, Guillem and {Osorio}, Mayra and {Stanke}, Thomas and {Muzerolle}, James and {Looney}, Leslie W. and {Yang}, Yao-Lun and {Klaassen}, Pamela and {Karnath}, Nicole and {Atnagulov}, Prabhani and {Brunken}, Nashanty and {Fischer}, William J. and {Furlan}, Elise and {Green}, Joel and {Habel}, Nolan and {Hartmann}, Lee and {Linz}, Hendrik and {Nazari}, Pooneh and {Pokhrel}, Riwaj and {Rahatgaonkar}, Rohan and {Rocha}, Will R.~M. and {Sheehan}, Patrick and {Slavicinska}, Katerina and {Stutz}, Amelia M. and {Tobin}, John J. and {Tychoniec}, Lukasz and {Wolk}, Scott},
  journal       = {\apjl},
  title         = {{Discovery of a Collimated Jet from the Low-luminosity Protostar IRAS 16253‑2429 in a Quiescent Accretion Phase with the JWST}},
  year          = {2024},
  month         = feb,
  number        = {1},
  pages         = {L16},
  volume        = {962},
  adsurl        = {https://ui.adsabs.harvard.edu/abs/2024ApJ...962L..16N},
  archiveprefix = {arXiv},
  doi           = {10.3847/2041-8213/ad1de3},
  eid           = {L16},
  eprint        = {2310.14061},
  primaryclass  = {astro-ph.SR},
}

@Article{Assani.Harsono.ea2024,
  author        = {{Assani}, K.~D. and {Harsono}, D. and {Ramsey}, J.~P. and {Li}, Z. -Y. and {Bjerkeli}, P. and {Pontoppidan}, K.~M. and {Tychoniec}, {\L}. and {Calcutt}, H. and {Kristensen}, L.~E. and {J{\o}rgensen}, J.~K. and {Plunkett}, A. and {van Gelder}, M.~L. and {Francis}, L.},
  journal       = {\aap},
  title         = {{The asymmetric bipolar [Fe II] jet and H$_{2}$ outflow of TMC1A resolved with the JWST NIRSpec Integral Field Unit}},
  year          = {2024},
  month         = aug,
  pages         = {A26},
  volume        = {688},
  adsurl        = {https://ui.adsabs.harvard.edu/abs/2024A&A...688A..26A},
  archiveprefix = {arXiv},
  doi           = {10.1051/0004-6361/202449745},
  eid           = {A26},
  eprint        = {2404.18334},
  primaryclass  = {astro-ph.SR},
}

@Article{Nisini.Navarro.ea2024,
  author        = {{Nisini}, Brunella and {Navarro}, Maria Gabriela and {Giannini}, Teresa and {Antoniucci}, Simone and {Kavanagh}, Patrick, J. and {Hartigan}, Patrick and {Bacciotti}, Francesca and {Caratti o Garatti}, Alessio and {Noriega-Crespo}, Alberto and {van Dishoeck}, Ewine F. and {Whelan}, Emma T. and {Arce}, Hector G. and {Cabrit}, Sylvie and {Coffey}, Deirdre and {Fedele}, Davide and {Eisl{\"o}ffel}, Jochen and {Palumbo}, Maria Elisabetta and {Podio}, Linda and {Ray}, Tom P. and {Schultze}, Megan and {Urso}, Riccardo G. and {Alcal{\'a}}, Juan M. and {Bautista}, Manuel A. and {Codella}, Claudio and {Greene}, Thomas P. and {Manara}, Carlo F.},
  journal       = {\apj},
  title         = {{PROJECT-J: JWST Observations of HH46 IRS and Its Outflow. Overview and First Results}},
  year          = {2024},
  month         = jun,
  number        = {2},
  pages         = {168},
  volume        = {967},
  adsurl        = {https://ui.adsabs.harvard.edu/abs/2024ApJ...967..168N},
  archiveprefix = {arXiv},
  doi           = {10.3847/1538-4357/ad3d5a},
  eid           = {168},
  eprint        = {2404.06878},
  primaryclass  = {astro-ph.GA},
}

@Article{Dionatos.Nisini.ea2009,
  author        = {{Dionatos}, O. and {Nisini}, B. and {Garcia Lopez}, R. and {Giannini}, T. and {Davis}, C.~J. and {Smith}, M.~D. and {Ray}, T.~P. and {DeLuca}, M.},
  journal       = {\apj},
  title         = {{Atomic Jets from Class 0 Sources Detected by Spitzer: The Case of L1448-C}},
  year          = {2009},
  month         = feb,
  number        = {1},
  pages         = {1-11},
  volume        = {692},
  adsurl        = {https://ui.adsabs.harvard.edu/abs/2009ApJ...692....1D},
  archiveprefix = {arXiv},
  doi           = {10.1088/0004-637X/692/1/1},
  eprint        = {0810.1651},
  primaryclass  = {astro-ph},
}

@Article{Gusdorf.PineaudesForets.ea2008,
  author   = {Gusdorf, A. and Pineau des For{\^e}ts, G. and Cabrit, S. and Flower, D. R.},
  journal  = {\aap},
  title    = {SiO line emission from interstellar jets and outflows: silicon-containing mantles and non-stationary shock waves},
  year     = {2008},
  month    = nov,
  pages    = {695-706},
  volume   = {490},
  doi      = {10.1051/0004-6361:200810443},
}

@Article{Giannini.Nisini.ea2019,
  author        = {{Giannini}, T. and {Nisini}, B. and {Antoniucci}, S. and {Biazzo}, K. and {Alcal{\'a}}, J. and {Bacciotti}, F. and {Fedele}, D. and {Frasca}, A. and {Harutyunyan}, A. and {Munari}, U. and {Rigliaco}, E. and {Vitali}, F.},
  journal       = {\aap},
  title         = {{GIARPS High-resolution Observations of T Tauri stars (GHOsT). I. Jet line emission}},
  year          = {2019},
  month         = nov,
  pages         = {A44},
  volume        = {631},
  adsurl        = {https://ui.adsabs.harvard.edu/abs/2019A&A...631A..44G},
  archiveprefix = {arXiv},
  doi           = {10.1051/0004-6361/201936085},
  eid           = {A44},
  eprint        = {1909.10392},
  primaryclass  = {astro-ph.SR},
}

@Article{Giannini.Antoniucci.ea2015,
  author        = {{Giannini}, T. and {Antoniucci}, S. and {Nisini}, B. and {Bacciotti}, F. and {Podio}, L.},
  journal       = {\apj},
  title         = {{Solving the Excitation and Chemical Abundances in Shocks: The Case of HH 1}},
  year          = {2015},
  month         = nov,
  number        = {1},
  pages         = {52},
  volume        = {814},
  adsurl        = {https://ui.adsabs.harvard.edu/abs/2015ApJ...814...52G},
  archiveprefix = {arXiv},
  doi           = {10.1088/0004-637X/814/1/52},
  eid           = {52},
  eprint        = {1510.06880},
  primaryclass  = {astro-ph.SR},
}

@Article{Podio.Bacciotti.ea2006,
  author        = {{Podio}, L. and {Bacciotti}, F. and {Nisini}, B. and {Eisl{\"o}ffel}, J. and {Massi}, F. and {Giannini}, T. and {Ray}, T.~P.},
  journal       = {\aap},
  title         = {{Recipes for stellar jets: results of combined optical/infrared diagnostics}},
  year          = {2006},
  month         = sep,
  number        = {1},
  pages         = {189-204},
  volume        = {456},
  adsurl        = {https://ui.adsabs.harvard.edu/abs/2006A&A...456..189P},
  archiveprefix = {arXiv},
  doi           = {10.1051/0004-6361:20054156},
  eprint        = {astro-ph/0606280},
  primaryclass  = {astro-ph},
}

@Article{Nisini.Bacciotti.ea2005,
  author        = {{Nisini}, B. and {Bacciotti}, F. and {Giannini}, T. and {Massi}, F. and {Eisl{\"o}ffel}, J. and {Podio}, L. and {Ray}, T.~P.},
  journal       = {\aap},
  title         = {{A combined optical/infrared spectral diagnostic analysis of the HH1 jet}},
  year          = {2005},
  month         = oct,
  number        = {1},
  pages         = {159-170},
  volume        = {441},
  adsurl        = {https://ui.adsabs.harvard.edu/abs/2005A&A...441..159N},
  archiveprefix = {arXiv},
  doi           = {10.1051/0004-6361:20053097},
  eprint        = {astro-ph/0507074},
  primaryclass  = {astro-ph},
}

@Article{Nisini.CarattioGaratti.ea2002,
  author   = {{Nisini}, B. and {Caratti o Garatti}, A. and {Giannini}, T. and {Lorenzetti}, D.},
  journal  = {\aap},
  title    = {{1-2.5 mu m spectra of jets from young stars: Strong Fe II emission in HH111, HH240-241 and HH120}},
  year     = {2002},
  month    = oct,
  pages    = {1035-1051},
  volume   = {393},
  adsurl   = {https://ui.adsabs.harvard.edu/abs/2002A&A...393.1035N},
  doi      = {10.1051/0004-6361:20021062},
}

@Article{CarattioGaratti.Ray.ea2024,
  author        = {{Caratti o Garatti}, A. and {Ray}, T.~P. and {Kavanagh}, P.~J. and {McCaughrean}, M.~J. and {Gieser}, C. and {Giannini}, T. and {van Dishoeck}, E.~F. and {Justtanont}, K. and {van Gelder}, M.~L. and {Francis}, L. and {Beuther}, H. and {Tychoniec}, {\L}. and {Nisini}, B. and {Navarro}, M.~G. and {Devaraj}, R. and {Reyes}, S. and {Nazari}, P. and {Klaassen}, P. and {G{\"u}del}, M. and {Henning}, Th. and {Lagage}, P.~O. and {{\"O}stlin}, G. and {Vandenbussche}, B. and {Waelkens}, C. and {Wright}, G.},
  journal       = {\aap},
  title         = {{JWST Observations of Young protoStars (JOYS): HH211: Textbook case of a protostellar jet and outflow}},
  year          = {2024},
  month         = nov,
  pages         = {A134},
  volume        = {691},
  adsurl        = {https://ui.adsabs.harvard.edu/abs/2024A&A...691A.134C},
  archiveprefix = {arXiv},
  doi           = {10.1051/0004-6361/202451350},
  eid           = {A134},
  eprint        = {2409.16061},
  primaryclass  = {astro-ph.SR},
}

@Article{Tychoniec.vanGelder.ea2024,
  author        = {{Tychoniec}, {\L}ukasz and {van Gelder}, Martijn L. and {van Dishoeck}, Ewine F. and {Francis}, Logan and {Rocha}, Will R.~M. and {Caratti o Garatti}, Alessio and {Beuther}, Henrik and {Gieser}, Caroline and {Justtanont}, Kay and {Linnartz}, Harold and {Le Gouellec}, Valentin J.~M. and {Perotti}, Giulia and {Devaraj}, Rangaswamy and {Tabone}, Beno{\^\i}t and {Ray}, Thomas P. and {Brunken}, Nashanty G.~C. and {Chen}, Yuan and {Kavanagh}, Patrick J. and {Klaassen}, Pamela and {Slavicinska}, Katerina and {G{\"u}del}, Manuel and {{\"O}stlin}, Goran},
  journal       = {\aap},
  title         = {{JWST Observations of Young protoStars (JOYS). Linked accretion and ejection in a Class I protobinary system}},
  year          = {2024},
  month         = jul,
  pages         = {A36},
  volume        = {687},
  adsurl        = {https://ui.adsabs.harvard.edu/abs/2024A&A...687A..36T},
  archiveprefix = {arXiv},
  doi           = {10.1051/0004-6361/202348889},
  eid           = {A36},
  eprint        = {2402.04343},
  primaryclass  = {astro-ph.SR},
}

\begin{appendix}

\section{PSF-subtraction}
\label{sec:app_psf}

Using the \texttt{stpsf} package \citep{Perrin.Long.ea2025}, we obtain the model of the PSF of the MIRI/MRS IFU. The procedure is illustrated in Fig. \ref{fig:fig:app_psf_images} and described below.

\begin{figure}
\begin{center}
\includegraphics[width=0.48\textwidth]{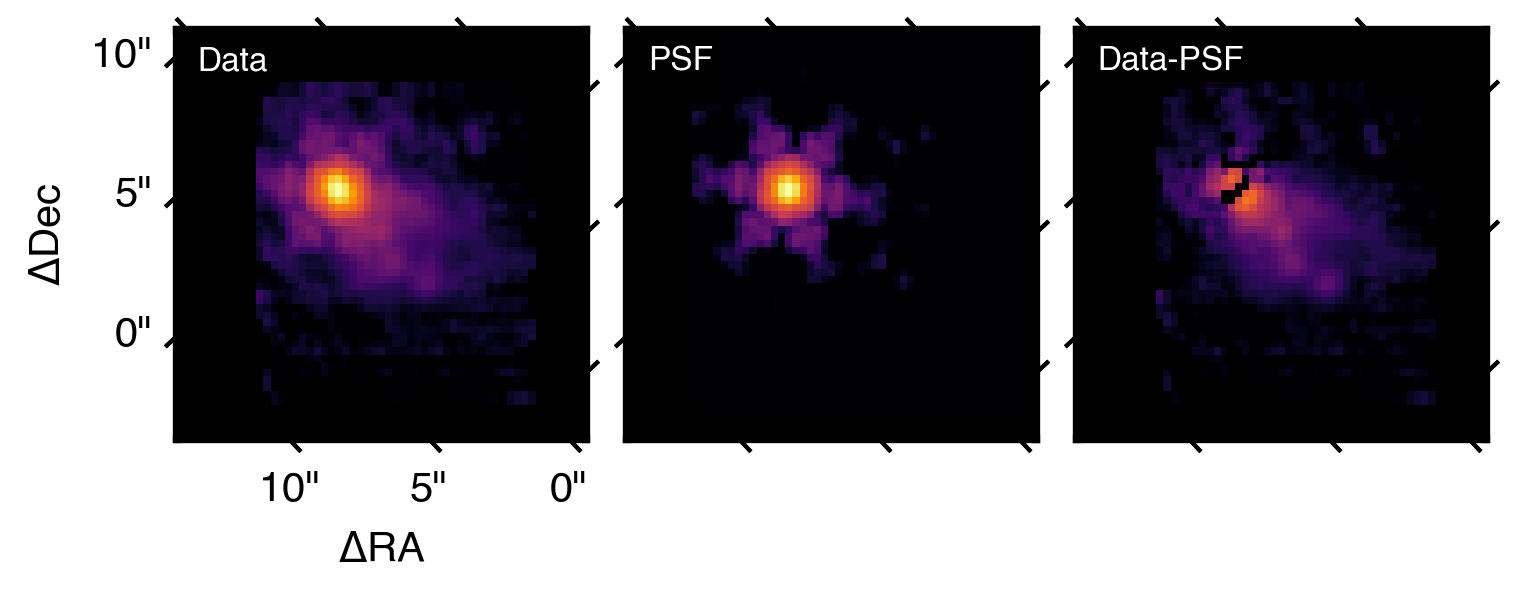}
\end{center}
\caption{Images showing an example of the PSF-subtraction. All images are integrated on the 19.3-19.9 $\mu$m range, and the color map is scaled to the same min and max values. {\it Left:} Original science cube created with \texttt{ifualign} step in the pipeline. {\it Middle:} PSF-model centered on the same position as the source centroid, resampled on the same pixel scale, normalized, and multiplied by the maximum value on the science cube per channel. {\it Right:} Data with PSF model subtracted. }
\label{fig:fig:app_psf_images}
\end{figure}

\begin{figure}
\begin{center}
\includegraphics[width=0.4\textwidth]{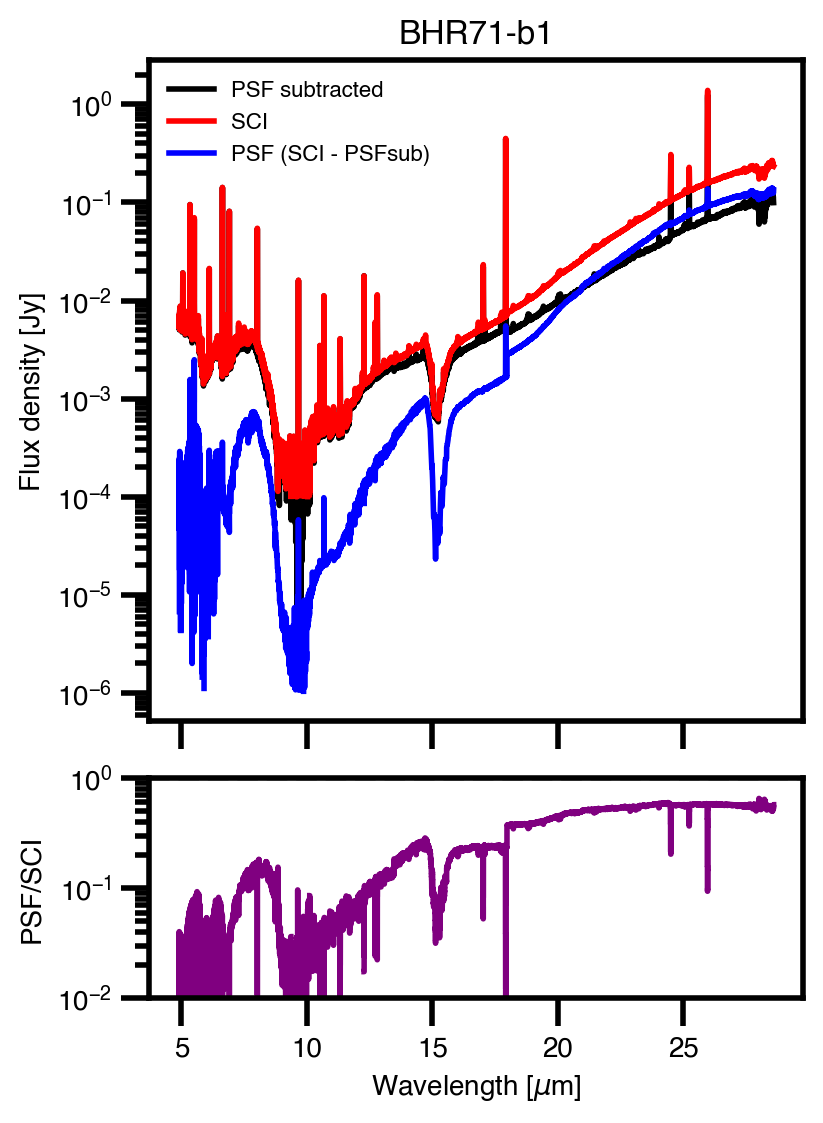}
\caption{Spectra extracted from the science data (red), from the PSF-subtracted data (black), and their difference, which is equivalent to the PSF-model spectrum (blue). In the bottom row, the ratio of flux between the PSF model and the science data. The sharp lines in the ratio plot are due to emission lines, which are prominent off-source, but usually fainter at the point source position.}
\label{fig:fig:app_psf_spectra}
\end{center}
\end{figure}

We generate the MIRI-MRS cube using \texttt{Spec3 ifualign} option in the pipeline, which sets the alignment of the resulting image with the instrument reference frame. This enables a direct comparison with the model PSF without introducing any field rotation. The example part of the cube integrated from 19.0-19.6 $\mu$m is shown in Fig. \ref{fig:fig:app_psf_images} (left). 

Next, the \texttt{stpsf} is used to generate a PSF model for each channel and band separately. The high-resolution model is first aligned to the source centroid on the detector and then returned as a cube resampled to the observation pixel size with distortions included in the model (\texttt{DET-DIST} column) (Fig. \ref{fig:fig:app_psf_images}, middle). 

Subsequently, the model cube is normalized, scaled by the maximum pixel value of the science cube per channel, and subtracted from the data cube (Fig. \ref{fig:fig:app_psf_images}, right). 

Spectra are extracted from the constructed PSF-subtracted cubes. In Fig. \ref{fig:fig:app_psf_spectra}, we present a comparison of two example regions: B1 and B4. It is clear that in B1 region close to the source (1\farcs7), the contribution at longer wavelengths is significant, with $\geq 40\%$ of flux at channel 4 ($\geq 18\ \mu$m ) coming from the point source in the unsubtracted spectra. At shorter wavelengths, the contribution is typically below 10$\%$ at this position. At B4, which is located 6\farcs5 away from the protostar, the contribution of PSF to the total flux is typically much lower than 10$\%$ for channels 1 - 3, but increases up to 30$\%$ in parts of channel 4 above 20 $\mu$m. This analysis confirms that the subtraction of PSF is crucial for the dust analysis, especially for fitting the SEDs, since the PSF contribution varies with wavelength.

\begin{figure}
    \begin{center}

    \includegraphics[width=0.45\textwidth]{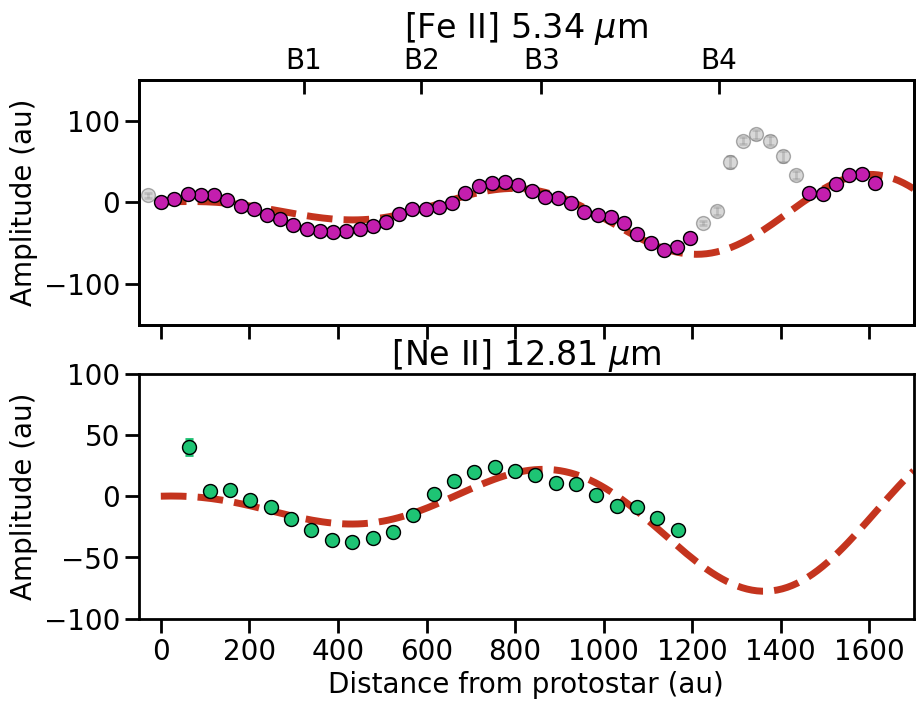}
 \caption{ Deprojected positions of the jet measured as a Gaussian centroid fitted to the spatial slice perpendicular to the jet for [Fe \textsc{ii}] at 5.34 $\mu$m and [Ne \textsc{ii}] 12.81 $\mu$m. The red curve shows the best fit to Eq. \ref{eq:wiggle}. Grey dots show the masked region where the expanding internal shock is likely affecting the centroid measurement. 
      }       
        \label{fig:SpatialFits_NeFe.png}
\end{center}
\end{figure}

\section{Modeling the wiggling of the jet}
\label{sec:appendix_wiggle}

The jet appears to wiggle, with a pattern consistent across all emission lines associated with the jet.  This is illustrated in Fig.   \ref{fig:SpatialFits_NeFe.png}, which shows the variation of the jet amplitude with distance from the protostar for [Fe II] 5.34 $\mu$m (top panel) and [Ne II] (bottom panel). The amplitude was determined from the peak of a Gaussian fit to the brightness profile of pixels measured perpendicular to the jet axis.
 Jet wiggling can be fit with a formula allowing to fit $\kappa$ - half-angle of the wiggling from the jet axis, $\delta_z$ - periodic length, and $\phi_0$ - phase offset, and $\eta$ - jet bending, which allows to describe wiggling along the z axis of the jet \citep{Lee.Hasegawa.ea2010,Anglada.Lopez.ea2007, Louvet.Dougados.ea2018}: 

 \begin{equation}
     y = -\kappa \cdot z \cdot \cos\left(\phi_0 - \frac{2\pi z}{\delta_z} + \frac{\pi}{2}\right) + \eta \cdot z, 
      \label{eq:wiggle}
 \end{equation}

\begin{figure*}
\begin{center}
\includegraphics[width=0.40\textwidth]{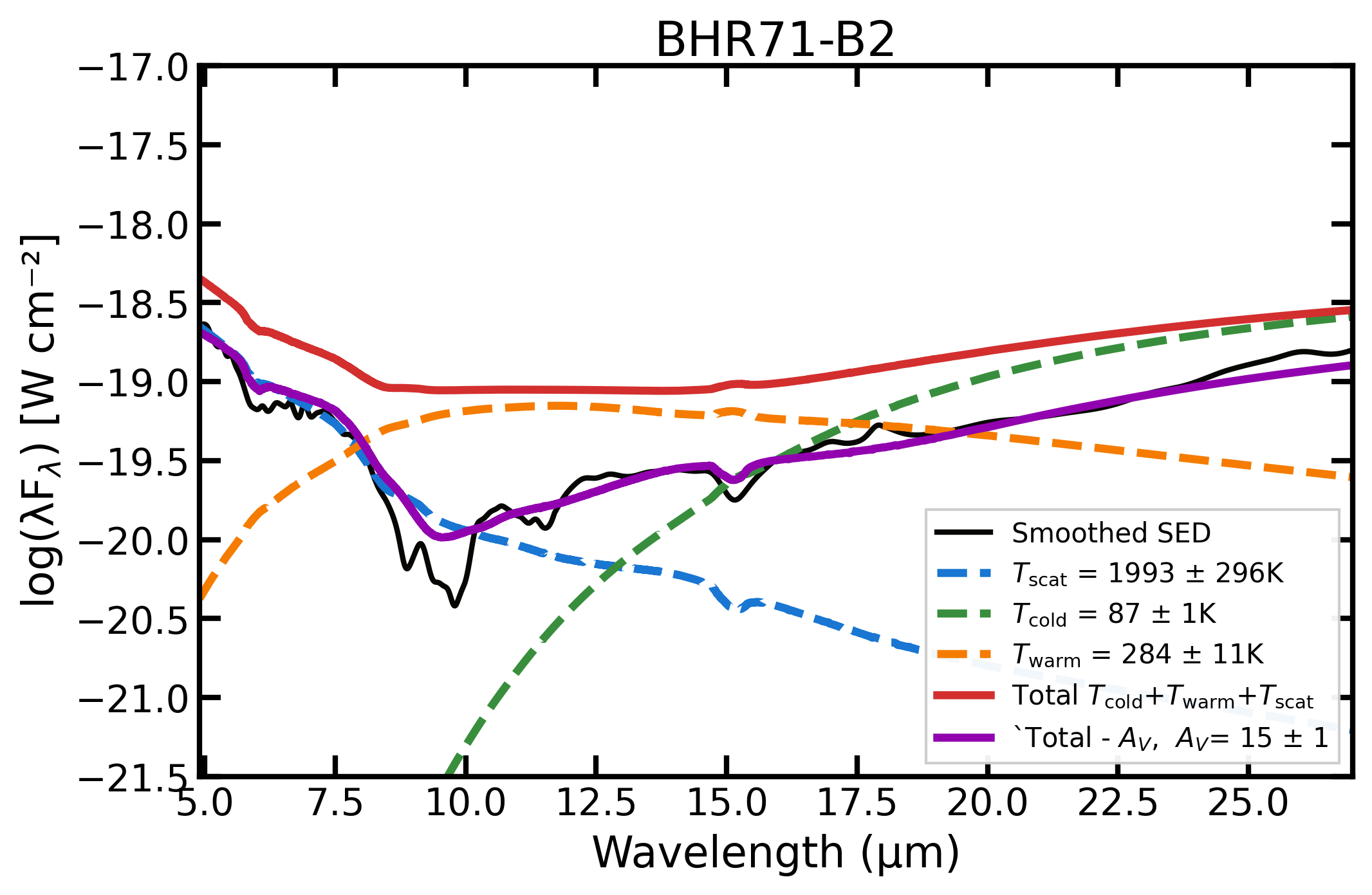}
\includegraphics[width=0.4\textwidth]{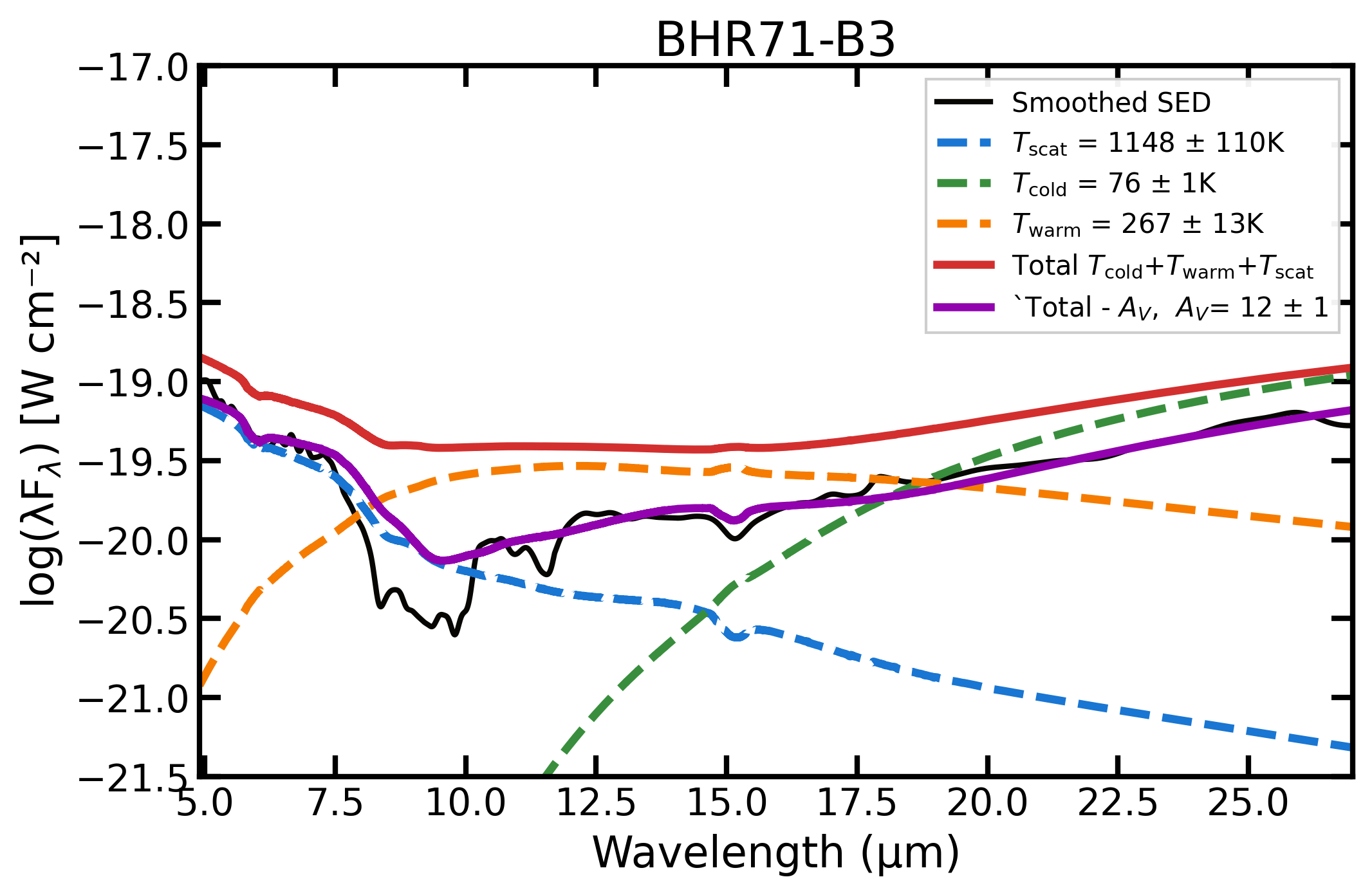}
\includegraphics[width=0.4\textwidth]{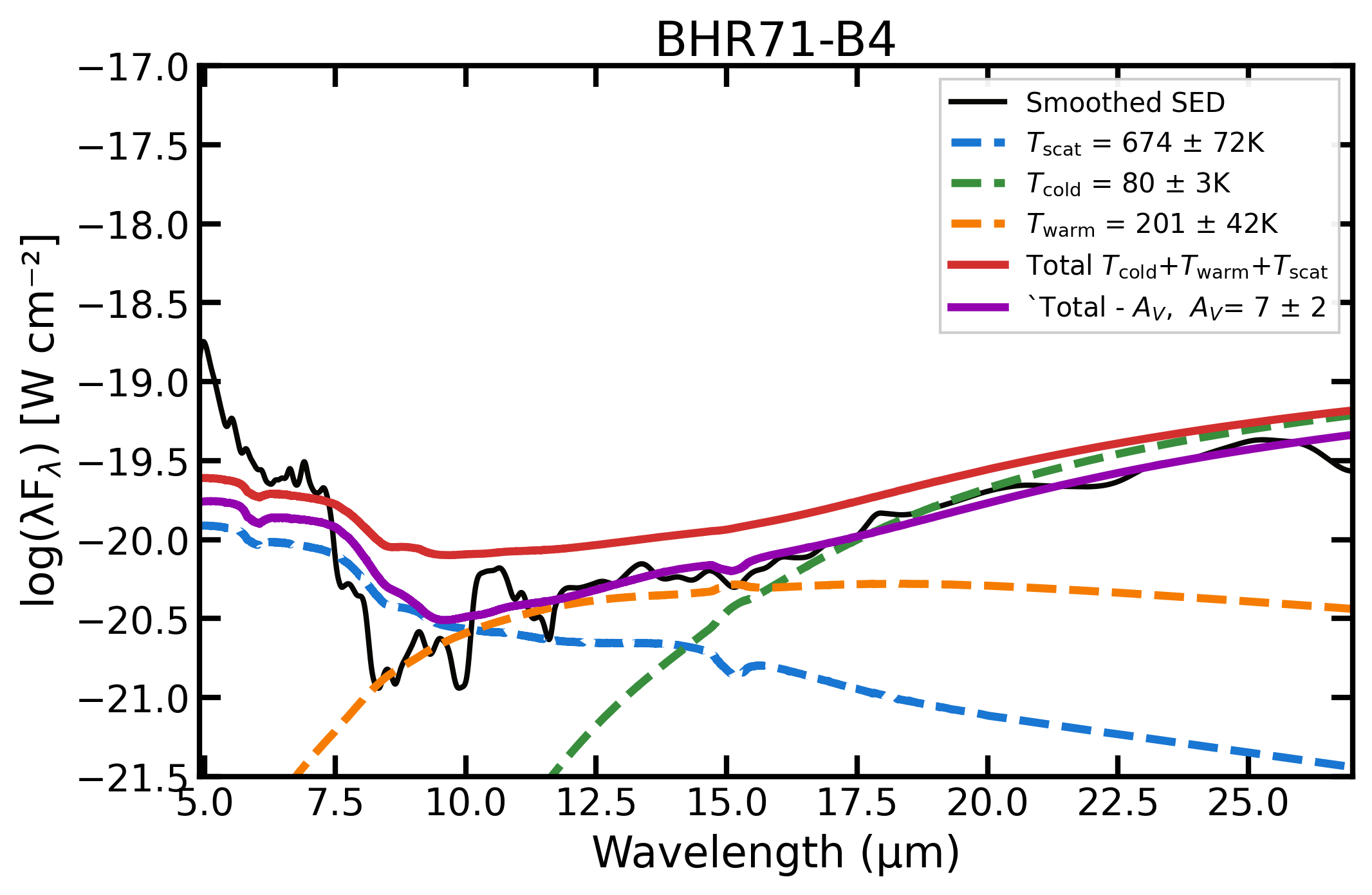}
\includegraphics[width=0.4\textwidth]{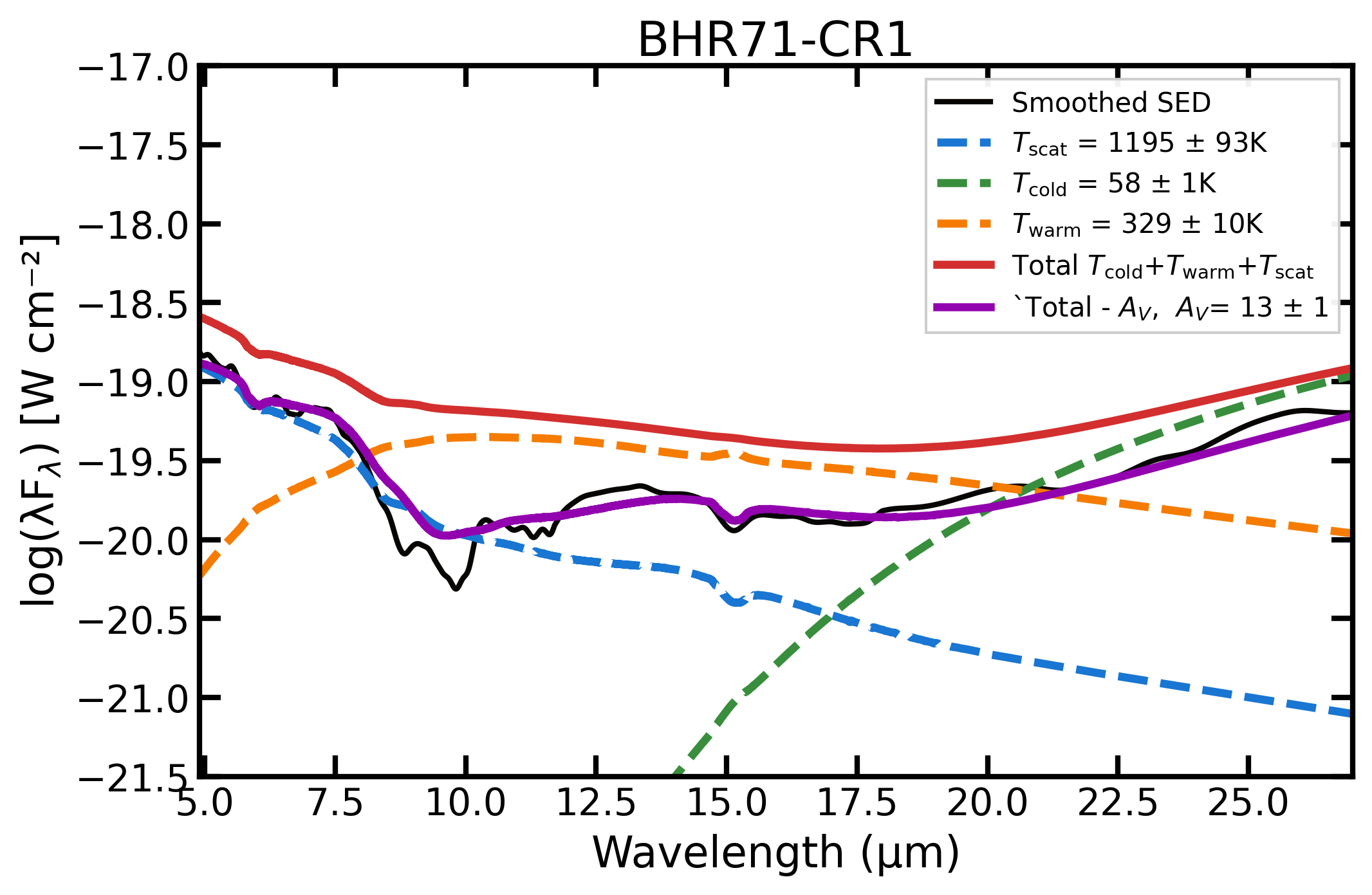}
\caption{Fits to spectral energy distributions at the positions from PSF-subtracted images. Absorption-subtracted spectra and a best-fit three-component blackbody (red) with components also indicated separately.}
\label{fig:A.3_SED}
\end{center}
\end{figure*}

 We mask the region at 1200-1400 au between B3 and B4, where the asymmetric bow-shock structure contaminates the analysis.
 We find values of $\kappa$ = 2.1 $\pm$ 0.1\degree, and $\delta_z$ = 810 $\pm$ 21 au for the [Fe \textsc{ii}]. and $\kappa$ = 2.4 $\pm$ 0.4\degree, and $\delta_z$ = 928 $\pm$ 125 for [Ne \textsc{ii}]. Knowing the deprojected jet velocity of 220 km s$^{-1}$, we can measure the period of precession to be 17.3$\pm$0.4 and 21$\pm$3 years for [Fe \textsc{ii}] and [Ne \textsc{ii}], respectively. The [Ne \textsc{ii}] also shows an offset in the phase of 40$\degree$ ahead of the [Fe \textsc{ii}]. This is interesting, as it could reveal shock-layering along the flow.
 
 Protostellar jets record the recent accretion history: jet knots are commonly attributed to episodic accretion and variability \citep{Raga.Canto.ea1990}. Here we summarize what the resolved JWST MIRI-MRS jet reveals about the velocity and spatial variability in BHR71‑IRS1. The observed oscillation of the jet is most naturally explained by precession of the inner (gaseous) disk where the jet is launched, as a result of either tidal interaction with a companion or by torques from a massive, misaligned disk. The orbital period of the perturbing body responsible for such precession is expected to be shorter than the precession period \citep{Lee.Hasegawa.ea2010,Anglada.Lopez.ea2007}; following Eq.~14 of \citet{Anglada.Lopez.ea2007}, the orbital period would be $\lesssim1.7$~yr. Adopting the protostellar mass upper limit from disk kinematics, $M_\star\lesssim0.46\,M_{\odot}$ \citep{Gavino.Joergensen.ea2024}, this orbital period corresponds to a semimajor axis of $\sim1.1$~au (Kepler's third law), well inside the ALMA‑resolved disk radius of $\sim80$~au. ALMA continuum imaging finds no companion down to $\sim5$~au scales \citep{Gavino.Joergensen.ea2024}, but the disk shows asymmetries consistent with a massive, potentially unstable disk \citep[disk mass $\sim0.1\,M_{\odot}$;][]{Gavino.Joergensen.ea2024}. Such disk mass and structure favor precession driven by disk–star interaction rather than an orbital motion of a massive stellar companion \citep{Kratter.Lodato2016,Speedie.Dong.ea2024} and would cause noticeable precession of the disk \citep{Vioque.Booth.ea2026}.

If instead the wiggle were due to orbital motion of the jet source itself, the observed single‑jet geometry implies the launching star would be more massive than any companion. Applying Eq.~7 of \citet{Lee.Hasegawa.ea2010} to the observed properties yields no physically plausible solution for a low‑mass system: equal‑mass binaries that could reproduce the wiggle would require total masses well in excess of a few solar masses, inconsistent with the kinematic mass limits for BHR71‑IRS1.

We note a localized discrepancy: the [Fe\,\textsc{ii}] trace departs from the global wiggle pattern at the position where [Ne\,\textsc{ii}] emission fades. This spatial coincidence may signal a local change in shock strength, ionization state, or extinction, and merits targeted follow‑up (spectro‑kinematic and shock diagnostics) to clarify its origin.

Alternative mechanisms can also produce episodic knots and apparent wiggling. For example, MHD models invoking magnetic reconnection or rotating reconnection sites can generate time‑dependent ejection events and apparent lateral oscillations \citep{Ouyed.Pudritz1997}. If such reconnection sites orbit the central source, they could naturally explain both the knot spacing and the wiggle symmetry.

In summary, the resolved JWST view of BHR71‑IRS1 demonstrates that detailed measurements of jet morphology and kinematics combined with high‑resolution ALMA disk kinematics \citep[e.g.,][]{Phuong.Lee.ea2025} offer a powerful probe of inner disk structure and dynamical state. Disentangling precession, orbital motion, and intrinsic MHD variability will require coordinated theoretical modeling and further multiwavelength observations.

\section{H$_2$ analysis: extinction and temperatures} \label{sec:app_H2analysis}

H$_2$ as the dominant molecule in the ISM is a valuable tracer of the physical conditions within the wide-angle wind and outflow. Its lowest-energy pure rotational $v=0$~--~$0$ transitions occur in the mid-IR. Mid-IR H$_2$ emission has been explored in the past with {\it ISO} \citep{Rosenthal.Bertoldi.ea2000}, {\it Spitzer} \citep{Lahuis.vanDishoeck.ea2010, Neufeld.Melnick.ea2006}, and more recently with JWST-MIRI \citep{Tychoniec.vanGelder.ea2024, Okoda.Yang.ea2025,Gieser.Beuther.ea2024, Schwarz.Henning.ea2024, Navarro.Nisini.ea2025, CarattioGaratti.Ray.ea2024, Skretas.Karska.ea2025, Francis.Tychoniec.ea2026}. Rotational lines of H$_2$ are typically a good measure of the gas kinetic temperature since the rotational levels at the ground vibrational state are not affected by UV pumping \citep{Black.vanDishoeck1987}; however, the UV irradiation contributes to the heating of the gas and affects the shock structure \citep{Kristensen.Godard.ea2023, Skretas.Karska.ea2025}.

\begin{table}
\tiny
\centering
\caption{Results of the H$_2$ line intensity analysis}
\label{tab:appE2_h2}
\setlength{\tabcolsep}{3pt}
\begin{tabular}{lccccccccc}
\hline \hline
Reg, & log($N_{\rm warm}$)  & log($N_{\rm hot}$)  & $\frac{N_{\rm hot}}{N_{\rm warm}}$  & T$_{\rm warm}$ & T$_{\rm hot}$ &  OPR \\
 & cm$^{-2}$ & cm$^{-2}$ & $\%$ & K & K &  \\\hline
B1 & 21.2 $\pm$ 0.2 & 19.2 $\pm$ 0.3 & 0.9 & 674 $\pm$ 75 & 2350 $\pm$ 390 & 1.9 $\pm$ 0.5\\
B2 & 20.2 $\pm$ 0.1 & 19.0 $\pm$ 0.2 & 6.0 & 881 $\pm$ 80 & 2270 $\pm$ 220 & 2.6 $\pm$ 0.3\\
B3 & 19.8 $\pm$ 0.1 & 18.4 $\pm$ 0.1 & 3.8 & 813 $\pm$ 52 & 2220 $\pm$ 160 & 2.8 $\pm$ 0.3\\
B4 & 19.6 $\pm$ 0.1 & 18.8 $\pm$ 0.1 & 13.4 & 847 $\pm$ 82 & 2080 $\pm$ 120 & 3.0 $\pm$ 0.3\\
CR1 & 19.5 $\pm$ 0.1 & 17.7 $\pm$ 0.4 & 1.4 & 684 $\pm$ 64 & 2590 $\pm$ 1320 & 2.8 $\pm$ 0.5\\
O5 & 19.4 $\pm$ 0.0 & 17.5 $\pm$ 0.1 & 1.4 & 570 $\pm$ 12 & 1770 $\pm$ 100 & 3.0 $\pm$ 0.1\\
\hline
\end{tabular}
\end{table}

Line intensities of the H$_2$ rotational transitions can be used to calibrate the extinction \citep[e.g.,][]{Rosenthal.Bertoldi.ea2000,Barsony.WolfChase.ea2010, Narang.Manoj.ea2024, Tychoniec.vanGelder.ea2024, Francis.vanDishoeck.ea2025}. For that, we use the S(3) line at 9.7 $\mu$m, affected by the silicate absorption peak \citep{Bertoldi.Timmermann.ea1999, Neufeld.Melnick.ea1998}.  We fit line intensities of the S(1)--S(4) H$_2$ 0--0 lines to a single temperature component with the extinction curve KP5 \citep{Pontoppidan.Evans.ea2024} with opacity magnitude as a free parameter ($\tau_{\rm S(3)}$). We choose this extinction curve as it is based on mid-IR measurements from {\it Spitzer} spanning a range of extinctions more relevant for the most embedded sources \citep{Chapman.Mundy.ea2008}.

H$_2$-based extinctions are reported in Table \ref{tab:appE3_regions}. We denote this as $\tau_{\rm S (3)}$ rather than $\tau_{\rm 9.7}$ to avoid confusion with directly measured optical depth of the silicate band. The values of measured extinction A$_v$ range from  $39.8\pm5.1$ to $9.4\pm0.5$, decreasing with distance from the protostar.

\begin{figure*}
\begin{center}
    \includegraphics[width=0.33\textwidth]{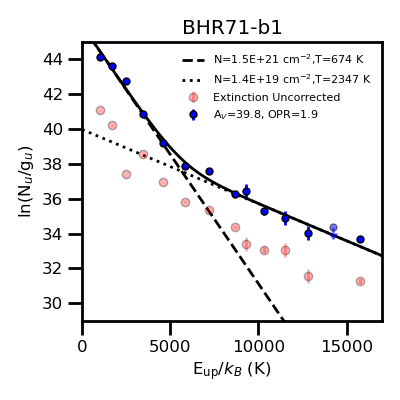}
        \includegraphics[width=0.33\textwidth]{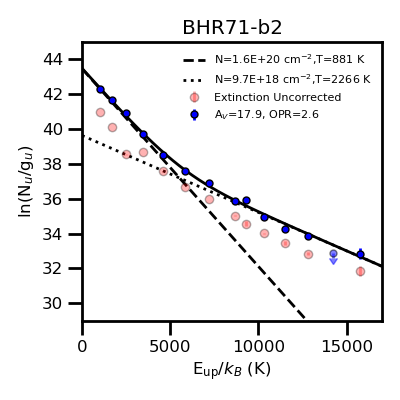}
        \includegraphics[width=0.33\textwidth]{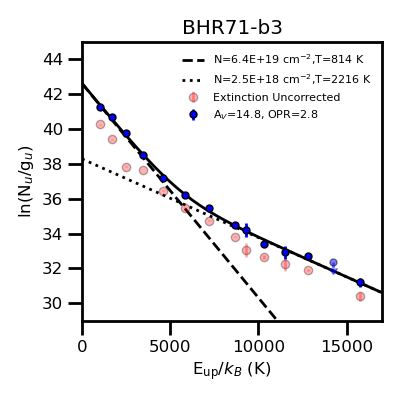}
        \includegraphics[width=0.33\textwidth]{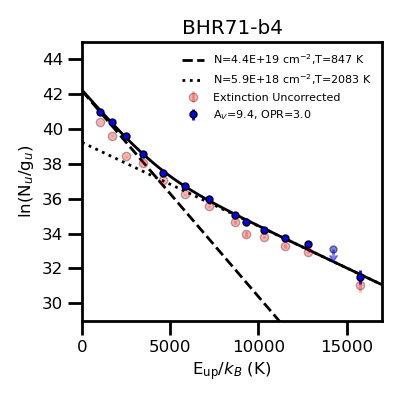}
        \includegraphics[width=0.33\textwidth]{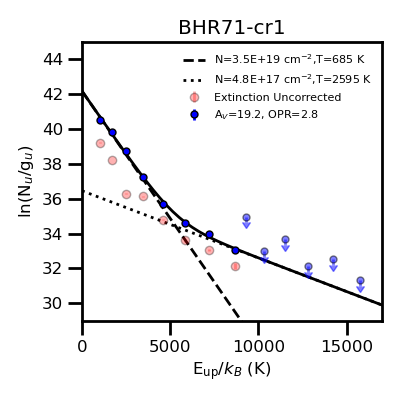}
        \includegraphics[width=0.33\textwidth]{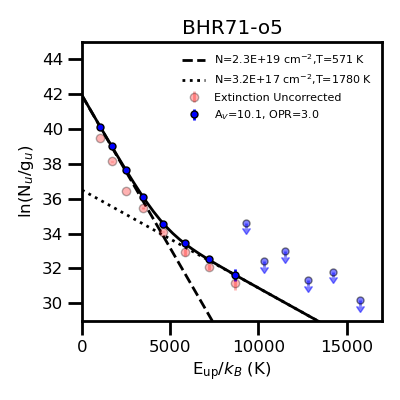}
    \caption{Rotational diagrams with extinction correction. Red dots show observed data, blue dots show column densities corrected for extinction following KP5 \cite{Pontoppidan.Evans.ea2024} curve. Dashed and dotted lines show fits to warm and hot components, respectively. The solid line shows the sum of the two components. Line intensities of H$_2$ are reported in Tab. \ref{tab:fluxes_atomic}.}
    \label{fig:H2RotDiag}
\end{center}
\end{figure*}

We create rotational diagrams from extinction-corrected H$_2$ line intensities at selected positions, presented in the Appendix Fig. \ref{fig:H2RotDiag}. The rotational diagrams of H$_2$ show curvature that can be fitted with two temperature components, which we label as warm and hot. We also set the ortho-to-para ratio (OPR) as a free parameter. The rotational temperatures, column densities, and OPR are presented in Table \ref{tab:appE2_h2}. 
See \cite{Francis.vanDishoeck.ea2025} and \cite{Gieser.Beuther.ea2024} for the details on the H$_2$ rotational diagrams from MIRI-MRS data.
Critical densities of H$_2$ are low, making the rotational diagrams reliable diagnostics of kinetic temperature \citep{Nisini.Giannini.ea2010}.

Rotational temperatures across different regions are of similar order of magnitude, with slightly lower temperatures for the warm component at the B1 spot ($674\pm75$ K) and higher ($>$~800~K) farther from the source. The hot component shows no significant changes across the flow  ($>$~2000~K) within its uncertainties. In the O5 region, located outside the main final shell associated with B4 positions, the temperatures are $570\ \pm 12$ K and $1773 \ \pm 103$ K for the warm and hot components, respectively.

Comparing the column densities across the extracted regions, we find a steady decrease with distance from the protostar for both warm and hot components. However, we observe an increasing contribution of the hot component to the total column density from bullets B1 to B4. Outside the main H$_2$ emission shells, the ratio decreases. One interpretation of this behavior is a decrease in contributions from warm H$_2$ gas, which may arise from the wide-angle wind, whereas the shock surrounding a jet stands out more in contrast to this wind at larger distances.

The ortho-to-para (OPR) ratio across the outflow appears to be well within the equilibrium value of 3, except for the B1 spot, which is the closest one to the source. It could be a signature of the time evolution of the shocked gas as the shock propagates forward: more material is heated, while close to the source, the gas is mostly released from the cold disk or envelope \citep{Neufeld.Melnick.ea1998, Neufeld.Melnick.ea2006}. While the higher temperature also means faster para-to-ortho conversion, which primarily occurs via reaction with atomic hydrogen with a high activation barrier of $\sim$4000 K, we would expect the gas at the highest temperature to reach equilibrium faster \citep{Sternberg.Neufeld1999, Wilgenbus.Cabrit.ea2000}, with H$_2$-H$_2$ and He-H$_2$ reactions also contributing \citep{LeBourlot.PineaudesForets.ea1999, Kristensen.Ravkilde.ea2007}. Given that temperatures found in the rotational diagrams appear similar, the time needed for the conversion of OPR is crucial in achieving equilibrium \citep[e.g.,][]{Maret.Bergin.ea2009}. It also shows that the H$_2$ gas detected in the outflow is lifted from the colder regions of the disk, not only from the innermost parts, and is likely not predominantly reformed in the shocks. In other words, the pattern we observe (OPR $\simeq3$ downstream, lower OPR at B1) supports a scenario in which gas near the source retains a lower, pre-shock OPR while more distant shocked gas has had time and/or conditions necessary for conversion to the equilibrium value.

\section{Shock models}
\label{sec:appendix_hm89}
In this section, we compare the obtained fluxes and flux ratios with the J-type shock model with a radiative precursor presented in \citetalias{Hollenbach.McKee1989}.Those models can be used to derive the pre-shock density $n_0$ and shock velocity $v_s$ along the jet. The model predicts line intensities for a range of shock velocities 30-150 km s$^{-1}$ and density bins of 10$^3$, 10$^4$, 10$^5$, 10$^6$ cm$^{-3}$. In this model, shock is treated as a discontinuity, and ambipolar diffusion is negligible. The initial elemental abundances are assumed to be depleted according to \cite{Harris.Gry.ea1984}.

First, we estimate shock properties using absolute extinction-corrected line intensities. The [Ne \textsc{ii}] 12.82 $\mu$m line is very sensitive to shock velocity but poorly constrains the pre-shock density. On the other hand, lines like [S \textsc{i}] 25.25 $\mu$m or [Cl \textsc{i}] at 11.34 $\mu$m  are good density tracers. We follow a similar method of estimating the shock conditions as in \cite{CarattioGaratti.Ray.ea2024}. Based on the [Ne \textsc{ii}] extinction-corrected line intensities, we find the best-match velocity for each density bin (Fig. \ref{fig:fig12_HM89models} top). Then, based on [S \textsc{i}] and [Cl \textsc{i}], we extrapolate two independent estimates of the pre-shock density Fig. \ref{fig:fig12_HM89models}, (middle and bottom, respectively). within the limits provided by [Ne \textsc{ii}]. This density value is then used to obtain the specific value of shock velocity from [Ne \textsc{ii}] interpolated to the density found for [S \textsc{i}] and [Cl \textsc{i}]. We note, however, that the shock model accounts for the fraction of the sulphur in the neutral and ionized form, so we can approximate that the comparison to [S \textsc{i}] is representative of the total sulphur budget.

\begin{figure}
    \begin{center}
    \includegraphics[width=0.34\textwidth]{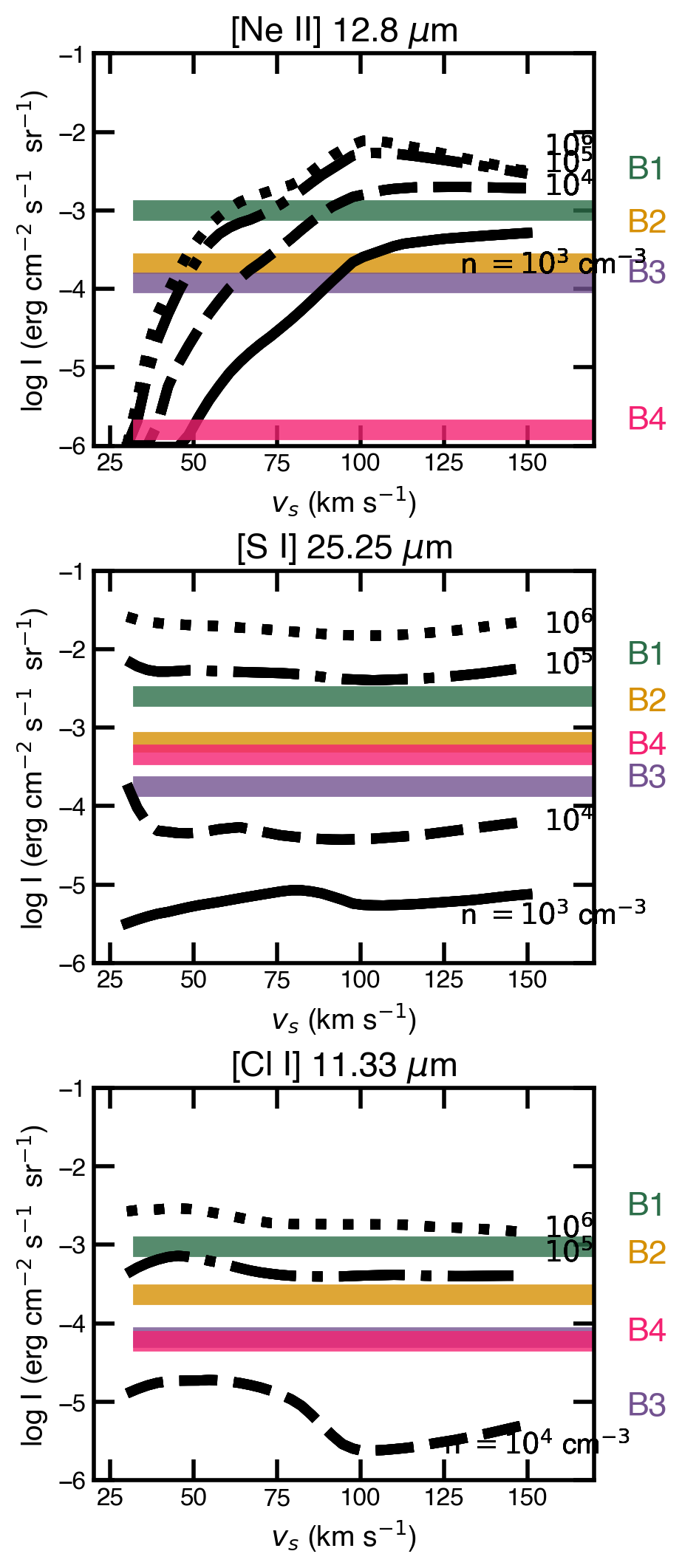}
        \caption{Extinction-corrected intensities of [Ne \textsc{ii}] (top), [S \textsc{i}] (middle), and [Cl \textsc{i}] (bottom) compared to the shock velocities (x-axis) and densities (black lines) of \citetalias{Hollenbach.McKee1989}. Intensities measured at each bullet are marked with the horizontal line.}
        \label{fig:fig12_HM89models}
\end{center}
\end{figure}

\begin{figure}
    \begin{center}
    \includegraphics[width=0.45\textwidth]{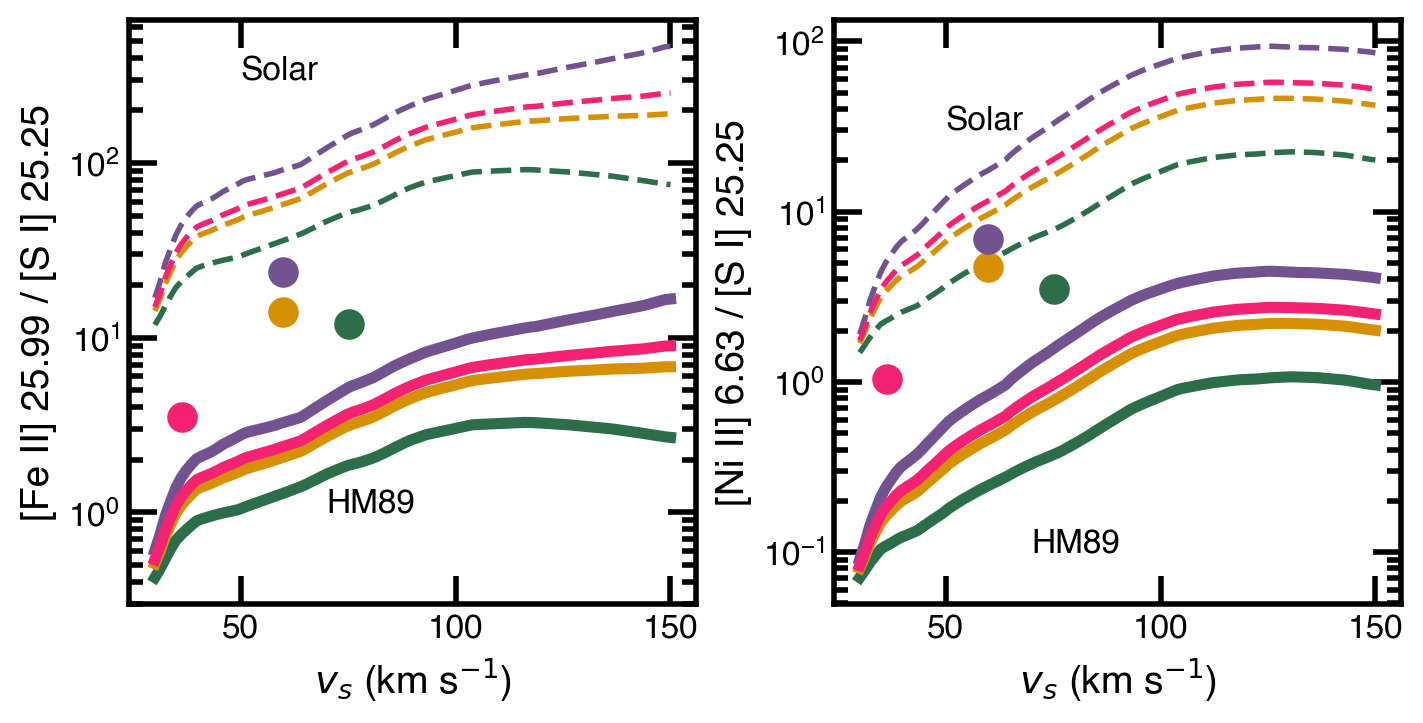}
        \caption{ Extinction-corrected line-intensity ratios shown as point for each region B1-B4 separately. Solid lines refer to the predictions by \citepalias{Hollenbach.McKee1989}, which assumed depleted values of the refractory species. Dashed lines show the model predictions rescaled to the Solar values.  This is used to estimate the abundances in the BHR71-IRS1 jet.}
        \label{fig:Hollenbach_abundances}
\end{center}
\end{figure}

\begin{figure}
    \begin{center}
    \includegraphics[width=0.34\textwidth]{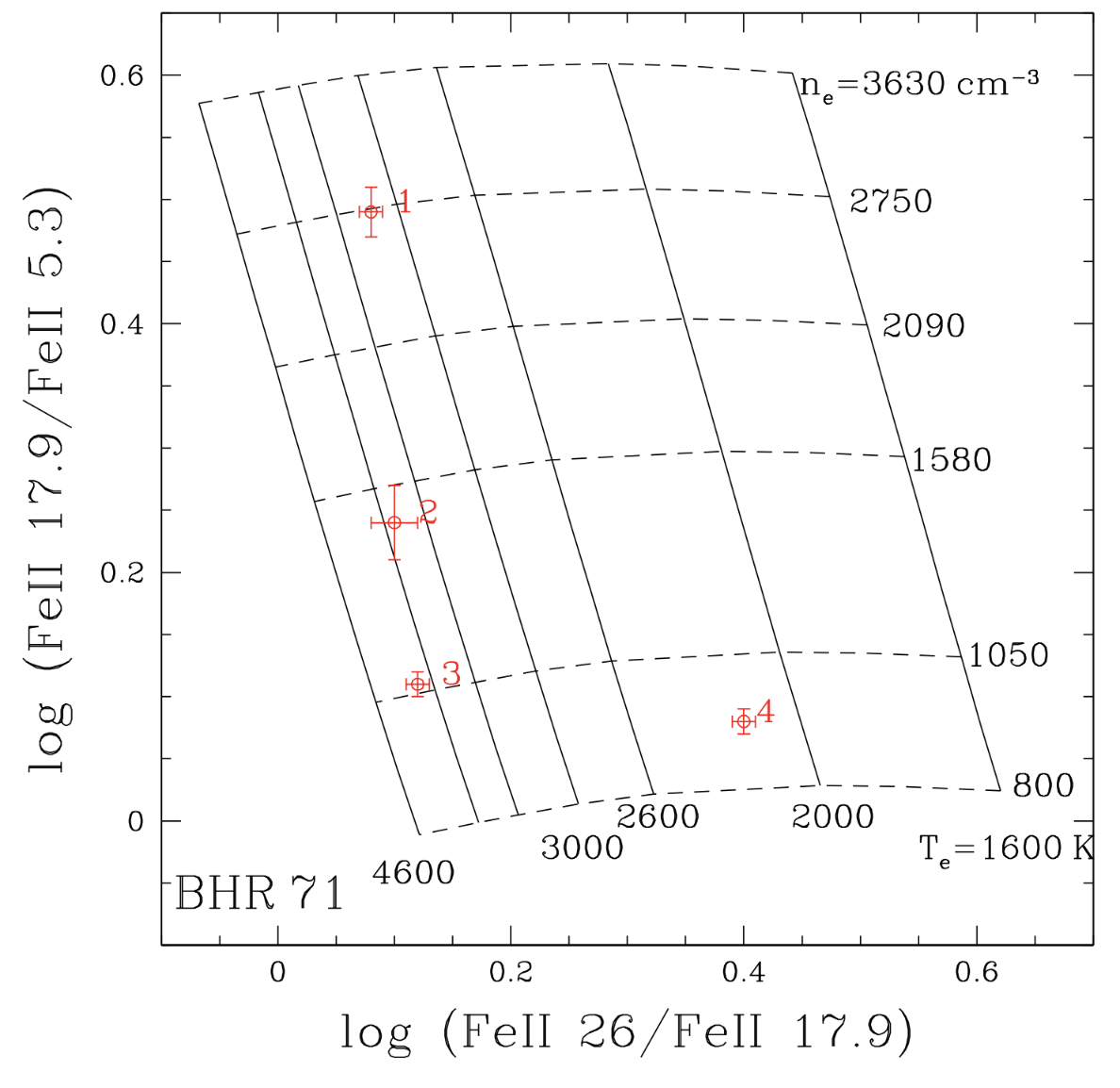}
        \caption{Extinction-corrected intensity ratios of [Fe II] lines for the B1-B4 bullets. The ratios are plot on a grid of electron temperature ($T_e$), and density ($n_e$) based on NLTE excitation model \citep{Giannini.Nisini.ea2013}.}
        \label{fig:D4}
\end{center}
\end{figure}

\begin{table}
\tiny
\centering
\caption{Hollenbach shock model results}
\label{tab:Hollenbach}
\begin{tabular}{lcccc}
\hline \hline
 & n$_{\rm 0}$ & v$_{\rm shock}$ & Fe/Fe$_\odot$ & Ni/Ni$_\odot$ \\
  & $\times\ 10^5 $ cm$^{-3}$ & km s$^{-1}$ & & \\
\hline
B1 & 0.9-7.0 & 75.2-66.3 & 0.23-0.36 & 0.45-0.76 \\
B2 & 0.6-0.9 & 59.8-53.9 & 0.24-0.45 & 0.49-1.13 \\
B3 & 0.3-0.6 & 59.9-54.8 & 0.26-0.48 & 0.39-0.91 \\
B4 & 0.5-0.4 & 36.4-37.2 & 0.10-0.08 & 0.26-0.20 \\
\hline
\end{tabular}
\tablefoot{Shock model parameters and abundance ratios derived from \citetalias{Hollenbach.McKee1989} models.}
\end{table}

Then, the depletion can be compared with the values adopted in the \citetalias{Hollenbach.McKee1989} shock model, which already assumes depletion based on \cite{Harris.Gry.ea1984} with  abundance per hydrogen atom: x$_{\textrm{Fe}}$=10$^{-6}$; x$_{\textrm{Cl}}$ = 1.4$\times 10^{-7}$; and x$_{\textrm{Ni}} = $7.6$\times 10^{-8}$
  or Solar values based on \citep{Allen1973} for undepleted species, x$_\textrm{S}$ = 10${^{-5}}$.

Results are reported in Fig. \ref{fig:Hollenbach_abundances}. The abundances found for the BHR71-71 shocks are predominantly lower than the solar values. For iron, the range is 0.1-0.4 of solar abundance; for nickel, 0.02-0.41; and for chlorine, 0.3-1.7. These results indicate a clear depletion of refractory species in the jet's gas phase, suggesting that most of them remain on dust grains launched with the jet, assuming they are launched at Solar abundances. 

Ratios of [Fe \textsc{ii}] emission lines are sensitive to electron temperature ($T_e$) and electron density ($n_e$). This has been explored in \cite{Giannini.Nisini.ea2013} and updated to include lower $E_{\rm up}$ in MIRI-MRS range in \cite{CarattioGaratti.Ray.ea2024}. In Fig. \ref{fig:D4} we show ratios of [Fe \textsc{ii}] 5.34, 17.9, and 25.9 $\mu$m lines on a grid of $T_e$ and $n_e$. We see that first three bullets are in the range of $T_e$ of 3000-4200 K with gradual increase from B1 to B3, while B4 shows a significant drop to less than 2500 K. At the same time $n_e$ is consistently dropping from 2750 cm$^{-3}$ in B1 to 900 cm$^{-3}$ in B4. Notably, in comparison with HH211 flow \citep{CarattioGaratti.Ray.ea2024}, the values for both $T_e$ and $n_e$ are higher in BHR71, where all jet knots present values below 3000 K and 400 cm$^{-3}$ for $T_e$ and $n_e$, respectively. The bow-shocks in HH211 show much wider range of values and those are more comparable to the BHR71 case. 

\section{Additional figures and tables}
\label{sec:app_figs_tabs}

\begin{figure}
    \begin{center}
    \includegraphics[width=0.45\textwidth]{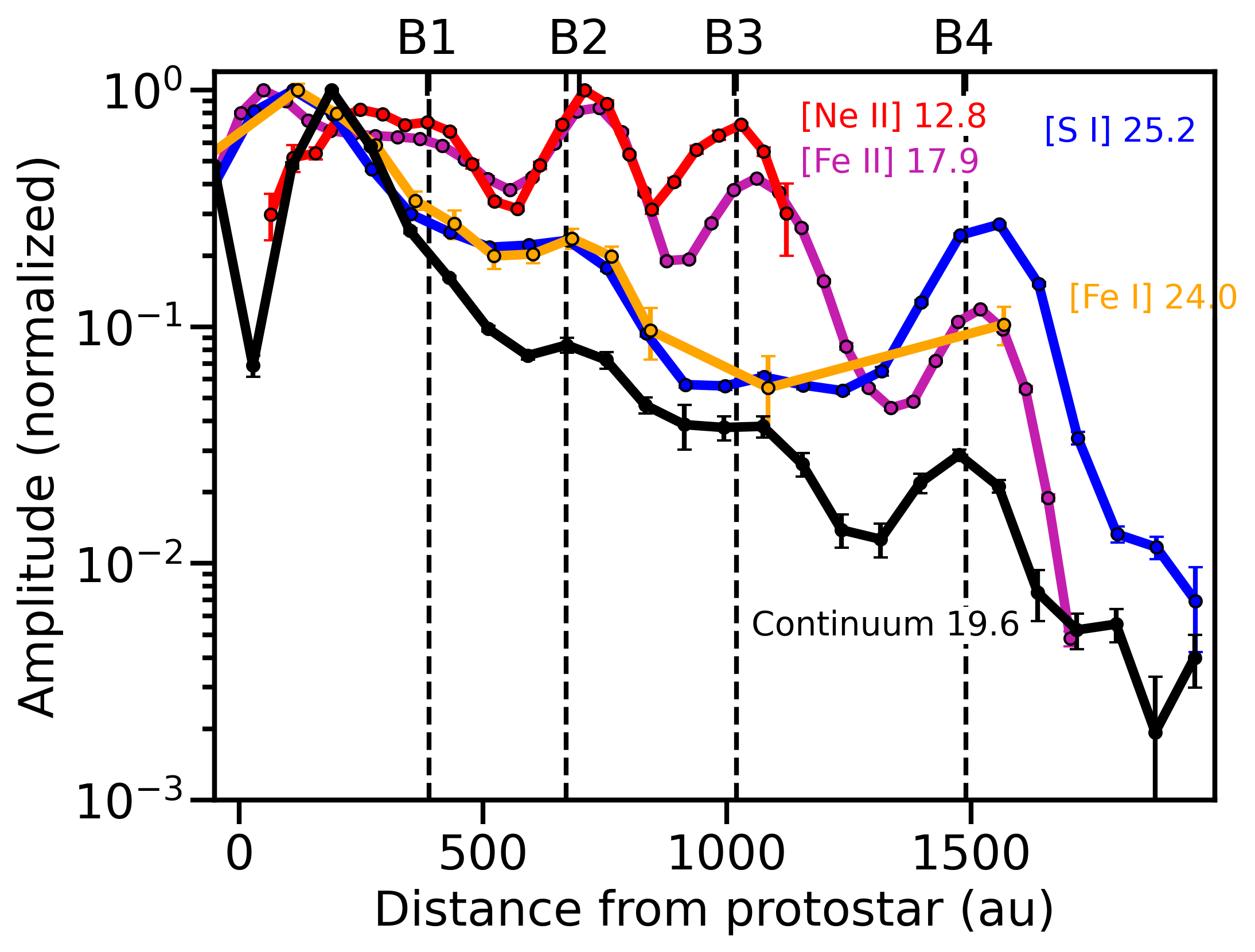}
        \caption{Peak brightness of emission line as a function of distance from the protostar, normalized to the maximum value for each line.}
        \label{fig:appE0_peaks}
\end{center}
\end{figure}

In Fig. \ref{fig:appE0_peaks}, we show line brightness as a function of a distance from a protostar for selected lines and for continuum at 19.6 $\mu$m.

In Fig. \ref{fig:appE1_h2}, we show all unblended lines of H$_2$ detected toward BHR71-IRS1. Fig. \ref{fig:appE2_spectra} presents spectra extracted at selected positions along the flow, as marked in Fig.  \ref{fig:Figure1_overviewmaps}.

In Tab. \ref{tab:appE1_Dust}, dust properties as measured in \ref{sec:extended_dust_4.1}. The range of values is provided since different values are obtained based on density estimates from S and Cl. Tab. \ref{tab:appE3_regions} provides a summary of the properties of regions used for spectral analysis.

\begin{table}
\tiny
\centering
\caption{Dust properties derived from blackbody fits}
\label{tab:appE1_Dust}
\begin{tabular}{lcccccc}
\hline \hline
Reg. & T$_{\rm warm}$  & T$_{\rm cold}$  & $\Sigma$ &  A$_V$\\
 & (K) & (K) & $10^{-3}$ g (cm$^{-2}$) & \\
\hline
 B1 & $406 \pm 6$ & $70.0 \pm 0.2$ & 1.80 $\pm$ 0.02	& $31.7\pm0.4$ \\
 B2 & $284 \pm 11$ & $87.0 \pm 0.9$ & $0.86\pm0.03$ & $15\pm1$	 \\
 B3 & $267 \pm 13$ & $76 \pm 1$ & $0.66\pm0.04$ & $12\pm1$ \\
 B4 & $201 \pm 42$  & $80 \pm 3$  & $0.38\pm0.09$ & $7\pm2$\\
CR1 & $329 \pm 10$  & $58.0 \pm 0.6$ &$0.74\pm0.03$  & $13\pm1$\\

\hline
\end{tabular}
\tablefoot{Temperature, mass, and column density for warm and cold dust components.}
\end{table}

\begin{table}
\tiny
\centering
\caption{Regions selected for the spectral analysis}
\label{tab:appE3_regions}
\begin{tabular}{lccccccccc}
\hline \hline

Reg. & R.A. & Dec & z &  z$_{\rm deprojected}$ & $\tau_{S(3)}$ & $A_V$ \\
 & (J2000) & (J2000) & " & au &   \\\hline
B1 & 12:01:36.47 & -65:08:50.99 & 1.7 & 390 $\pm$ 90 & 5.6$\pm$0.7 & 39.8$\pm$5.1\\
B2 & 12:01:36.43 & -65:08:52.36 & 3.0 & 670 $\pm$ 160 & 2.5$\pm$0.3 & 17.9$\pm$2.0 \\
B3 & 12:01:36.48 & -65:08:53.74 & 4.4 & 1020 $\pm$ 230 & 2.1$\pm$0.1& 14.8$\pm$0.5\\
B4 & 12:01:36.42 & -65:08:55.78 & 6.5 & 1490 $\pm$ 330 & 1.3$\pm$0.1 & 9.4$\pm$0.5 \\
CR1 & 12:01:36.12 & -65:08:53.21 & 4.4 & 1010 $\pm$ 230 & 2.7$\pm$0.1 & 19.2$\pm$0.1 \\
O5 & 12:01:36.51 & -65:08:58.36 & 9.1 & 2080 $\pm$ 460 & 1.4$\pm$0.5 & 10.1$\pm$3.4 \\
\hline
\end{tabular}
\tablefoot{Uncertainty in the deprojected distance is a result of error propagation, including error on the distance estimate and uncertainty in the inclination angle.}\end{table}

 \onecolumn

 \onecolumn

\begin{figure*}
    \begin{center}

    \includegraphics[width=0.75\textwidth]{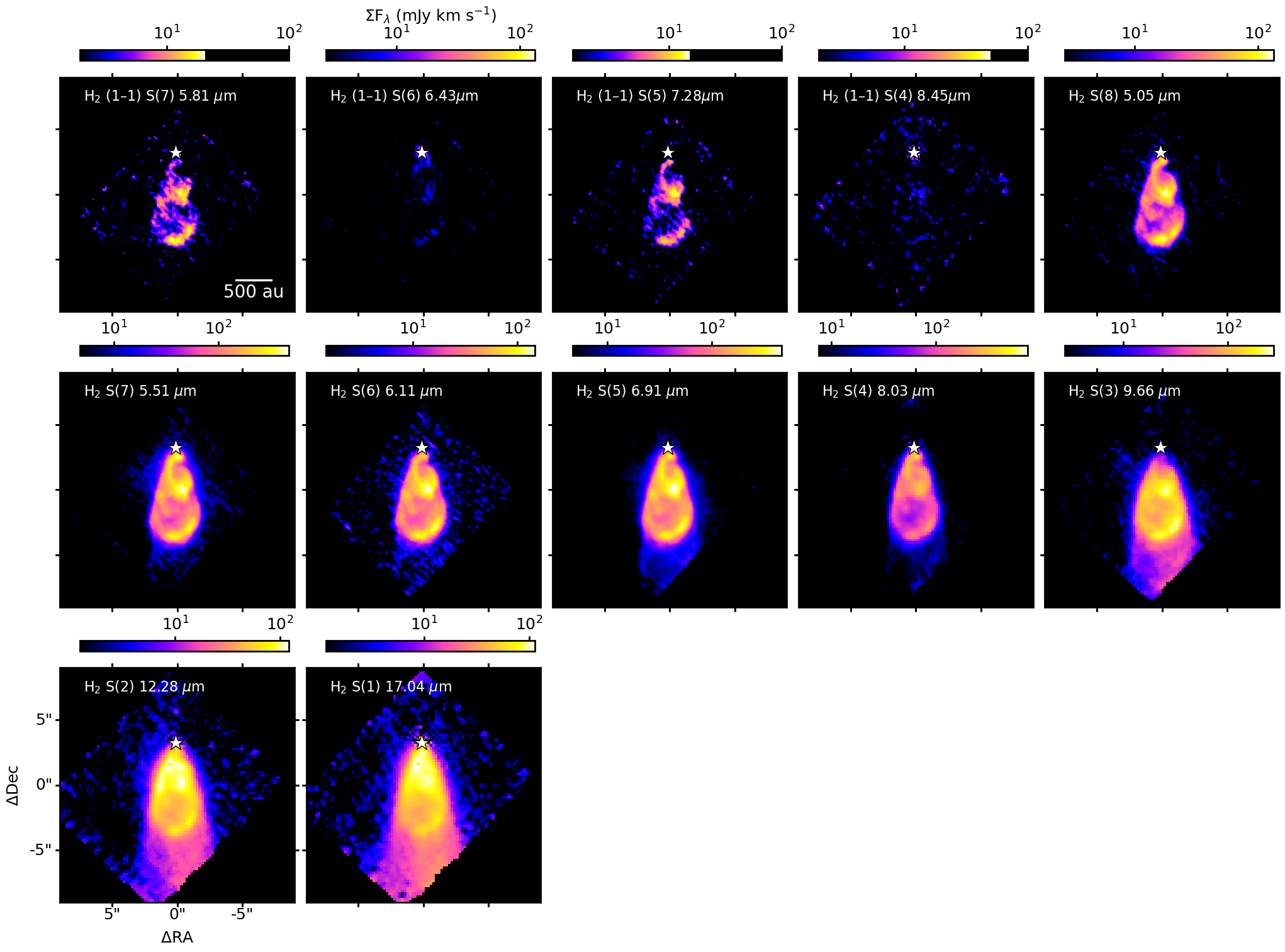}
        \caption{
All detected H$_2$ transitions are presented, except H$_2$ 1-1 S(9), which is contaminated by CO ro-vibrational forest.}
        \label{fig:appE1_h2}
\end{center}
\end{figure*}

\begin{figure*}
    \begin{center}
    \includegraphics[width=0.92\textwidth]{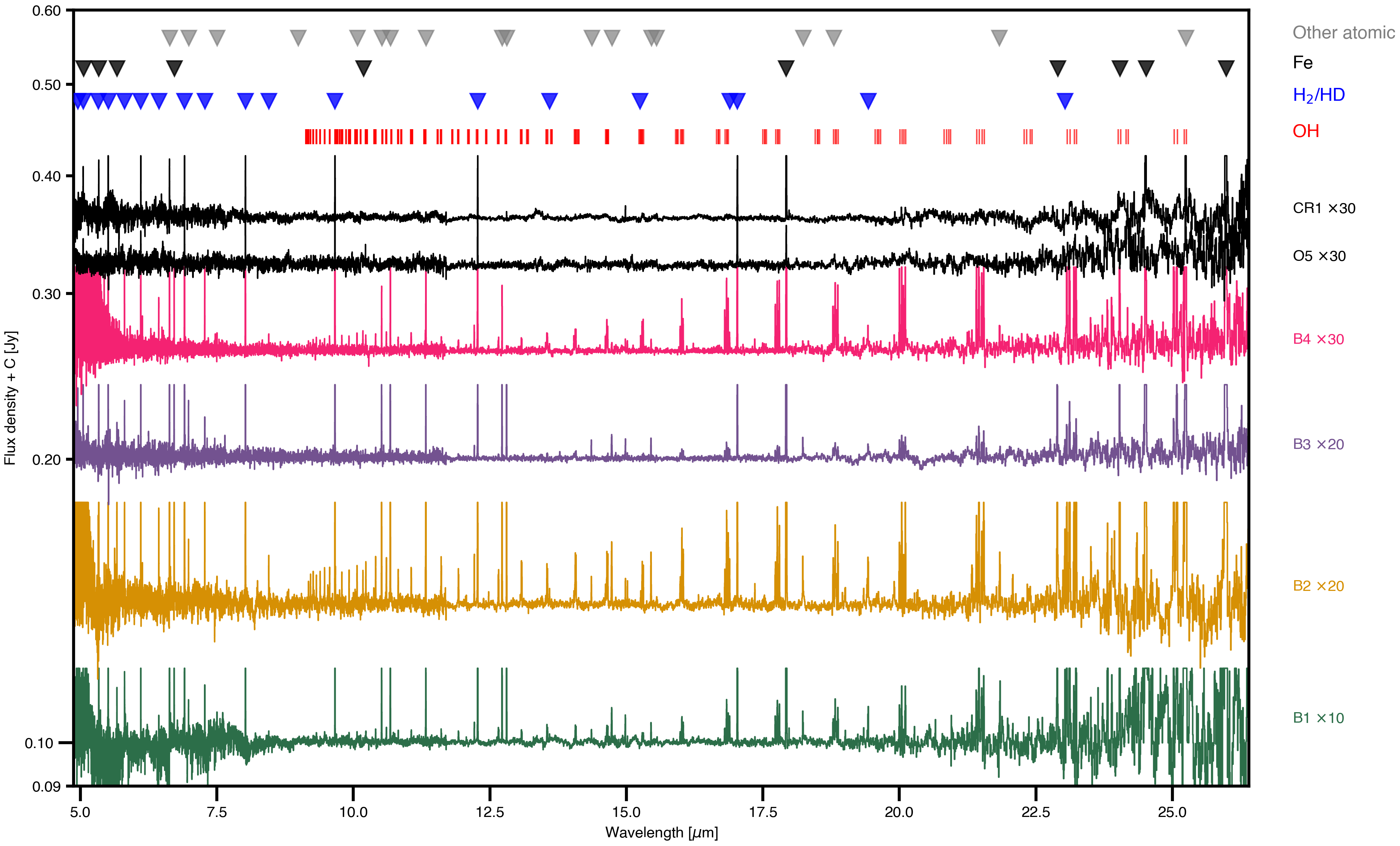}
\caption{Spectra extracted form 1\farcs0 diameter at positions shown in Fig. \ref{fig:Figure1_overviewmaps} (top right) and listed in \ref{tab:appE3_regions}. H$_2$  and HD lines are indicated in blue markers, [Fe \textsc{ii}] lines in black markers, and remaining lines are indicated with grey markers and labeled. Spectra between channels were stitched with an additive factor, and the baseline was removed with a manual spline fit. Brightest lines are truncated.}\label{fig:appE2_spectra}
\end{center}
\end{figure*}
\begin{sidewaystable}
    \tiny

\caption{Measured emission line properties}
\label{tab:fluxes_atomic}
\begin{tabular}{p{2.5cm}p{0.7cm}p{0.4cm}p{0.4cm}p{0.4cm}p{1.6cm}p{1.6cm}p{1.6cm}p{1.6cm}p{1.6cm}p{1.6cm}p{1.6cm}p{1.6cm}p{1.6cm}p{2.0cm}}
\centering
 & &  & & & BHR71-b1 & & BHR71-b2 & & BHR71-b3 & & BHR71-b4 & \\
\hline
 Line & \multicolumn{1}{c}{$\lambda$} & \multicolumn{1}{c}{$E_{\rm up}$}  & \multicolumn{1}{c}{I.P$_{\rm l}$}  & \multicolumn{1}{c}{I.P$_{\rm u}$} &\multicolumn{1}{c} {I$_\lambda$} & $v$ & \multicolumn{1}{c} {I$_\lambda$} & $v$ & \multicolumn{1}{c} {I$_\lambda$} & $v$ & \multicolumn{1}{c} {I$_\lambda$} & $v$ \\
 & \multicolumn{1}{c}{$\mu$m} &  \multicolumn{1}{c}{K} & \multicolumn{1}{c}{eV} & \multicolumn{1}{c}{eV} & $10^{-5}$ erg cm$^{-2}$ s$^{-1}$ sr$^{-1}$ & km s$^{-1}$ & $10^{-5}$ erg cm$^{-2}$ s$^{-1}$ sr$^{-1}$ & km s$^{-1}$ & $10^{-5}$ erg cm$^{-2}$ s$^{-1}$ sr$^{-1}$ & km s$^{-1}$ & $10^{-5}$ erg cm$^{-2}$ s$^{-1}$ sr$^{-1}$ & km s$^{-1}$  \\
$[$Fe \textsc{i}$]$ a$^{5}$D$_{4}$-a$^{5}$D$_{3}$ & 24.042 & 598 & 0.0 & 7.9 & $21.0\pm2.9$&$-91.1\pm7.4$   & $5.0\pm0.4$&$-145.3\pm7.4$    & $1.7\pm0.3$&$-203.5\pm9.0$  & $2.1\pm0.1$&$-161.7\pm3.5$  \\
 $[$Fe \textsc{ii}$]$ a$^{6}$D$_{3/2}$-a$^{4}$F$_{5/2}$ & 5.062 & 4083 & 7.9 & 16.2 & $<16.7 $ & -   & $<10.2 $ & -    & $<2.0 $ & -  & $<14.0 $ & -  \\
 $[$Fe \textsc{ii}$]$ a$^{6}$D$_{9/2}$-a$^{4}$F$_{9/2}$ & 5.340 & 2694 & 7.9 & 16.2 & $799.7\pm17.7$&$-195.4\pm1.5$   & $416.9\pm22.8$&$-237.4\pm3.1$    & $241.2\pm3.2$&$-258.7\pm0.7$  & $61.2\pm0.9$&$-196.4\pm0.7$  \\
 $[$Fe \textsc{ii}$]$ a$^{6}$D$_{5/2}$-a$^{4}$F$_{7/2}$ & 5.674 & 3496 & 7.9 & 16.2 & $24.8\pm5.6$&$-259.5\pm20.4$   & $6.4\pm0.8$&$-264.9\pm8.2$    & $1.8\pm0.4$&$-263.3\pm9.8$  & $1.0\pm0.3$&$-103.2\pm17.8$  \\
 $[$Fe \textsc{ii}$]$ a$^{6}$D$_{7/2}$-a$^{4}$F$_{9/2}$ & 6.721 & 2694 & 7.9 & 16.2 & $31.1\pm1.6$&$-193.8\pm3.3$   & $19.8\pm0.8$&$-226.9\pm2.4$    & $13.0\pm0.4$&$-255.4\pm1.8$  & $3.1\pm0.2$&$-187.9\pm2.6$  \\
 $[$Fe \textsc{ii}$]$ a$^{4}$F$_{9/2}$-a$^{4}$F$_{7/2}$ & 17.936  & 3496 & 7.9 & 16.2 & $2467.2\pm61.7$&$-216.7\pm1.5$   & $725.0\pm19.0$&$-244.4\pm1.5$    & $317.7\pm3.4$&$-247.4\pm0.6$  & $62.4\pm0.2$&$-182.9\pm0.2$  \\
 $[$Fe \textsc{ii}$]$ a$^{4}$D$_{7/2}$-a$^{4}$D$_{5/2}$ & 22.902 & 12076 & 7.9 & 16.2 & $16.5\pm1.3$&$-222.0\pm5.1$   & $4.7\pm0.4$&$-221.1\pm5.4$    & $2.0\pm0.2$&$-227.6\pm5.5$  & $0.3\pm0.1$&$-79.4\pm64.9$  \\
 $[$Fe \textsc{ii}$]$ a$^{4}$F$_{7/2}$-a$^{4}$F$_{5/2}$ & 24.519 & 4083 & 7.9 & 16.2 & $571.0\pm7.6$&$-198.1\pm1.1$   & $152.1\pm1.8$&$-219.0\pm0.9$    & $67.9\pm0.5$&$-222.7\pm0.6$  & $14.0\pm0.2$&$-172.9\pm1.0$  \\
 $[$Fe \textsc{ii}$]$ a$^{6}$D$_{9/2}$-a$^{6}$D$_{7/2}$ & 25.988 & 554 & 7.9 & 16.2 & $2991.2\pm27.7$&$-174.3\pm0.8$   & $904.4\pm19.2$&$-214.7\pm1.9$    & $422.0\pm5.6$&$-238.9\pm1.1$  & $158.0\pm0.6$&$-196.5\pm0.2$  \\
 $[$Fe \textsc{iii}$]$ $^{5}$D$_{4}$-$^{5}$D$_{3}$ & 22.925 & 627 & 16.2 & 30.7 & $<9.3 $ & -   & $<2.4 $ & -    & $<1.0 $ & -  & $<0.7 $ & -  \\
 $[$Ni \textsc{i}$]$ $^{3}$F$_{4}$-$^{3}$F$_{3}$ & 7.507 & 1917 & 0.0 & 7.6 & $14.1\pm3.1$&$-100.2\pm56.0$   & $<1.2 $ & -    & $<1.0 $ & -  & $<0.6 $ & -  \\
 $[$Ni \textsc{i}$]$ $^{3}$D$_{3}$-$^{3}$D$_{2}$ & 14.814 & 1266 & 0.0 & 7.6 & $<3.9$&		-&			$<0.5$&			-			&			$0,4 \pm 0,1$\quad &$-390,6 \pm 69,2$\quad &$ <0,2$\quad &$-$\\
 $[$Ni \textsc{ii}$]$ $^{2}$D$_{5/2}$-$^{2}$D$_{3/2}$ & 6.636 & 2168 & 7.6 & 18.2 & $880.7\pm29.8$&$-224.2\pm2.1$   & $303.9\pm11.9$&$-251.6\pm2.2$    & $121.5\pm3.8$&$-251.6\pm1.7$  & $46.4\pm1.0$&$-194.2\pm1.0$  \\
 $[$Ni \textsc{ii}$]$ $^{4}$F$_{9/2}$-$^{4}$F$_{7/2}$ & 10.682  & 13424 & 7.6 & 18.2 & $422.2\pm11.3$&$-225.2\pm1.5$   & $72.7\pm3.1$&$-242.3\pm2.3$    & $31.4\pm0.4$&$-240.3\pm0.7$  & $8.1\pm0.2$&$-186.5\pm1.0$  \\
 $[$Ni \textsc{ii}$]$ $^{4}$F$_{7/2}$-$^{4}$F$_{5/2}$ & 12.729  & 14554 & 7.6 & 18.2 & $52.0\pm1.0$&$-199.7\pm1.2$   & $12.9\pm0.4$&$-218.2\pm1.9$    & $5.1\pm0.1$&$-217.2\pm1.0$  & $1.5\pm0.1$&$-156.5\pm1.5$  \\
 $[$Ni \textsc{ii}$]$ $^{4}$F$_{5/2}$-$^{4}$F$_{3/2}$ & 18.241  & 15343 & 7.6 & 18.2 & $5.2\pm0.8$&$-200.5\pm10.6$   & $1.2\pm0.1$&$-199.0\pm8.5$    & $0.6\pm0.0$&$-195.6\pm7.3$  & $0.3\pm0.1$&$-172.9\pm172$  \\
 $[$Ni \textsc{iii}$]$ $^{3}$F$_{3}$-$^{3}$F$_{2}$ & 11.002 & 3265 & 18.2 & 35.2 & $<6.9 $ & -   & $<1.2 $ & -    & $<0.8 $ & -  & $<0.4 $ & -  \\
 $[$Co \textsc{ii}$]$ a$^{3}$F$_{4}$-a3f3 & 10.521 & 1368 & 7.9 & 17.1 & $158.8\pm4.6$&$-135.3\pm2.0$   & $22.1\pm1.0$&$-171.3\pm2.7$    & $8.2\pm0.2$&$-168.5\pm1.5$  & $2.3\pm0.1$&$-114.7\pm1.1$  \\
 $[$Co \textsc{ii}$]$ a$^{5}$F$_{5}$-a$^{5}$F$_{4}$ & 14.736 & 5797 & 7.9 & 17.1 & $9.4\pm1.0$&$-141.0\pm11.8$   & $2.6\pm0.2$&$-145.4\pm7.1$    & $1.1\pm0.1$&$-144.3\pm9.1$  & $0.2\pm0.1$&$-86.1\pm30.8$  \\
 $[$Co \textsc{ii}$]$ a$^{3}$F$_{3}$-a3f2 & 15.456 & 2298 & 7.9 & 17.1 & $3.6\pm0.7$&$-108.0\pm11.4$   & $1.7\pm0.1$&$-133.0\pm4.2$    & $0.6\pm0.1$&$-143.5\pm5.7$  & $0.3\pm0.0$&$-89.7\pm9.9$  \\
 $[$Co \textsc{ii}$]$ a$^{5}$F$_{4}$-a$^{5}$F$_{3}$ & 18.801 & 6562 & 7.9 & 17.1 & $3.9\pm1.6$&$-335.3\pm76.9$   & $1.4\pm0.6$&$-326.4\pm67.2$    & $0.4\pm0.2$&$-315.2\pm86.4$  & $<1.2 $ & -  \\
 $[$Co \textsc{ii}$]$ a$^{5}$F$_{3}$-a$^{5}$F$_{2}$ & 25.685 & 7122 & 7.9 & 17.1 & $<14.8 $ & -   & $0.4\pm0.1$&$-333.6\pm37.7$    & $<1.7 $ & -  & $<1.5 $ & -  \\
 $[$Co \textsc{ii}$]$ b$^{3}$F$_{4}$-b$^{3}$F$_{3}$ & 11.165 & 15407 & 7.9 & 17.1 & $<5.9 $ & -   & $<1.1 $ & -    & $<1.0 $ & -  & $<0.4 $ & -  \\
 $[$Co \textsc{ii}$]$ b$^{3}$F$_{3}$-b$^{3}$F$_{2}$ & 16.297 & 16290 & 7.9 & 17.1 & $<0.8 $ & -   & $<0.4 $ & -    & $<0.2 $ & -  & $<0.1 $ & -  \\
 $[$Ne \textsc{ii}$]$ $^{4}$P$_{3/2}$-$^{4}$P$_{1/2}$ & 12.814 & 1123 & 21.6 & 41.0 & $99.9\pm0.7$&$-241.3\pm0.4$   & $21.0\pm0.2$&$-254.1\pm0.4$    & $11.9\pm0.1$&$-275.4\pm0.4$  & $0.2\pm0.1$&$-223.6\pm19.8$  \\
 $[$Ne \textsc{iii}$]$ $^{3}$P$_{2}$-$^{3}$P$_{1}$ & 15.555 & 925 & 41.0 & 63.5 & $<3.8 $ & -   & $0.1\pm0.0$&$3.1\pm25.0$    & $0.2\pm0.1$&$-119.3\pm78.6$  & $<0.2 $ & -  \\
 $[$Ar \textsc{ii}$]$ $^{2}$P$_{3/2}$-$^{2}$P$_{1/2}$ & 6.985 & 2060 & 15.8 & 27.6 & $5.3\pm0.9$&$-256.1\pm9.0$   & $1.9\pm0.2$&$-247.9\pm5.2$    & $2.2\pm0.3$&$-274.5\pm6.4$  & $<0.6 $ & -  \\
 $[$Ar \textsc{iii}$]$ $^{3}$P$_{2}$-$^{3}$P$_{1}$ & 8.991 & 1600 & 27.6 & 40.7 & $<9.3 $ & -   & $<1.5 $ & -    & $0.9\pm0.3$&$-95.4\pm20.7$  & $<0.4 $ & -  \\
 $[$Ar \textsc{iii}$]$ $^{3}$P$_{1}$-$^{3}$P$_{0}$ & 21.830 & 2259 & 27.6 & 40.7 & $<8.8 $ & -   & $<0.8 $ & -    & $<0.5 $ & -  & $<0.5 $ & -  \\
 $[$S \textsc{i}$]$ $^{3}$P$_{2}$-$^{3}$P$_{1}$ & 25.249 & 570 & 0.0 & 10.4 & $251.2\pm7.6$&$-91.9\pm2.8$   & $64.6\pm1.9$&$-118.6\pm2.8$    & $17.6\pm0.5$&$-172.9\pm3.0$  & $44.9\pm0.9$&$-168.9\pm1.5$  \\
 $[$S \textsc{iii}$]$ $^{3}$P$_{1}$-$^{3}$P$_{2}$ & 18.713  & 1199 & 23.3 & 34.8 & $<3.8 $ & -   & $<1.1 $ & -    & $<0.5 $ & -  & $<0.4 $ & -  \\
 $[$Cl \textsc{i}$]$ $^{2}$P$_{3/2}$-$^{2}$P$_{1/2}$ & 11.333 & 1270 & 0.0 & 13.0  & $94.3\pm4.9$&$-221.5\pm3.3$   & $23.0\pm1.7$&$-236.6\pm5.4$    & $6.6\pm0.3$&$-240.0\pm2.1$  & $5.9\pm0.3$&$-183.9\pm2.2$  \\
 $[$Cl \textsc{ii}$]$ $^{3}$P$_{2}$-$^{3}$P$_{0}$ & 10.035 & 1434 & 13.0 & 23.8 & $<19.4 $ & -   & $<2.2 $ & -    & $<1.5 $ & -  & $<0.6 $ & -  \\
 $[$Cl \textsc{ii}$]$ $^{3}$P$_{2}$-$^{3}$P$_{1}$ & 14.368 & 1001 & 13.0 & 23.8 & $1.9\pm0.2$&$-222.2\pm7.9$   & $0.8\pm0.1$&$-274.6\pm4.5$    & $0.6\pm0.1$&$-309.5\pm7.0$  & $<0.1 $ & -  \\
 H$_2$ (1–1)S(9) & 4.953 & 15721 & 0.0 & 4.5 & $3.8\pm0.7$&$-51.9\pm8.7$   & $6.6\pm2.0$&$-13.2\pm17.7$    & $1.6\pm0.4$&$-24.9\pm13.5$  & $3.0\pm1.2$&$-19.8\pm46.3$  \\
 H$_2$ (1–1)S(8) & 5.330 & 14220 & 0.0 & 4.5 &  $<2.5 $ & -   & $<1.6 $ & -    & $<1.0 $ & -  & $<2.9 $ & -  \\
 H$_2$ (1–1)S(7) & 5.811 & 12817 & 0.0 & 4.5 &  $1.5\pm0.6$&$-72.5\pm26.0$   & $5.2\pm0.3$&$-115.4\pm3.1$    & $2.0\pm0.1$&$-101.9\pm4.2$  & $5.7\pm0.3$&$-154.2\pm3.8$  \\
 H$_2$ (1-1)S(6) & 6.438 & 11521 & 0.0 & 4.5 &  $1.0\pm0.4$&$-101.9\pm28.0$   & $1.5\pm0.1$&$-115.7\pm3.9$    & $0.4\pm0.2$&$-40.9\pm19.6$  & $1.3\pm0.1$&$-144.4\pm6.8$  \\
 H$_2$ (1–1)S(5) & 7.281 & 10341 & 0.0 & 4.5 &  $1.2\pm0.2$&$-84.0\pm8.9$   & $3.1\pm0.1$&$-124.9\pm1.4$    & $0.8\pm0.1$&$-107.2\pm5.8$  & $2.5\pm0.1$&$-155.0\pm3.5$  \\
 H$_2$ (1–1)S(4) & 8.453 & 9286 & 0.0 & 4.5 &  $0.2\pm0.1$&$-51.9\pm19.4$   & $0.6\pm0.1$&$-78.9\pm6.3$    & $0.1\pm0.1$&$-115.1\pm26.8$  & $0.3\pm0.1$&$-106.9\pm12.3$  \\
 H$_2$ (0–0)S(8) & 5.053 & 8677 & 0.0 & 4.5 & $18.2\pm0.9$&$-71.6\pm3.0$   & $34.8\pm1.5$&$-91.3\pm2.0$    & $10.4\pm0.4$&$-81.2\pm2.0$  & $23.5\pm2.7$&$-123.2\pm6.5$  \\
 H$_2$ (0–0)S(7) & 5.511 & 7197 & 0.0 & 4.5 &  $73.1\pm1.4$&$-61.4\pm0.9$   & $137.4\pm1.4$&$-84.3\pm0.5$    & $40.0\pm0.8$&$-71.7\pm1.0$  & $91.4\pm2.0$&$-107.7\pm1.3$  \\
 H$_2$ (0–0)S(6) & 6.109 & 5830 & 0.0 & 4.5 &  $18.4\pm0.4$&$-67.4\pm1.0$   & $42.5\pm0.4$&$-97.0\pm0.5$    & $12.9\pm0.3$&$-74.2\pm1.0$  & $27.8\pm0.8$&$-111.7\pm1.8$  \\
 H$_2$ (0–0)S(5) & 6.910 & 4586 & 0.0 & 4.5 &  $68.3\pm1.3$&$-59.6\pm0.9$   & $125.4\pm1.0$&$-76.1\pm0.4$    & $41.2\pm0.9$&$-63.5\pm1.1$  & $75.1\pm1.8$&$-97.5\pm1.4$  \\
 H$_2$ (0–0)S(4) & 8.025 & 3474 & 0.0 & 4.5 &  $38.1\pm0.4$&$-41.7\pm0.5$   & $43.0\pm0.4$&$-62.0\pm0.5$    & $15.1\pm0.2$&$-48.1\pm0.8$  & $22.3\pm0.4$&$-88.2\pm1.2$  \\
 H$_2$ (0–0)S(3) & 9.665 & 2504 & 0.0 & 4.5 &  $9.4\pm0.1$&$-48.1\pm0.5$   & $29.1\pm0.3$&$-67.7\pm0.5$    & $14.5\pm0.2$&$-48.5\pm0.7$  & $25.9\pm0.4$&$-86.6\pm0.9$  \\
 H$_2$ (0–0)S(2) & 12.279 & 1682 & 0.0 & 4.5 & $9.3\pm0.1$&$-29.7\pm0.7$   & $8.5\pm0.2$&$-61.4\pm1.1$    & $4.2\pm0.1$&$-39.0\pm1.2$  & $5.2\pm0.1$&$-87.9\pm1.3$  \\
 H$_2$ (0–0)S(1) & 17.035 & 1015 & 0.0 & 4.5 & $6.5\pm0.0$&$-40.7\pm0.2$   & $5.9\pm0.0$&$-58.3\pm0.2$    & $2.8\pm0.0$&$-39.8\pm0.6$  & $3.2\pm0.0$&$-84.5\pm0.5$  \\
\hline
\hline
\end{tabular}
\caption{I$_{\lambda}$ is line intensity corrected for extinction reported in Table \ref{tab:appE3_regions}. $v$ is velocity offset of the fitted line from the rest value ($\lambda$) offset from the $v_{LSR} = -4.7$ and corrected for inclination of 50$\degree$. I.P. - ionization potential of the current level (l) and needed to ionize to the next level (u). For H$_2$, the dissociation energy is listed instead \citep{Herzberg.Monfils1961}.}
\end{sidewaystable}
\end{appendix}
\end{document}